%% file: solid_shell_30_arxiv_2025.tex
\documentclass[11pt]{article}
\usepackage{verbatim,amsmath,amssymb}
\usepackage{epsfig,float,color}
\usepackage{geometry}
\usepackage{setspace}
\usepackage{natbib}
\usepackage{amsmath,amssymb,amsthm,mathrsfs,amsfonts,dsfont} 
\usepackage{longtable}
\usepackage{hyperref}
\usepackage{multirow}
\usepackage[bottom]{footmisc}

\definecolor{darkblue}{rgb}{0,0,1}
\definecolor{darkgreen}{rgb}{0,0.7,0}
\hypersetup{pdftex=true, colorlinks=true, breaklinks=true, linkcolor=darkblue, citecolor=darkblue, menucolor=darkblue, pagecolor=darkblue, urlcolor=darkblue}

\input{neco_updated.tex}

\newcounter{remark} 	

\begin{document}

\begin{center}
\Large{\bf{A new rotation-free isogeometric thin shell formulation and a corresponding continuity constraint for patch boundaries}}\\

\end{center}

\begin{center}
\large{Thang X. Duong, Farshad Roohbakhshan and Roger A. Sauer
\footnote{corresponding author, email: roger.sauer@rub.de}}\\
\vspace{4mm}

\small{\textit{Aachen Institute for Advanced Study in Computational Engineering Science (AICES), RWTH Aachen
University, Templergraben 55, 52056 Aachen, Germany}}

\vspace{4mm}

Published\footnote{This pdf is the personal version of an article whose final publication is available at \href{http://dx.doi.org/10.1016/j.cma.2016.04.008}{www.sciencedirect.com}}
in \textit{Computer Methods in Applied Mechanics and Engineering}, \\
\href{http://dx.doi.org/10.1016/j.cma.2016.04.008}{DOI: 10.1016/j.cma.2016.04.008} \\
Submitted on 29.~December 2015, Revised on 7.~April 2016, Accepted on 8.~April 2016

\end{center}


\makebox[\textwidth][s]{\textbf{{This  is a corrected version of the published  journal article.\,\,All corrections  are mark-}}} \par 
\vspace{-2mm}
\makebox[\textwidth][s]{\textbf{{ed in \textcolor{darkgreen}{green}. They were typographical errors  that did not affect simulation results.}} \par}

\vspace{5mm}
%

\rule{\linewidth}{.15mm}
{\bf Abstract:}
This paper presents a general non-linear computational formulation for rotation-free thin shells based on isogeometric finite elements. It is a displacement-based formulation that admits general material models. The formulation allows for a wide range of constitutive laws, including both shell models that are extracted from existing 3D continua using numerical integration and those that are directly formulated in 2D manifold form, like the Koiter, Canham and Helfrich models. Further, a unified approach to enforce the $G^1$-continuity between patches, fix the angle between surface folds, enforce symmetry conditions and prescribe rotational Dirichlet boundary conditions, is presented using penalty and Lagrange multiplier methods. The formulation is fully described in the natural curvilinear coordinate system of the finite element description, which facilitates an efficient computational implementation. It contains existing isogeometric thin shell formulations as special cases. Several classical numerical benchmark examples are considered to demonstrate the robustness and accuracy of the proposed formulation. The presented constitutive models, in particular the simple mixed Koiter model that does not require any thickness integration, show excellent performance, even for large deformations.


\vspace{-4mm}
\rule{\linewidth}{.15mm}
%
\section*{List of important symbols}
\begin{longtable}[l]{ l l }
$ \bone $ & identity tensor in $\bbR^3$ \\
$ a $ & determinant of matrix $[a_{\alpha\beta}]$ \\
$ A $ & determinant of matrix $[A_{\alpha\beta}]$ \\
$\ba_\alpha$ & co-variant tangent vectors of surface $\sS$ at point $\bx$; $\alpha = 1, 2$ \\
$\bA_\alpha$ & co-variant tangent vectors of surface $\sS_0$ at point $\bX$; $\alpha = 1, 2$ \\
$ \ba_{\alpha,\beta}$ & parametric derivative of $\ba_\alpha$ w.r.t.~$\xi^\beta$ \\
$ \ba_{\alpha;\beta}$ & co-variant derivative of $\ba_\alpha$ w.r.t.~$\xi^\beta$ \\
$ a_{\alpha\beta} $ & co-variant metric tensor components of surface $\sS$ at point $\bx$ \\
$ A_{\alpha\beta}$ & co-variant metric tensor components of surface $\sS_0$ at point $\bX$ \\
$ a^{\alpha\beta\gamma\delta}$ & contra-variant components of the derivative of $ a^{\alpha\beta} $ w.r.t.~$a_{\gamma\delta}$ \\
$ b $ & determinant of matrix $[b_{\alpha\beta}]$  \\
$ B $ & determinant of matrix $[B_{\alpha\beta}]$ \\
$ \bb $ & curvature tensor of surface $\sS$ at point $\bx$ \\
$ \bb_\mathrm{0} $ & curvature tensor of surface $\sS_0$ at point $\bX$ \\
$ \bB $ & left Cauchy-Green tensor of the shell mid-surface\\
$ b_{\alpha\beta}$ & co-variant curvature tensor components of surface $\sS$ at point $\bx$ \\
$ B_{\alpha\beta}$ & co-variant curvature tensor components of surface $\sS_0$ at point $\bX$ \\
$ \tilde{b}^{\alpha\beta} $ & contra-variant components of the adjugate tensor of  $b_{\alpha\beta}$ \\ 
$ \mB $ & matrix of the coefficients of the Bernstein polynomials for element $\Omega^e$ \\
$ c $ & bending stiffness \\
$ c^{\alpha\beta\gamma\delta} $ & contra-variant components of the derivative of $\tau^{\alpha\beta}$ w.r.t.~$a_{\gamma\delta}$ \\
$ \bC $ & right Cauchy-Green tensor of the shell mid-surface\\
$ \tilde\bC $ & right Cauchy-Green tensor of a {3D} continuum \\
$ \bC^{*} $ & right Cauchy-Green tensor of a shell layer \\
$\mC^e$ & B{\'e}zier extraction operator for finite element $\Omega^e$ \\
$ \bd $ & shell director vector \\
$ d^{\alpha\beta\gamma\delta} $ & contra-variant components of the derivative of $ \tau^{\alpha\beta} $  w.r.t.~$ b_{\gamma\delta} $ \\
$ \delta ... $ & variation of $...$ \\
$ \epsilon $ & penalty parameter \\
$ \bE $ & Green-Lagrange strain tensor of the shell mid-surface \\
$ e^{\alpha\beta\gamma\delta} $ & contra-variant components of the derivative of $ M_0^{\alpha\beta} $ w.r.t.~$ a_{\gamma\delta} $ \\ 
$ \bff $ & prescribed surface loads \\
$ f^\alpha $ & in-plane components of $ \bff$ \\
$ f^{\alpha\beta\gamma\delta} $ & contra-variant components of the derivative of $ M_0^{\alpha\beta} $ w.r.t.~$ b_{\gamma\delta} $ \\
$ \mf^e_\bullet $ & discretized finite element force vector \\
$ \bF $ & deformation gradient of the shell mid-surface\\
$ \tilde\bF $ & deformation gradient of a {3D} continuum \\
$ g $ & determinant of matrix $[g_{\alpha\beta}]$ \\
$ G $ & determinant of matrix $[G_{\alpha\beta}]$ \\
$ \bg_\alpha $, $ \bg_3 $ & current tangent and normal vectors of a shell layer; $\alpha = 1, 2$ \\
$ \bG_\alpha $, $ \bG_3 $ & reference tangent vectors and normal of a shell layer; $\alpha = 1, 2$ \\
$ g_\mrc $ & $G^1$-continuity and symmetry constraints \\
$ g_{\alpha\beta}$ & co-variant components of the metric tensor of $\sS^*$ \\
$ G_{\alpha\beta}$ & co-variant components of the metric tensor of $\sS_0^*$ \\
$ G_\mathrm{ext} $ & external virtual work \\
$ G_\mathrm{int} $ & internal virtual work \\
$ \Gamma^\gamma_{\alpha\beta} $ & Christoffel symbols of the second kind \\
$ H $ & mean curvature of surface $\sS$ at $\bx$ \\
$ H_0 $ & mean curvature of surface $\sS_0$ at $\bX$ \\
$ \bi $ & identity tensor in $\sS$ \\
$ \bI $ & identity tensor in $\sS_0$ \\
$ I_1 $ & first invariant of $ \bC $ \\
$\tilde{I}_1$ & first invariant of $ \tilde\bC $ \\
$ J $ & surface area change \\
$\tilde{ J} $ & volume change of a {3D} continuum \\
$ J_a $ & Jacobian of the mapping $\sP \rightarrow \sS$ \\
$ J_A $ & Jacobian of the mapping $\sP \rightarrow \sS_0$ \\
$ \mk^e_\bullet $ & finite element tangent matrices \\ 
$ \kappa $ & Gaussian curvature of surface $\sS$ at $\bx$ \\
$ \boldsymbol{\kappa} $ &  pull-back of the curvature tensor $\bb$ \\
$ \bK $ & relative curvature tensor  \\
$ K $ & surface bulk modulus \\
$ \tilde K $ & bulk modulus of {3D} continua \\
$\sL $ & patch boundary on which edge rotation conditions are prescribed \\
$\lambda_i$ & principal stretches of surface $\sS$ at $\bx$ \\
$ \mu $ & surface shear modulus \\
$ \tilde\mu $ & shear modulus of {3D} continua \\
$ \boldsymbol{\mu}_0 $ & moment tensor in the reference configuration \\
$ {\boldsymbol{\mu}} $ & moment tensor in the current configuration \\
$ M^{\alpha\beta} $, $ M_0^{\alpha\beta} $ & contra-variant bending moment components \\
$ \bn $, $ \tilde\bn $ & surface normals of $\sS$ at $\bx$ \\
$ \bN $, $ \tilde\bN $ & surface normals of $\sS_0$ at $\bX$ \\
$ \mN $ & array of NURBS-based shape functions \\
$ \hat{\mN}^e $ & array of B-spline basis functions in terms of the Bernstein polynomials \\
$ \hat{N}^e_A $ & B-spline basis function of the $A^\mathrm{th}$ control point; $A = 1,~...,~n$ \\
$ N^{\alpha\beta} $ & total, contra-variant components of $\bsig$ \\
$ \boldsymbol{\bnu} $ & unit normal on $\partial\mathcal{S}$ \\
$ \xi $ &  out-of-plane coordinate \\
$ \xi^\alpha $ & in-plane coordinates; $\alpha = 1, 2$ \\
$ p $ & external pressure \\
$ \sP $ & parametric domain spanned by $\xi^1$ and $\xi^2$ \\
$ P $ & shell material point \\
$ \Pi_\sL $ & potential of the constraint used to enforce edge rotation conditions \\
$ \bvphi $ & deformation map of surface $\sS$ \\
$ q $ & Lagrange multiplier for the continuity constraint \\
$ \rho $ & areal density of surface $\sS$ \\
$\sS$ & current configuration of the shell surface \\
$\sS_0$ & reference configuration of the shell surface \\
$\sS^*$ & current configuration of a shell layer \\
$\sS_0^*$ & reference configuration of a shell layer \\
$\partial\sS$ & boundary of $\sS$ \\
$ \bS $ & second Piola-Kirchhoff stress tensor of the shell \\
$ \tilde\bS $ & second Piola-Kirchhoff stress tensor of a {3D} continuum \\
$ S^\alpha $ & contra-variant, out-of-plane shear stress components \\
$ \bsig $ &  Cauchy stress tensor of the shell \\
$ \tilde\bsig $ & Cauchy stress tensor of a {3D} continuum \\
$ \sigma^{\alpha\beta} $ & stretch related, contra-variant components of $\bsig$ \\
$ t $ & current shell thickness \\
$ T $ & reference shell thickness \\
$ \bT $ & traction acting on a cut $\partial \sS$ normal to $\bnu$ \\
$\bT^\alpha$ & traction acting on a cut $\partial \sS$ normal to $ \ba^\alpha $ \\
$ \btau $ & unit direction along a surface boundary \\
$ \tilde\btau $ & Kirchhoff stress tensor of a {3D} continuum \\
$ \tau^{\alpha\beta} $ & contra-variant components of the Kirchhoff stress tensor of the shell \\
$ \tilde{\tau}^{\alpha\beta} $ & in-plane components of $ \tilde\btau $ \\
$ \bv $ & velocity, i.e. the material time derivative of $\bx$ \\ 
$ \sV $ & space for admissible variations $\delta\bx$ \\
$ w_A $ & NURBS weight of the $A^\mathrm{th}$ control point ($\hat{=}$ FE node); $A = 1,~...,~n$ \\ 
$ W $ & strain energy density function per reference area \\
$ \tilde{W} $ & strain energy density function per reference volume \\
$ \bx $ & current position of a surface point on $\sS$ \\
$ \bX $ & initial position of $\bx$ on the reference surface $\sS_0$ \\
$ \tilde\bx $ & current position of a material point of a {3D} continuum \\
$ \tilde\bX $ & reference position of a material point of a {3D} continuum \\
$ \mx_e $ & array of all nodal positions for finite element $\Omega^e$ \\
$ \mX_e $ & array of all nodal positions for finite element $\Omega^e_0$ \\
$\Omega^e$ & current configuration of finite element $e$ \\
$\Omega^e_0$ & reference configuration of finite element $e$ \\
\end{longtable}
\section{Introduction}\label{s:intro}

This work presents a new rotation-free isogeometric finite element formulation for general shell structures. The focus here is on solids, even though the formulation generally also applies to liquid shells. The formulation is based on \citet{shelltheo}, who provide a theoretical framework for Kirchhoff--Love shells under large deformations and nonlinear material behavior suitable for both solid and liquid shells. 

From the computational point of view, among the existing shell theories, the rotation-free Kirchhoff--Love shell theory is attractive since it only requires displacement degrees of freedom in order to describe the shell behavior. The necessity of at least $C^1$-continuity across shell elements and their boundaries is the main reason why this formulation is not widely used in practical finite element analysis. Although various efforts have been made for imposing $C^1$-continuity on Lagrange elements (see e.g.~\citet{onate00,brunet06,stolarski13,munglani15} and references therein), the proposed computational formulations are usually either expensive or difficult to implement. Due to this cost and complexity, finite shell elements derived from Reissner--Mindlin theory, which require only $C^0 $-continuity but need additional rotational degrees of freedom, are more widely used \citep{simo89,simo90,bischoff97,yang00,bischoff04,wriggers-fee}. It is worth noting that there are some other formulations that are different from the above prevailing approaches, like extended rotation-free shells including transverse shear effects \citep{zarate12}, rotation-free thin shells with subdivision finite elements \citep{cirak00,cirak01,green05,cirak10}, meshfree Kirchhoff--Love shells \citep{ivannikov14} and discontinuous Galerkin method for Kirchhoff--Love shells \citep{noels08,becker11}. 

Isogeometric analysis (IGA), initially introduced by \citet{hughes05}, has become a promising tool for the computational modeling of shells.  For instance, \citet{benson10,thai12,dornisch13,dornisch14,kang15,lei15} study various Reissner--Mindlin shells with isogeometric analysis. \citet{uhm09} introduce a Reissner--Mindlin shell described by T-splines. The hierarchic family of isogeometric shell elements presented by \citet{echter13} includes 3-parameter (Kirchhoff--Love), 5-parameter (Reissner--Mindlin) and 7-parameter (three-dimensional shell) models. Solid-shell elements based on isogeometric NURBS are investigated by \cite{bouclier13,bouclier13b,hosseini13,hosseini14,bouclier15,du215}. The shell formulation of \citet{benson13} blends both Kirchhoff--Love and Reissner--Mindlin theories. 

Particularly for rotation-free thin shells, which are the focus of this research, isogeometric analysis is a great help. This is due to the fact that the IGA discretization can provide smoothness of any order across elements, allowing an efficient yet accurate surface description, which is suitable for thin shell structures. The first work on combining IGA with Kirchhoff--Love theory is presented by \citet{kiendl09}. Later, \citet{kiendl10} use the bending strip method to impose the $C^1$-continuity of Kirchhoff--Love shell structures comprised of multiple patches. \citet{thanh11} then propose PHT-splines for rotation-free shells. The approach is tested for a linear shell formulation and it demonstrates the advantage of providing $C^1$-continuity for thin shells. In the rotation-free shell formulation suggested by \citet{benson11}, the Kirchhoff--Love assumptions are satisfied only at discrete points, so that the required continuity can be lower. \citet{nagy13} propose an isogeometric design framework for composite Kirchhoff--Love shells with anisotropic material behavior. \citet{goyal13} investigate the dynamics of Kirchhoff--Love shells discretized by NURBS. \citet{nguyen15} propose an extended isogeometric element formulation for the analysis of through-the-thickness cracks in thin shell structures based on Kirchhoff--Love theory. \citet{deng15} suggest a rotation-free shell formulation equipped with a damage model. For thin biological membranes, a thin shell formulation is developed by \citet{tepole15}. \citet{riffnaller16} present a discretization of Kirchhoff--Love thin shells based on a subdivision algorithm. Weak Dirichlet boundary conditions of isogeometric rotation-free thin shells are considered by \citet{guo15b}. \citet{lei15b} introduce a penalty and a static condensation method to enforce the $C^0/G^1$-continuity for NURBS-based meshes with multiple patches. Recently, isogeometric collocation methods have been also introduced for Kirchhoff--Love and Reissner--Mindlin plates as an alternative for isogeometric Galerkin approaches \citep{kiendl15b,reali15}.

Recently, \citet{kiendl15} extended the proposed formulation of \citet{kiendl09} to non-linear material models. By using numerical integration over the thickness of 3D continua, the extended formulation admits arbitrary nonlinear material models. However, the formulation of \citet{kiendl15} leads to special relations for the membrane stresses and bending moments. In general, such a (fixed) relation is rather restricted.  An example is cell membranes composed of lipid bilayers (see e.g.~\citet{shelltheo}). In this case, the material behaves like a fluid in the in-plane direction, i.e. without elastic resistance, while in the out-of-plane direction the material behaves like a solid with elastic resistance. Hence, the membrane and bending response may range from fully decoupled to very complicated relations. Additionally, for fluid materials, it is desired to include other conditions such as area incompressibility and stabilization techniques. Therefore, a further extension of the formulation of \citet{kiendl15} is needed. 

Besides, in computation of thin shells, it is beneficial to accommodate material models that are directly constructed in surface strain energy form, like the Koiter model \citep{ciarlet05}, Canham model \citep{canham70} or Helfrich model \citep{helfrich73}. In these models, contrary to some of the approaches mentioned above, no numerical integration is required such that the computational time reduces drastically.

In this paper, we develop a general nonlinear IGA thin shell formulation. The surface formulation, presented in Sec.~\ref{s:consti}, admits any (non)linear material law with arbitrary relation between bending and membrane behavior, while the models of \citet{kiendl09,kiendl10,thanh11,echter13,guo15,lei15} are based on linear {2D} stress-strain relationships. The {3D} formulation, presented in Sec.~\ref{s:consituive3D}, can be reduced to \citet{benson11,kiendl15} as special cases. However, the model of \citet{benson11} is restricted to single patch shells and the continuity constraint of \citet{kiendl15} is different than our constraint. Further, our approach to model symmetry and clamped boundaries is distinct from all the existing rotation-free IGA shell formulations. In summary, our formulation contains the following new items:
\vspace{-\topsep}
\bitm
\setlength{\itemsep}{0pt}
\setlength{\parskip}{0pt}
\setlength{\parsep}{0pt}
\item The bending and membrane response can be flexibly defined at the constitutive level.
\item It admits both constitutive laws obtained by numerical integration over the thickness of 3D material models and those constructed directly in surface energy form. 
\item It can thus be used for both solid and liquid shells.
\item It is fully described in curvilinear coordinates, which includes the constitutive laws, FE weak form and corresponding FE matrices. 
\item It includes a consistent treatment for the application of boundary moments.
\item It includes an efficient finite element implementation of the formulation.
\item It includes a unified treatment of edge rotation conditions such as the $G^1$-continuity between patches, symmetry conditions and rotational boundary conditions.
\eitm \vspace{-\topsep}

The remaining part of this paper is organized as follows: Sec.~\ref{s:theory} summarizes the theory of rotation-free thin shells, including the kinematics, balance laws, strong and weak forms of the governing equations, as well as remarks on constitutive laws. In Sec.~\ref{s:consituive3D}, we present a concise and systematic procedure to extract shell constitutive relations from existing 3D material models. Sec.~\ref{s:fe} discusses the finite element discretization as well as the treatment of symmetry, surface folds and $ G^1 $-continuity constraints for multi-patch NURBS. Several linear and nonlinear benchmark tests are presented in Sec.~\ref{s:examples} to illustrate the capabilities of the new model. These examples consider solid shells. Liquid shells will be presented in future work \citep{liquidshell}. Sec.~\ref{s:con} concludes the paper.
\section{Summary of rotation-free thin shell theory}\label{s:theory}

This section summarizes the theory of nonlinear shells in the unified framework presented in \citet{shelltheo} and references therein. Here, the shells are treated mathematically as {2D} manifolds. Later, in Sec.~\ref{s:consituive3D}, the kinematics and constitutive formulations are defined in a {3D} setting considering different shell layers through the thickness. These two different approaches are schematically illustrated in Fig.~\ref{f:mapping}.  	

\subsection{Thin shell kinematics and deformation}\label{s:kin}
\begin{figure}[ht]
\vspace{-0.3cm}
\begin{center} \unitlength1cm
\includegraphics[height=70mm]{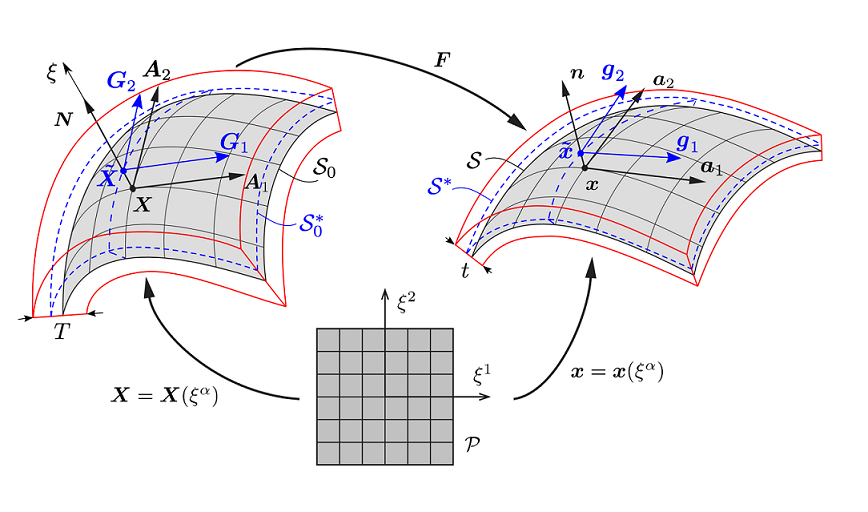}
\caption{Mapping between parameter domain $\sP$, reference surface $\sS_0$ and current surface $\sS$ of a Kirchhoff--Love shell. The boundaries of the physical shell are shown by solid red lines. A shell layer is denoted by $\sS^*$ and $\sS_0^*$ in the current and reference configuration, respectively, and is shown by dashed blue lines.}
\label{f:mapping}
\end{center}
\vspace{-0.3cm}
\end{figure}
As shown in Fig.~\ref{f:mapping}, let the mid-surface $\sS$ of a thin shell be described by a general mapping
\eqb{l}
\bx=\bx(\xi^\alpha)~, ~\alpha=1,\,2,
\eqe
where $\bx$ denotes the surface position in 3D space and $\xi^\alpha$ are curvilinear coordinates embedded in the surface. They can be associated with a parameter domain $\sP$. Differential geometry allows us to determine the co-variant tangent vectors $\ba_\alpha=\partial{\bx}/\partial\xi^\alpha$ on $\sS$, their dual vectors $\ba^\alpha$ defined by $\ba^\alpha\cdot\ba_\beta  = \delta^\alpha_\beta$, and the surface normal $\bn = (\ba_1\times\ba_2)/||\ba_1\times\ba_2||$, so that the metric tensor with co-variant components $a_{\alpha\beta}=\ba_\alpha\cdot\ba_\beta$ and contra-variant components $[a^{\alpha\beta}] = [a_{\alpha\beta}]^{-1}$ of the first fundamental form are defined. The relation between area elements on $\sS$ and the parameter domain is $\dif a = J_a\,\dif\xi^1\,\dif\xi^2$ with $J_a=\sqrt{\det a_{\alpha\beta}}$.

The bases $\{\ba_1,\, \ba_2,\, \bn\}$ and $\{\ba^1,\, \ba^2,\, \bn\}$ define the usual identity tensor $\bone$ in $\bbR^3$ as $\bone=\bi+\bn\otimes\bn$ with $\bi=\ba_\alpha\otimes\ba^\alpha=\ba^\alpha\otimes\ba_\alpha$. The components of the curvature tensor $\bb = b_{\alpha\beta}\,\ba^\alpha\otimes\ba^\beta$ are given by the Gauss--Weingarten equation
\eqb{l}
b_{\alpha\beta}=\bn\cdot\ba_{\alpha,\beta}=-\bn_{,\beta}\cdot\ba_\alpha~,
\label{e:bab}\eqe
where a comma denotes the parametric derivative $\ba_{\alpha,\beta} = \partial\ba_\alpha/\partial\xi^\beta$. With the definition of the Christoffel symbol $\Gamma^\gamma_{\alpha\beta}=\ba^\gamma\cdot\ba_{\alpha,\beta}$, the so-called co-variant derivative is defined as $\ba_{\alpha;\beta} := \ba_{\alpha,\beta} - \Gamma^\gamma_{\alpha\beta}\,\ba_\gamma = (\bn\otimes\bn)\,\ba_{\alpha,\beta}$. The mean and Gaussian curvature of $\sS$ can be computed from the first and second invariants of the curvature tensor $\bb$, respectively, as
\begin{equation}
	H := \ds\frac{1}{2}\tr{\bb} = \frac{1}{2}\,b^\alpha_\alpha = \frac{1}{2}\,a^{\alpha\beta}\,b_{\alpha\beta}~,
\label{e:meanH}	
\end{equation}
and
\begin{equation}
\kappa := \det{\bb} = \frac{b}{a}~,
\label{e:kappa}\end{equation}
where
\begin{equation}
	a = \det[a_{\alpha\beta}]~, \quad
	b = \det[b_{\alpha\beta}]~.
\end{equation}

The variations of the above quantities such as $\delta\auab$, $\delta\aab$, $\delta \buab$, $\delta\bab $, $\delta H$ and $\delta \kappa$ can be found e.g.~in \citet{shelltheo}.

To characterize the deformation of $\sS$ due to loading, the initially undeformed configuration is chosen as a reference configuration and is denoted by $\sS_0$. It is described by the mapping $\bX=\bX(\xi^\alpha)$, from which we have $\bA_\alpha=\partial{\bX}/\partial\xi^\alpha$, $A_{\alpha\beta}=\bA_\alpha\cdot\bA_\beta$, $[A^{\alpha\beta}] = [A_{\alpha\beta}]^{-1}$, $\bA^\alpha = A^{\alpha\beta}\,\bA_\beta$, $\bN = (\bA_1\times\bA_2)/||\bA_1\times\bA_2||$, the initial identity tensor $\bI:=\bA_\alpha\otimes\bA^\alpha = \bA^\alpha\otimes\bA_\alpha$, where $\bone=\bI+\bN\otimes\bN$, and the initial curvature tensor $\bb_0:= B_{\alpha\beta}\,\bA^\alpha\otimes\bA^\beta$.

Having the definition of $\auab$ and $\buab$, the deformation map $\bx=\bvphi(\bX)$ is fully characterized 
by the two following quantities: \\
1. The surface deformation gradient, which is defined as
\eqb{l}
\bF:=\ba_\alpha\otimes\bA^\alpha~.
\eqe
Accordingly, the right Cauchy-Green surface tensor, 
$\bC=\bF^T\bF=a_{\alpha\beta}\,\bA^\alpha\otimes\bA^\beta$, its inverse $\bC^{-1}=a^{\alpha\beta}\,\bA_\alpha\otimes\bA_\beta$, and the left Cauchy-Green surface tensor, $\bB=\bF\bF^T=A^{\alpha\beta}\,\ba_\alpha\otimes\ba_\beta$, are defined. They have the two invariants $I_1=  \bC : \bI = \bB: \bi = \Aab\,\auab$ and $J^2 = \det\bC = \det\bB$. Therefore, similar to the volumetric case, the surface Green-Lagrange strain tensor can be defined as
\eqb{l}
\bE = E_{\alpha\beta}\,\bA^\alpha\otimes\bA^\beta  := \ds\frac{1}{2}(\bC - \bI) = \ds\frac{1}{2}(\auab - \Auab)\,\bA^\alpha\otimes\bA^\beta~,
\label{e:Eten}
\eqe
which represents \textit{the change of the metric tensor} due to the surface deformation.\\
2. The symmetric relative curvature tensor, which is defined as 
\eqb{l}
\bK = K_{\alpha\beta}\,\bA^\alpha\otimes\bA^\beta:= \boldsymbol{\kappa} - \bb_0 =  (b_{\alpha\beta} - B_{\alpha\beta})\,\bA^\alpha\otimes\bA^\beta,
\label{e:Kten}
\eqe
with
\eqb{l}
\boldsymbol{\kappa} := \bF^T\,\bb\,\bF = b_{\alpha\beta}\,\bA^\alpha\otimes\bA^\beta~,
\eqe
which furnishes \textit{the change of the curvature tensor field}, following the terminology of \citet{steigmann99}.

\subsection{Stress and moment tensors}\label{s:stress}
In order to define stresses at a material point $\bx$ of the shell surface $\sS$, the shell is cut into two at $\bx$ by the parametrized curve $\mathcal{C}(s)$.\footnote{$s$ denotes the arc length such that $\dif s = \norm{\dif \bx}$.}  Further, let $\btau := \partial\bx / \partial s$ be the unit tangent and $\bnu := \btau\times\bn$ be the unit normal of $\mathcal{C}$ at $\bx$. Then, one can define the traction vector $\bT=\bT^\alpha\,\nu_\alpha$ and the moment vector $\bM=\bM^\alpha\,\nu_\alpha$ on each side of $\mathcal{C}(s)$ at $\bx$, where $\nu_\alpha = \bnu\cdot\ba_\alpha$. According to Cauchy's theorem, these vectors can be linearly mapped to the normal $\bnu$ by second order tensors as
\eqb{lll}
\bT := \bsig^\mrT\bnu~,\\[2mm]
\bM := \bmu^\mrT\,\bnu~,
\label{e:T}\eqe
where 
\eqb{lll}
\bsig := N^{\alpha\beta} \, \ba_{\alpha} \otimes \ba_{\beta} + S^{\alpha} \, \ba_{\alpha} \otimes \bn~,\\[2mm]
\bmu := -M^{\alpha\beta}\,\ba_\alpha\otimes\ba_\beta~
\label{e:sigmaM}
\eqe
are the Cauchy stress and moment tensors, respectively. 

It should be noted that the Cauchy stress $\bsig$ in Eq.~(\ref{e:sigmaM}.1) is generally not symmetric. Furthermore, as shown e.g.~in \citet{shelltheo}, a suitable work conjugation (per reference area) arising in the presented theory is the quartet
\eqb{l}
\dot{W}_\mathrm{int} :=\ \tauab\dot{a}_{\alpha\beta}/2 + \Mab_0\dot{b}_{\alpha\beta}~,
\eqe
where $\dot{W}_\mathrm{int}$ is the local power density and we have defined
\eqb{lll}
\tau^{\alpha\beta} \dis J\sigma^{\alpha\beta}~, \\[1mm] 
M_0^{\alpha\beta} \dis JM^{\alpha\beta}~.
\label{e:tauM0}\eqe
Further,
\eqb{lll}
\sigma^{\alpha\beta} :=N^{\alpha\beta}  - b^\beta_\gamma \, M^{\gamma\alpha}
\label{e:sigmah}
\eqe
are the components of the membrane stress responding to the change of the metric tensor.\footnote{Note that $\sigma^{\alpha\beta} \neq \ba^\alpha\cdot\bsig\,\ba^\beta$, instead $\ba^\alpha\cdot\bsig\,\ba^\beta = N^{\alpha\beta}$.} 

We also note that the mathematical quantities introduced in Eqs.~\eqref{e:T} and \eqref{e:sigmaM} stem from the Cosserat theory for thin shells \citep{steigmann99}. As shown e.g.~by \citet{steigmann99b} and \citet{shelltheo}, these quantities can be related to the effective traction $\bt$ and physical moment $\bm$ transmitted across $\mathcal{C}(s)$. Namely for $\bm$, the relation is given by
\eqb{l}
\bm:=\bn\times\bM = m_\nu\,\bnu + m_\tau\,\btau,
\label{e:m}\eqe
where
\eqb{lll}
m_\nu \dis M^{\alpha\beta}\,\nu_\alpha\,\tau_\beta~, \\[1mm]
m_\tau \dis - M^{\alpha\beta}\,\nu_\alpha\,\nu_\beta~,
\label{e:m_comp}\eqe
are the normal and tangential bending moment components along $\mathcal{C}(s)$. For the effective traction $\bt$, one can show that \citep{steigmann99,shelltheo} 
\eqb{l}
\bt := \bT -\left(m_\nu\,\bn\right)'~,
\label{e:bt}
\eqe
where $(\bullet)'$ denotes the derivative with respect to $s$.

\subsection{Balance laws}
To derive the governing equations, the surface $\sS$ is subjected to the prescribed body force  $\bff = f^\alpha\,\vaua + p\,\bn$ on $ \sS$ and the boundary conditions
\eqb{rllll}
\bu \is \bar\bu & $on$~\partial_u\sS~, \\[1mm]
\bn \is \bar\bn & $on$~\partial_n\sS~, \\[1mm]
\bt \is \bar\bt & $on$~\partial_t\sS~, \\[1mm]
m_\tau \is \bar{m}_\tau & $on$~\partial_m\sS~,
\eqe
where $ \bar\bu $ is a prescribed displacement; $\bar\bn$ is a prescribed rotation, $\bar\bt:=\bar t^\alpha\,\ba_\alpha$ is a prescribed boundary traction and $\bar{m}_\tau$ is a prescribed bending  moment. The equilibrium of the shell is then governed by the balance of linear momentum  together with mass conservation, which gives
\eqb{lll}
\bT^{\alpha}_{;\alpha} + \bff =
\rho\,\dot\bv\quad\forall\,\bx\in\sS~,
\label{e:sf}
\eqe
while the balance of angular momentum leads to the condition that $\sigma^{\alpha\beta}$, defined in Eq.~\eqref{e:sigmah}, is symmetric and the shear stress is related to the bending moment via
\eqb{lll}
S^\alpha = - M^{\beta\alpha}_{~~;\beta}~.
\label{e:Sa}
\eqe
We note that $M^{\alpha\beta}$ is also symmetric but  $N^{\alpha\beta}$ is generally not symmetric. 

\subsection{Weak form}
The weak form of Eq.~\eqref{e:sf} can be obtained by contracting Eq.~\eqref{e:sf} with the admissible variation $\delta\bx\in\sV$ and
integrating over $\sS$, (see \citet{shelltheo}). This results in
\eqb{l}
G_\mathrm{in} + G_\mathrm{int} - G_\mathrm{ext} = 0 \quad\forall\,\delta\bx\in\sV~,
\label{e:wfu}\eqe
where
\eqb{lll}
G_\mathrm{in} 
\is \ds\int_{\sS_0} \delta\bx\cdot\rho_0\,\dot\bv\,\dif A~, \\[4mm]
G_\mathrm{int} \is  \ds\int_{\sS_0} \frac{1}{2}\,\delta a_{\alpha\beta} \, \tau^{\alpha\beta} \, \dif A  + \int_{\sS_0} \delta b_{\alpha\beta} \, M_0^{\alpha\beta} \, \dif A~ = \ds\int_{\sS_0} \delta E_{\alpha\beta} \, \tau^{\alpha\beta} \, \dif A 
+ \int_{\sS_0} \delta K_{\alpha\beta} \, M_0^{\alpha\beta} \, \dif A~,\\[4mm]
G_\mathrm{ext} \is \ds\int_{\sS}\delta\bx\cdot\bff\,\dif a 
+ \ds\int_{\partial_t\sS} \delta\bx\cdot\bt\,\dif s +  \ds\int_{\partial_m\sS}\delta\bn\cdot m_\tau\,\bnu\,\dif s + [\delta\bx\cdot m_\nu\,\bn\big]~.
\label{e:Giie}\eqe

\refstepcounter{remark}
\textbf{Remark \arabic{remark}:} Here, the last term in $G_\mathrm{ext}$ is the virtual work of the point load $m_\nu\,\bn$ at corners of $\partial_m\sS $ (in case $m_\nu\neq0$), and the second last term  in $G_\mathrm{ext}$ denotes the virtual work of the moment $m_\tau\,\bnu$. It is worth noting that this is the consistent way to apply bending moments on the boundary of Kirchhoff--Love shells.

\subsection{Linearization of the weak form}
For solving the nonlinear equation \eqref{e:wfu} by the Newton--Raphson method, its linearization is needed. For the kinetic virtual work, one gets
\eqb{lll}
\Delta G_\mathrm{in} 
\is \ds\int_{\sS_0} \delta\bx\cdot\rho_0\,\Delta \dot\bv\,\dif A~,
\eqe
in  which $\Delta \dot\bv$ depends on the time integration scheme used. For dead loading of $\bff$, $\bt$ and $m_\tau\,\bnu$, one finds\footnote{For live $m_\tau\,\bnu$, see Eqs.~\eqref{e:DGextM} and \eqref{e:kextM}.}
\eqb{l}
\Delta G_\mathrm{ext} = \ds\int_{\partial\sS}m_\tau\,\delta\ba_\alpha\cdot\big(\nu^\beta\,\bn\otimes\ba^\alpha + \nu^\alpha\,\ba^\beta\otimes\bn\big)\,\Delta\ba_\beta\,\dif s
\eqe
and for the internal virtual work term
\eqb{llllll}
\Delta G_\mathrm{int} 
= \ds\int_{\sS_0}\Big(\hspace{-2ex}&c^{\alpha\beta\gamma\delta}\,\frac{1}{2}\delta\auab\,\frac{1}{2}\Delta\augd 
\plus d^{\alpha\beta\gamma\delta}\,\frac{1}{2}\delta\auab\,\Delta\bugd 
\plus \tauab\,\frac{1}{2}\Delta\delta\auab \;+ \\[3mm]
& e^{\alpha\beta\gamma\delta}\,\delta\buab\,\frac{1}{2}\Delta\augd 
\plus f^{\alpha\beta\gamma\delta} \,\delta\buab\,\Delta\bugd
\plus \Mab_0\,\Delta\delta\buab \Big)\,\dif A~,
\label{e:dxdxWS}\eqe
where the material tangent matrices are defined as
\eqb{lllll}
c^{\alpha\beta\gamma\delta}\dis 2\ds\frac{\partial \tau^{\alpha\beta}}{\partial a_{\gamma\delta}}~, \quad ~
d^{\alpha\beta\gamma\delta}\dis \ds\frac{\partial \tau^{\alpha\beta}}{\partial b_{\gamma\delta}}~, \\[2mm]
e^{\alpha\beta\gamma\delta}\dis 2\ds\frac{\partial M_0^{\alpha\beta}}{\partial a_{\gamma\delta}}~,\quad
f^{\alpha\beta\gamma\delta}\dis \ds\frac{\partial M_0^{\alpha\beta}}{\partial b_{\gamma\delta}}~.
\eqe
They are given in \citet{shelltheo} for various material models. Here it is noted that $d^{\alpha\beta\gamma\delta} = e^{\gamma\delta\alpha\beta}$ posses minor symmetries, while $c^{\alpha\beta\gamma\delta}$ and $f^{\alpha\beta\gamma\delta}$ posses both minor and major symmetries.

\subsection{Constitution}\label{s:consti}
In this framework, $\sigma^{\alpha\beta}$ and $M^{\alpha\beta}$ are determined from constitutive relations for stretching and bending. For hyperelastic shells, we assume there exists a stored energy function in the form
\eqb{l}
W = W(\bE,\bK),
\label{e:WEK}
\eqe
where $\bE$ and $\bK$ are defined in Eqs.~\eqref{e:Eten} and \eqref{e:Kten}. The Koiter model (see Eq.~\eqref{e:Koiter}) is a simple example of this form. It is easily seen that $W$ can be equivalently expressed as a function of $\bC$ and $\boldsymbol{\kappa}$, or as a function of $a_{\alpha\beta}$ and $b_{\alpha\beta}$, since $\bA^\alpha$ is constant. If the material has symmetries, e.g.~isotropy, the stored energy function can also be expressed as a function of the invariants $I_1$, $J$, $H$ and $\kappa$, defined in Sec.~\ref{s:kin} (see e.g.~\citet{steigmann99b}). Thus, the following functional forms are equivalent to Eq.~\eqref{e:WEK}.   
\eqb{l}
W = W(\bC,\boldsymbol{\kappa}) = W(a_{\alpha\beta},b_{\alpha\beta})= W(I_1,J,H, \kappa)~.
\label{e:WEK2}
\eqe
Given the stored energy function $W$, the common argument by \citet{coleman64} leads to the constitutive equations (see \citet{shelltheo})
\eqb{llrlrlr}
\tau^{\alpha\beta} \is \ds2\,\pa{W}{a _{\alpha\beta}}~,\\[4mm]
M_0^{\alpha\beta} \is \ds\pa{W}{b _{\alpha\beta}}~.
\label{e:consticom}\eqe

\refstepcounter{remark}
\label{r:rm2}
\textbf{Remark \arabic{remark}:} The constitutive equations \eqref{e:consticom} are defined in curvilinear coordinates. Therefore, they either give the components of the classical second Piola-Kirchhoff stress and moment 
\eqb{lll}
\bS \dis \ds \pa{W(\bE,\bK)}{\bE} =\ds 2\,\pa{W}{\auab}\,\bA_\alpha\otimes\bA_\beta =:  \tau^{\alpha\beta}\,\bA_\alpha\otimes\bA_\beta~, \\[4mm]
\boldsymbol{{\mu}_0}\dis -\ds\pa{W(\bE,\bK)}{\boldsymbol{\bK}} = - \ds \pa{W}{\buab}\,\bA_\alpha\otimes\bA_\beta =: -M_0^{\alpha\beta}\,\bA_\alpha\otimes\bA_\beta,
\label{e:piola}
\eqe
of the surface manifold. Alternatively, they give the components of the Kirchhoff stress and moment defined by pushing forward $\bS$ and $\boldsymbol{{\mu}_0}$ as
\eqb{lll}
\bF\,\bS\,\bF^T \dis  \tau^{\alpha\beta}\,\ba_\alpha\otimes\ba_\beta~, \\[4mm]
\bF\,\boldsymbol{\mu_0}\,\bF^T \dis - M_0^{\alpha\beta}\,\ba_\alpha\otimes\ba_\beta =  J\,\boldsymbol{\mu}.
\eqe
Here, we note that $\pa{(\bullet)}{\bX}$ in Eq.~\eqref{e:piola} denotes the derivative of a scalar-valued function by an arbitrary second order tensor $\bX$ (see e.g.~\citet{itskov2009}).

\refstepcounter{remark}
\textbf{Remark \arabic{remark}:}  In this framework, the stress $\sigma^{\alpha\beta}$ and moment $M^{\alpha\beta}$  are both constitutively determined from a surface stored energy function $W$.  The advantage of this setup is that it can accept general constitutive relations, which implies that the relation between the membrane and bending response can be flexibly realized at the constitutive level and it is not restricted within the formulation. The stretching and bending response may range from fully decoupled, like in the Koiter material model (Eq.~\eqref{e:Koiter}), to a coupled relation, such as in the Helfrich material model (see e.g.~\citet{shelltheo}). Due to this flexibility, the presented formulation is suitable for both solid and liquid shells and is also convenient for adding kinematic constraints (e.g.~area constraint) or stabilization potentials \citep{liquidshell}.

\refstepcounter{remark}
\textbf{Remark \arabic{remark}:} In the presented formulation, we note that the definition of the shell thickness is not needed for solving Eq.~\eqref{e:wfu}, which is often referred to as a ``zero-thickness'' formulation. Consequently, the stress and moment in Eq.~\eqref{e:consticom} can be computed without defining a thickness. However, this does not imply that thickness effects have been neglected or approximated. Instead, they are in some sense hidden in the constitutive law of Eq.~\eqref{e:WEK}. 
If desired, they can be determined as noted in Remark~\ref{r:koi}. But this connection is not a requirement as it is in the case of constitutive laws derived from {3D} material models (see Sec.~\ref{s:koiterst}).

In the following, we consider some example constitutive models suitable for solid shells to demonstrate the flexibility of the formulation. The application of the formulation to liquid shells is considered in future work \citep{liquidshell}. 

\subsubsection{Initially planar shells} \label{s:canham}

For initially planar shells, one can consider the model of \citet{canham70} for the bending contribution, while for the membrane contribution, the stretching response can be modeled with a nonlinear Neo-Hookean law. This gives
\eqb{l}
W =  \ds\frac{\Lambda}{4}(J^2 - 1 - 2\ln J) + \ds\frac{\mu}{2}(I_1 - 2 - 2\ln J) + \ds c\,J\,\big(2\,H^2 - \kappa\big) ~,
\label{e:WmsF}
\eqe
where $c$, $\Lambda$ and $\mu$ are {2D} material constants. From Eqs.~\eqref{e:consticom} and \eqref{e:WmsF}, we find the stress
\eqb{l}
\tauab = \ds\frac{\Lambda}{2}\big(J^2-1\big)\aab + \mu\Big(\Aab - \aab\Big) + c\,J\, (2H^2+\kappa)\,a^{\alpha\beta} - 4c\,J\, H\,b^{\alpha\beta}  ~,
\label{e:sig_bs}\eqe
and the moment
\eqb{l}
M^{\alpha\beta}_0 = c\,J\,b^{\alpha\beta} ~,
\label{e:M_bs}\eqe
which is linear w.r.t.~the curvature. The tangents for this are given in \citet{shelltheo}. We note that since Eq.~\eqref{e:WmsF} is given in surface strain energy form, the stress and moment can be computed directly from Eqs.~\eqref{e:sig_bs} and~\eqref{e:M_bs} without needing any thickness integration.

\subsubsection{Initially curved solid shell}\label{s:koiter}
For initially curved shells, the surface strain energy model proposed by Koiter \citep{ciarlet05,steigman13} can be considered. It is defined in tensor notation as
\eqb{l}
W(\bE,\bK) = \ds\frac{1}{2}\, \bE : \bbC : \bE 
+ \ds\frac{1}{2}\,\bK : \bbF:\bK ~,
\label{e:Koiter}\eqe
with the constant fourth order tensors
\eqb{lll}
\bbC \is \Lambda\,\bI\odot \bI + 2\, \mu\, (\bI \otimes\bI)^\mrs ~, \\[2mm]
\bbF\is \ds\frac{T^2}{12}\,\bbC~.
\label{e:Ckoi}
\eqe

Since $\bI \odot \bI : \bX  = (\tr\bX)\,\bI $ and $(\bI\otimes \bI)^\mrs : \bX  = \mathrm{sym}(\bX)$, with $\bX$ being an arbitrary second order tensor, it follows from Eqs.~\eqref{e:Koiter} and \eqref{e:Ckoi} that
\eqb{lll}
\tau^{\alpha\beta} \is \Lambda\,\tr\bE\,\Aab + 2\,\mu\,E^{\alpha\beta} ~, \\[2mm]
M_0^{\alpha\beta} \is  \ds\frac{T^2}{12}(\Lambda\,\tr\bK\,\Aab + 2\,\mu\, K^{\alpha\beta}),
\label{e:stresKoi}
\eqe
where  $K^{\alpha\beta}:= A^{\alpha\gamma}\,K_{\gamma\delta}\, A^{\delta\beta}$,  $E^{\alpha\beta}:= A^{\alpha\gamma}\,E_{\gamma\delta}\, A^{\delta\beta}$, and
\eqb{lll}
\tr\bK = K^{\alpha\beta}\,A_{\alpha\beta}~,\quad \tr\bE = E^{\alpha\beta}\,A_{\alpha\beta}~.
\eqe
From Eq.~\eqref{e:stresKoi}, we further find $d^{\alpha\beta\gamma\delta} = e^{\alpha\beta\gamma\delta} = 0 $ and
\eqb{lllll}
c^{\alpha\beta\gamma\delta} \is \Lambda\,A^{\alpha\beta}\,A^{\gamma\delta}
+ \mu\big(A^{\alpha\gamma}\,A^{\beta\delta}+A^{\alpha\delta}\,A^{\beta\gamma} \big)~, \\[2mm]
f^{\alpha\beta\gamma\delta}  \is \ds\frac{T^2}{12}\,c^{\alpha\beta\gamma\delta}~.
\eqe

\refstepcounter{remark}\label{r:koi}
\textbf{Remark \arabic{remark}:} The set of parameters $\Lambda$ and $\mu$ can be determined in different ways. Firstly, they can be determined directly from experiments, i.e.~without explicitly considering the thickness. Secondly, they may be obtained by analytical integration over the thickness of the simple {3D} Saint Venant--Kirchhoff model (see e.g.~\citet{ciarlet05} and Sec.~\ref{s:koiterst}). In this case, $\mu$ and $\Lambda$ are given by 
\eqb{lll}
\Lambda := \ds T\,\frac{2\,\tilde{\Lambda}\,\tilde{\mu}}{\tilde{\Lambda} + 2\,\tilde{\mu}},\quad \mu:= T\,\tilde{\mu},
\label{e:rel2D3D}
\eqe
where $\tilde{\Lambda}$ and $\tilde{\mu}$ are the classical {3D} Lam{\'e} constants in linear elasticity. Thirdly, they can be determined by numerical integration over the thickness of a general {3D} material model.

\refstepcounter{remark}
\textbf{Remark \arabic{remark}:} Note that, in  Eq.~\eqref{e:Koiter}, the stretching and the bending behavior can be specified separately. For instance, the first term can be replaced by a nonlinear Neo-Hooke model (see Eq.~\eqref{e:WmsF}), i.e.~   
\eqb{l}
W = \ds\frac{\Lambda}{4}(J^2 - 1 - 2\ln J) + \ds\frac{\mu}{2}(I_1 - 2 - 2\ln J) + \ds\frac{1}{2}\,\bK : \bbF:\bK ~,
\label{e:KoiterNH}\eqe
in order to capture large membrane strains. In this case $\tau^{\alpha\beta}$ follows from the front part of Eq.~\eqref{e:sig_bs}, while $M_0^{\alpha\beta}$ is given by Eq.~(\ref{e:stresKoi}.2).

\section{Shell constitution derived from 3D constitutive laws}\label{s:consituive3D}
Provided the displacement across the shell thickness complies with the Kirchhoff--Love assumption, a constitutive relation $W=W(\bE,\bK)$ for shells can also be extracted, without approximation, from classical three-dimensional constitutive models of the form $\tilde W=\tilde W(\tilde\bC)$, where $\tilde\bC$ is the right Cauchy-Green tensor for 3D continua. This procedure corresponds to a projection of {3D} models onto surface $\sS$, or to an extraction of {2D} surface models out of {3D} ones. This approach goes back to \citet{hughes83,de91,dvorkin95,klinkel02, kiendl15}, and it is presented here to show how it relates to our formulation.

\vspace{-0.3cm}
\subsection{Extraction procedure}

As shown in Fig.~\ref{f:mapping}, a shell material point $P$ can be described w.r.t.~the mid-surface in the reference configuration as \citep{wriggers-fee}
\eqb{l}
\tilde\bX(\xi^{\alpha},\xi_0) = \bX(\xi^\alpha) + \xi_0\,\bN(\xi^{\alpha})~,
\eqe
and in the current configuration as
\eqb{l}
\tilde\bx(\xi^{\alpha},\,\xi) = \bx(\xi^\alpha) + \xi\,\bd(\xi^{\alpha})~,
\eqe
where $\xi_0\in[-T/2,\,T/2]$ is the thickness coordinate of the shell and $\bd$ is the director vector which has three unknown components in general. For Kirchhoff--Love theory, which is considered here, $\bd:=\bn$ and $\xi = \lambda_3\,\xi_0$, where $\lambda_3$ denotes the stretch in the normal direction. Thus, the tangent vectors at $P$ are expressed w.r.t.~the basis formed by the tangent vectors on the mid-surface as 
\eqb{llll}
\bg_\alpha \dis \tilde\bx_{,\alpha} = \ba_\alpha - \xi\,b_\alpha^\gamma\,\ba_\gamma~,\\[2mm]
\bg_3 \dis \tilde\bx_{,\xi} = \bd~,\\[2mm]
\bG_\alpha \dis \tilde\bX_{,\alpha} = \bA_\alpha - \xi_0\,B_\alpha^\gamma\,\bA_\gamma~,\\[2mm]
\bG_3 \dis \tilde\bX_{,{\xi_0}} = \bN = \bG^3~,
\label{e:gij}
\eqe
and the metric tensors at $P$ can also be expressed in terms of the metric tensors on the mid-surface as
\eqb{llll}
g_{\alpha\beta} \dis \bg_\alpha\cdot\bg_\beta = g_a\,\auab + g_b\,\buab~,\\[4mm]
g^{\alpha\beta} \dis \bg^\alpha\cdot\bg^\beta = g^a\,\aab + g^b\,\bab~,\\[4mm]
G_{\alpha\beta} \dis \bG_\alpha\cdot\bG_\beta = G_A\,\Auab + G_B\,B_{\alpha\beta}~,\\[4mm]
G^{\alpha\beta} \dis \bG^\alpha\cdot\bG^\beta = G^A\,\Aab + G^B\,B^{\alpha\beta}~,
\label{e:metricG}
\eqe
where $\bg^\alpha=g^{\alpha\beta}\,\bg_\beta$, $[g^{\alpha\beta}]=[g_{\alpha\beta}]^{-1}$ and likewise $\bG^\alpha$. Due to Eq.~\eqref{e:gij}, the coefficients in Eq.~\eqref{e:metricG} are found to be 
\eqb{lllll}
g_a \dis 1 - \xi^2\,\kappa~,\quad\quad g_b \dis  -2\,\xi + 2\,H\,\xi^2~,\\[3mm]
G_A \dis 1 - \xi_0^2\,\kappa_0~,\hfill G_B \dis -2\,\xi_0 + 2\,H_0\,\xi_0^2~,
\label{e:coefG}
\eqe
and
\eqb{lllll}
g^a \dis s^{-2}\,( g_a + 2\,H\, g_b ) ~,\quad\quad\quad g^b \dis -s^{-2}\,g_b~,\\[3mm]
G^A \dis s_0^{-2}\,( G_A + 2\,H_0\, G_B )~,\hfill G^B \dis -s_0^{-2}\,G_B~.
\label{e:coefG2}
\eqe
Here,
\eqb{llll}
s := \ds\sqrt{g/a} = 1 - 2\,H\,\xi + \kappa\,\xi^2~,\quad s_0 := \ds\sqrt{G/A} = 1 - 2\,H_0\,\xi_0 + \kappa_0\,\xi_0^2~,
\label{e:shift}
\eqe
denote the so-called shifters, which are the determinants of the shifting tensors $\bi - \xi\,\bb$ and $\bI - \xi_0\,\bB$, respectively, and we have defined 
\eqb{l}
G := \det[G_{\alpha\beta}],\quad
A := \det[A_{\alpha\beta}],\quad   
B := \det[B_{\alpha\beta}],\quad
g := \det[g_{\alpha\beta}]~.
\eqe
From Eqs.~\eqref{e:coefG}, \eqref{e:coefG2}, and \eqref{e:shift}, we thus find the variations (considering $\xi$ fixed)
\eqb{lll}
\delta g_a \is \xi^2\,\kappa\,\agd \,\delta\augd - \xi^2\,\tilde{b}^{\gamma\delta}\,\delta\bugd~,\\[2mm]
\delta g_b \is - \xi^2\,\bgd \,\delta\augd + \xi^2\,\agd\,\delta\bugd~,
\eqe
where $ \tilde{b}^{\alpha\beta} := 2\,H\,\aab - \bab$. With these we get
\eqb{llll}
 \delta g_{\alpha\beta} =  g_a\,\delta\auab + g_b\,\delta\buab + \auab\,\delta g_a + \buab\,\delta g_b~.
 \label{e:deltgab}
\eqe

Further, we can represent the right 3D Cauchy-Green tensor w.r.t.~the basis $\{\bG^1,~\bG^2, ~\bN\}$ as (e.g.~see \citet{wriggers-fee})
\eqb{l}
\tilde\bC = \tilde C_{\alpha\beta}\,\bG^\alpha\otimes\bG^\beta + \tilde C_{\alpha3}\,\big(\bG^\alpha\otimes\bN+\bN\otimes\bG^\alpha\big) + \tilde C_{33}\,\bN\otimes\bN~.
\label{e:tildC}
\eqe
We note here that $\bG_\alpha$ accounts for the surface stretch due to the initial shell curvature (see Eq.~(\ref{e:gij}.3)) and  the basis $\{\bG^1,~\bG^2, ~\bN\}$ on a shell layer at $\xi$ defines the usual identity tensor in $\bbR^3$ as
\eqb{l}
\bone = \bI + \bN\otimes\bN~,\quad \bI:=\bG^\alpha\otimes\bG_\alpha=\bA^\alpha\otimes\bA_\alpha.
\eqe
Further, the {3D} Kirchhoff stress tensor can be written as 
\eqb{l}
\tilde\btau = \tilde{\tau}^{\alpha\beta}\,\bg_\alpha\otimes\bg_\beta + \tilde{\tau}^{\alpha3}\,\big(\bg_\alpha\otimes\bg_3+\bg_3\otimes\bg_\alpha\big) + \tilde{\tau}^{33}\,\bg_3\otimes\bg_3~.
\eqe
For the Kirchhoff--Love shell we have $\tilde C_{\alpha\beta} = g_{\alpha\beta}$, while $\tilde C_{\alpha3}=g_{\alpha 3} = \bg_{\alpha}\cdot\bg_3 =0$ and $\tilde C_{33} = \lambda_3^2$. Thus the variation of $\tilde\bC$ is
\eqb{l}
\delta\tilde{\bC} = \delta {\bC}^*+ 2\,\lambda_3\,\delta\lambda_3\,\bN\otimes\bN~, \quad\quad \bC^*:= g_{\alpha\beta}\, \bG^\alpha\otimes\bG^\beta~.
\label{e:cstar}
\eqe

With this, the total strain energy of the shell can be expressed w.r.t.~the mid-surface as
\eqb{l}
\mathcal{W} = \ds\int_{\Omega_0} \tilde{W} \dif V = \ds\int_{\sS_0} \int_{- \frac {T}{2}}^{\frac{T}{2}} \tilde{W}\,\dif \xi_0\,\dif G = \ds\int_{\sS_0} \int_{- \frac {T}{2}}^{\frac{T}{2}} s_0\, \tilde{W}\,\dif \xi_0\,\dif A ~,
\eqe
where $\dif G$ is the infinitesimal area element at layer $\xi$ along the thickness, and $\dif A$ is the infinitesimal area element on the mid-surface. Thus, the surface energy of the shell (per unit area) follows from the thickness integration
\eqb{l}
W = \ds\int_{- \frac {T}{2}}^{\frac{T}{2}} s_0\, \tilde{W} \dif\xi_0 = W(a_{\alpha\beta},\,b_{\alpha\beta})~,
\label{e:inWt}
\eqe
and its variation reads
\eqb{l}
\delta W = \ds\int_{- \frac {T}{2}}^{\frac{T}{2}} s_0\, \delta \tilde{W} \dif\xi_0 =  \ds\int_{-\frac{T}{2}}^{\frac{T}{2}} s_0\, \left( \ds\frac{1}{2}\tilde\tau^{\alpha\beta}\, \delta g_{\alpha\beta}  +  \tilde\tau^{33}\,\lambda_3\,\delta\lambda_3 \right) \,\dif\xi_0~,
 \label{e:varaW}
\eqe
where
\eqb{l}
\tilde{\tau}^{\alpha\beta} := \ds 2\pa{\tilde{W}}{g_{\alpha\beta}}; \quad \tilde{\tau}^{33} := \ds\frac{1}{\lambda_3}\,\ds\pa{\tilde{W}}{\lambda_3}~.
\label{e:tau}
\eqe
Here, similar to Remark~\ref{r:rm2}, $\tilde{\tau}^{\alpha\beta}$ are the components of either the second Piola-Kirchhoff stress tensor $\tilde{\bS}$ or the Kirchhoff stress $\tilde{\boldsymbol{\tau}}$, as
\eqb{rlcll}
\tilde{\tau}^{\alpha\beta} \is \bg^\alpha\cdot\tilde{\boldsymbol{\tau}}\,\bg^\beta \is\bG^\alpha\cdot\tilde{\boldsymbol{\bS}}\,\bG^\beta ~.
\eqe
For shells, since it is common to assume a plane stress state, we have the condition
 \eqb{l}
 \tilde{\tau}^{33} = \ds\frac{1}{\lambda_3}\,\ds\pa{\tilde{W}}{\lambda_3} = 0~.
 \label{e:planstr}
\eqe
Substituting Eqs.~\eqref{e:deltgab} and \eqref{e:planstr} into Eq.~\eqref{e:varaW} we get
\eqb{lll}
\delta W \is \ds\frac{1}{2}\int_{-\frac{T}{2}}^{\frac{T}{2}} s_0\,\left[ g_a\,\tilde\tau^{\alpha\beta} + \varepsilon_a\, \xi^2\,\kappa\,\aab - \varepsilon_b\,\xi^2\,\bab\right] \dif\xi_0\,\delta\auab~, \\[5mm]
\plus \ds\frac{1}{2}\int_{-\frac{T}{2}}^{\frac{T}{2}} s_0\, \left[ g_b\,\tilde\tau^{\alpha\beta} - \varepsilon_a\, \xi^2\, \tilde{b}^{\alpha\beta} + \varepsilon_b\,\xi^2\,\aab\right]\dif\xi_0 \,\delta\buab~,
\eqe
with $\varepsilon_a : = \tilde{\tau}^{\alpha\beta}\,\auab$ and  $\varepsilon_b : = \tilde{\tau}^{\alpha\beta}\,\buab$. Since
\eqb{l}
\delta W = \ds\pa{W}{a_{\alpha\beta}}\, \delta a_{\alpha\beta} + \pa{W}{b_{\alpha\beta}}\, \delta b_{\alpha\beta}~,
\eqe
we find
\eqb{lll}
\tau^{\alpha\beta} \is 2\,\ds\pa{W}{\auab} = \ds\int_{-\frac{T}{2}}^{\frac{T}{2}} s_0\,\left[ g_a\,\tilde{\tau}^{\alpha\beta} + \varepsilon_a\, \xi^2\,\kappa\,\aab - \varepsilon_b\,\xi^2\,\bab\right]\,\dif\xi_0~,\\[5mm]
M^{\alpha\beta}_0 \is \quad\ds\pa{W}{\buab} = \frac{1}{2}\,\ds\int_{-\frac{T}{2}}^{\frac{T}{2}} s_0\, \left[g_b\,\tilde{\tau}^{\alpha\beta} - \varepsilon_a\, \xi^2\, \tilde{b}^{\alpha\beta} + \varepsilon_b\,\xi^2\,\aab\right] \,\dif\xi_0.
 \label{e:M2Dab}
\eqe
It should be noted here that $s_0$, $g_a$, $g_b$, $\varepsilon_a$ and $ \varepsilon_b$ are all functions of $\xi_0$. 

\refstepcounter{remark}
\textbf{Remark \arabic{remark}:} So far, Eq.~\eqref{e:M2Dab} is exact, since we have not made any approximations apart from the Kirchhoff--Love hypothesis, i.e. $\bd = \bn$, and the plane stress assumption. For a curved thin shell ($T \ll R $, where $T$ is the thickness and $R$ is the radius of curvature) one may approximate the variations $\delta g_a \approx 0$ and $\delta g_b \approx 0 $ in Eq.~\eqref{e:deltgab}. In this case the two rear terms in Eq.~\eqref{e:M2Dab} vanish so that
\eqb{lll}
\tau^{\alpha\beta} \is \ds\int_{-\frac{T}{2}}^{\frac{T}{2}} s_0\, (1 - \xi^2\,\kappa)\,\tilde{\tau}^{\alpha\beta} \dif\xi_0~,\\[5mm]
M^{\alpha\beta}_0 \is  \ds\int_{-\frac{T}{2}}^{\frac{T}{2}} s_0\,(-\xi + H\, \xi^2)\,\tilde{\tau}^{\alpha\beta} \dif\xi_0~.
 \label{e:M2Dab2}
\eqe
This case will be examined in the examples in Sec.~\ref{s:examples}. If only the leading terms  in front of $\tilde{\tau}^{\alpha\beta}$ in Eq.~\eqref{e:M2Dab2} are considered, it is further simplified into
\eqb{lll}
\tau^{\alpha\beta} \is \ds\int_{-\frac{T}{2}}^{\frac{T}{2}} \tilde{\tau}^{\alpha\beta} \dif\xi_0~,\\[5mm]
M^{\alpha\beta}_0 \is -\ds\int_{-\frac{T}{2}}^{\frac{T}{2}} \xi \,\tilde{\tau}^{\alpha\beta} \dif\xi_0~.
 \label{e:M2Dab3}
\eqe
These expressions are commonly used in shell and plate formulations, see e.g.~\citet{kiendl15,thanh11}. Here the sign convention for $M_0^{\alpha\beta}$ follows \citet{shelltheo,steigmann99}.

\refstepcounter{remark}
\textbf{Remark \arabic{remark}:} The stretch through the thickness, $\lambda_3$, can be determined in various ways. It can be included in the formulation as a separate degree of freedom. Another common and straightforward approach is to derive it from the plane stress condition \eqref{e:planstr}. The resulting equation is usually nonlinear and can be solved numerically (see e.g.~\citet{hughes83,de91,dvorkin95,kiendl15}), or analytically for some special cases (e.g.~see Sec.~\ref{s:newo} for a Neo-Hookean model). In particular, if incompressible models are used, i.e.~$\tilde{J} = 1$, then $\lambda_3$ can be determined analytically (e.g.~see Sec.~\ref{s:incNH}). Sometimes also $\lambda_3\approx 1$ is assumed. In this case, condition \eqref{e:planstr} is generally not satisfied anymore. Instead $\tau^{33}$ should be treated as the Lagrange multiplier to the constraint $\lambda_3=1$.

\refstepcounter{remark}
\textbf{Remark \arabic{remark}:} For an efficient implementation, we consider that $\tilde{\tau}^{\alpha\beta}$ is expressible in the form
\eqb{l}
\tilde{\tau}^{\alpha\beta}= \tilde{\tau}_g\,g^{\alpha\beta} + \tilde{\tau}_G\,G^{\alpha\beta} ~. 
\label{e:tauNeo}
\eqe 
Substituting this into Eq.~\eqref{e:M2Dab}, and taking into account Eq.~\eqref{e:metricG}, we get the surface stress and moment written in the form
\eqb{lll}
\tau^{\alpha\beta} \is \tau_a\, \aab + \tau_b\, \bab + \tau_A\, \Aab + \tau_B\, B^{\alpha\beta} ~,\\[4mm] 
M^{\alpha\beta}_0 \is  M^0_a\, \aab + M^0_b\, \bab + M^0_A\, \Aab + M^0_B\, B^{\alpha\beta} ~, 
\eqe
so that only the (scalar) coefficients need to be computed by thickness integration as
\eqb{llllll}
\tau_a \is  \ds\int_{-\frac{T}{2}}^{\frac{T}{2}} s_0\, \left( g_a\,\tilde{\tau}_g\, g^a +\, \varepsilon_a\,\xi^2\,\kappa\right)  \,\dif\xi_0~, \quad \quad
&\tau_A \is  \ds\int_{-\frac{T}{2}}^{\frac{T}{2}} s_0\, g_a\,\tilde{\tau}_G\, G^A \,\dif\xi_0~,\\[5mm]
\tau_b \is  \ds\int_{-\frac{T}{2}}^{\frac{T}{2}} s_0\, \left( g_a\,\tilde{\tau}_g\, g^b - \,\varepsilon_b\,\xi^2\right)  \,\dif\xi_0~,\quad \quad
&\tau_B \is \ds\int_{-\frac{T}{2}}^{\frac{T}{2}} s_0\, g_a\,\tilde{\tau}_G\, G^B  \,\dif\xi_0~,
\eqe
and 
\eqb{llllll}
M^0_a \is  \ds\frac{1}{2}\ds\int_{-\frac{T}{2}}^{\frac{T}{2}} s_0\, \left[ g_b\,\tilde{\tau}_g\, g^a + (\varepsilon_b - 2\,H\,\varepsilon_a)\,\xi^2\right]  \,\dif\xi_0~, \quad \quad
&M^0_A \is  \ds\frac{1}{2}\ds\int_{-\frac{T}{2}}^{\frac{T}{2}}\,s_0\,g_b\,\tilde{\tau}_G\, G^A  \,\dif\xi_0~,\\[5mm]
M^0_b \is \ds\frac{1}{2}\ds\int_{-\frac{T}{2}}^{\frac{T}{2}} s_0\, \left( g_b\,\tilde{\tau}_g\, g^b + \varepsilon_a\,\xi^2\right)  \,\dif\xi_0~,\quad \quad
&M^0_B \is \ds\frac{1}{2}\ds\int_{-\frac{T}{2}}^{\frac{T}{2}}\,s_0\,g_b\,\tilde{\tau}_G\, G^B \,\dif\xi_0~.
\eqe

The linearization then follows the representation of Eq.~\eqref{e:unifcdef} with the scalar coefficients computed by  integration over the thickness. 

\refstepcounter{remark}
\textbf{Remark \arabic{remark}:}  We note that since the integrands in  Eq.~\eqref{e:M2Dab} and even in its reduced forms~\eqref{e:M2Dab2}, \eqref{e:M2Dab3}, are rather complex w.r.t.~$\xi$, we need numerical integration in general. Analytical integration may only be possible for special cases, such as the Saint Venant--Kirchhoff constitutive model discussed in Sec.~\ref{s:koiterst}. Thus, it is obvious that the models constructed in 2D manifold form, such as the Koiter (Sec.~\ref {s:koiter}), Canham (Sec.~\ref{s:canham}) and Helfrich constitutive models (see e.g.~\citet{liquidshell}), are much more efficient than models requiring numerical thickness integration.

In the following, we will present the extraction example for Neo-Hookean materials to demonstrate the procedure. We will also show that by analytical integration through the thickness of the 3D Saint Venant--Kirchhoff strain energy, the Koiter model is recovered.

\subsection{Extraction example: compressible Neo-Hookean materials}\label{s:newo}

The classical {3D} Neo-Hookean strain-energy per reference volume is considered in the form
\eqb{l}
\tilde{W}(\tilde{I}_1, \tilde{J}) = 
\ds \frac{\tilde\Lambda}{4}(\tilde{J}^2 - 1 -2 \ln \tilde{J}) + \ds\frac{\tilde{\mu}}{2}(\tilde{I}_1 - 3 -2\, \ln \tilde{J}),
\label{e:NeoH3D}
\eqe
where $\tilde{I}_1$ and $\tilde{J}$ are the invariants of  the 3D Cauchy-Green tensor $\tilde{\bC}$. Due to Eq.~\eqref{e:tildC}, they can also be expressed in terms of the kinematic quantities on the mid-surface, i.e. 
\eqb{lll}
\tilde{I}_1 \dis  \tilde\bC : \bone = I^*_1 + \lambda_3^2, \quad\quad \tilde{J} :=  \sqrt{\det\tilde\bC}  = J^*\,\lambda_3~,
\eqe
where
\eqb{lll}
I^*_1:= \bC^*:\bI = g_{\alpha\beta}\,G^{\alpha\beta} ~,\quad\quad J^* :=\sqrt{\det\bC^*} = \ds\sqrt{\frac{g}{G}}
\eqe
denote the invariants of $\bC^*$ defined in Eq.~\eqref{e:cstar}. Therefore, their variations  are
\eqb{lll}
\delta\tilde{I}_1 = \delta I^*_1  + 2\,\lambda_3\,\delta\lambda_3~, \quad \delta I^*_1  = G^{\alpha\beta} \,\delta g_{\alpha\beta}~, 
\eqe
and
\eqb{lll}
\delta\tilde{J} =  \lambda_3\,\delta J^* + J^*\,\delta\lambda_3~,\quad  \delta J^* = \ds\frac{{J^*}}{2}\,g^{\alpha\beta} \,\delta g_{\alpha\beta}.
\eqe

$\lambda_3$ can be determined using the plane stress condition Eq.~\eqref{e:planstr}, 
\eqb{lll}
\ds\pa{\tilde{W}}{\lambda_3} =\ds\frac{\tilde\Lambda}{2}\,\left[ J^{*2}\,\lambda_3 - \frac{1}{\lambda_3}\right] + \tilde{\mu}\left(\lambda_3 - \frac{1}{\lambda_3}\right) = 0,
\eqe
which is solvable analytically w.r.t.~$\lambda_3$ as
\eqb{lll}
\lambda_3^2 = \ds\frac{\tilde\Lambda + 2\,\tilde{\mu}} {\tilde\Lambda\,J^{*2} + 2\,\tilde{\mu}}~.
\eqe
From Eq.~(\ref{e:tau}.1) we then find
\eqb{l}
\tilde{\tau}^{\alpha\beta}= \tilde{\mu}\,G^{\alpha\beta} -  \tilde{\mu}\,\ds\frac{\tilde\Lambda + 2\,\tilde{\mu}} {\tilde\Lambda\,J^{*2} + 2\,\tilde{\mu}}\,g^{\alpha\beta}.
\label{e:case2NH}
\eqe 
The stress in Eq~\eqref{e:case2NH} is then substituted into Eq.~\eqref{e:M2Dab}, \eqref{e:M2Dab2} or \eqref{e:M2Dab3}. Numerical integration is still required to evaluate those. The corresponding linearized quantities and tangent matrices are provided in Appendix~\ref{s:linproj}.

\subsection{Extraction example: incompressible Neo-Hookean materials }\label{s:incNH}
The Neo-Hookean strain-energy per reference volume in this case is defined as
\eqb{l}
\tilde{W}(\tilde{I}_1, \tilde{J}, p) =  \ds\frac{\tilde{\mu}}{2}(\tilde{I}_1 - 3 ) + p\,(1 -  \tilde{J}),
\label{e:NeoH3DInc}
\eqe
where $p$ is the Lagrange multiplier associated with the volume constraint. From Eq.~\eqref{e:NeoH3DInc}, we thus find
\eqb{l}
 \tilde{\tau}^{\alpha\beta}=  \ds 2\,\pa{\tilde{W}(g_{\alpha\beta}, \lambda_3, p)}{g_{\alpha\beta}} = \tilde{\mu} G^{\alpha\beta} - p\, \tilde{J} g^{\alpha\beta}.
 \label{e:stressp}
\eqe 
Further, the plane stress condition implies $\partial\tilde W/\partial\lambda_3 = 0$ and the incompressibility constraint requires $\lambda_3 := 1/J^*$, which together allow us to determine the Lagrange multiplier analytically as
\eqb{l}
p = \ds\frac{\tilde{\mu}}{J^{*2}}.
\eqe 
Substituting this into Eq.~\eqref{e:stressp} we obtain
 \eqb{l}
 \tilde{\tau}^{\alpha\beta}= \ds\tilde{\mu} \, \left( G^{\alpha\beta} - \frac{1}{J^{*2}} \,g^{\alpha\beta} \right),
 \eqe 
for Eq.~\eqref{e:M2Dab} or~\eqref{e:M2Dab2},~\eqref{e:M2Dab3}.  Numerical integration is also required here.
 
\subsection{Extraction example: Saint Venant-Kirchhoff model}\label{s:koiterst}
We now consider the Saint Venant--Kirchhoff model,
\eqb{llll}
\tilde{W}(\tilde{\bC}) := \ds\frac{1}{2}\, \tilde{\bE} : \tilde{\bbC} : \tilde{\bE} ~,\quad
\tilde{\bbC} := \tilde\Lambda\,\bone \odot \bone + 2\, \tilde{\mu}\, (\bone \otimes\bone)^\mrs ~.
\label{e:stVN}
\eqe
In this case, the Koiter model of Eq.~\eqref{e:Koiter} can be recovered by using Eq.~\eqref{e:inWt} and Eq.~\eqref{e:planstr} together with the following ingredients:
\vspace{-\topsep}
\bitm
\setlength{\itemsep}{0pt}
\setlength{\parskip}{0pt}
\setlength{\parsep}{0pt}
\item $s_0 \approx 1$, $\xi\approx\xi_0$, and only the leading terms w.r.t.~$\xi$ in Eq.~\eqref{e:coefG} and  Eq.~\eqref{e:coefG2} are taken into account, so that Eq.~\eqref{e:stVN} reduces to
\eqb{llll}
\tilde{W}({\bC^*}) := \ds\frac{1}{2}\, {\bE^*} : {\bbC}^* : {\bE}^* ~,\quad
{\bbC}^* := \tilde\Lambda\,\bI \odot \bI + 2\, \tilde{\mu}\, (\bI \otimes \bI)^\mrs ~,
\label{e:stVN3}
\eqe
where
\eqb{llll}
\bE^* := \bE - \xi\,\bK.
\eqe
\item The bending and stretching response is considered fully decoupled, so that all the mixed products, such as $\tr\bE\,\tr\bK$ and $\tr(\bE\bK)$, are disregarded in the strain energy function. Eq.~\eqref{e:stVN3} then further reduces to
\eqb{llll}
\tilde{W}({\bE,\bK,\xi}) = \ds\frac{\tilde\Lambda}{2}\,(\tr\bE)^2 + \tilde{\mu}\,\tr(\bE^2) + \xi^2\,\ds\frac{\tilde\Lambda}{2}\,(\tr\bK)^2 + \xi^2\,\tilde{\mu}\,\tr(\bK^2)~.
\eqe
Integrating the above energy over the thickness $[-T/2,~T/2]$ gives
\eqb{llll}
W({\bE,\bK,T}) = \ds \frac{T}{2}\, \tilde\Lambda\,(\tr\bE)^2 + T\,\tilde{\mu}\,\tr(\bE^2) + \ds\frac{T^3}{24}\,\tilde\Lambda\,(\tr\bK)^2 + \frac{T^3}{12}\,\tilde{\mu}\,\tr(\bK^2)~.
\label{e:Wfrom3D}
\eqe
\item Finally, the plane stress assumption is used throughout the thickness. Accordingly, the 3D Lam{\'e} constant $\tilde{\Lambda}$ in Eq.~\eqref{e:Wfrom3D} is replaced by its plane-stress counterpart $(2\,\tilde\Lambda\,\tilde\mu)/(\tilde\Lambda + 2\,\tilde\mu)$ (see e.g.~\cite{Klaus10}), which results in the Koiter model of Eqs.~\eqref{e:Koiter}~and~\eqref{e:rel2D3D}.
\eitm\vspace{-\topsep}
Alternatively \citep{steigman13}, the Koiter model can also be derived systematically as the leading-order model from the Taylor  expansion of $W$ with the aid of Leibniz' rule for small thickness.
\section{FE discretization}\label{s:fe}

In this section, we first discuss the discretization of the weak form \eqref{e:wfu} and its linearization on the basis of IGA to obtain FE forces and tangent matrices. We will then discuss the continuity constraint between patches, patch folds, symmetry and rotational Dirichlet boundary conditions.

\subsection{$C^1$-continuous discretization}
In order to solve Eq.~\eqref{e:wfu} by the finite element method, the surface $\sS$ is discretized using the isogeometric analysis technique proposed by \cite{hughes05}. Thanks to the  B{\'e}zier extraction operator $\mC^e$ introduced by \cite{borden11},  the usual finite element structure is recovered for NURBS basis function $\{N_A\}_{A=1}^n$, where $n$ is the number of control points defining an element $e$. Namely,
\eqb{lll}
 N_A(\xi,\eta) = \ds\frac{w_A\,\hat{N}_A^e(\xi,\eta)}{\sum_{A=1}^n w_A\,\hat{N}_A^e(\xi,\eta)},
\label{e:shpfct}\eqe
where  $\{\hat{N}_A^e\}_{A=1}^n$ is the B-spline basis function expressed in terms of Bernstein polynomials as
\eqb{lll}
\hat{\mN}^e(\xi,\eta) = \mC^e_\xi\,\mB(\xi)\,\otimes\,\mC_\eta^e\,\mB(\eta),
\eqe
with $\hat{N}_A^e$ to be the corresponding entries of matrix $\hat{\mN}^e$. For T-splines, the construction of the isoparametric element  based on the  B{\'e}zier extraction operator can be found in \cite{scott11}.

\subsection{FE approximation}
The geometry within an undeformed element $\Omega^e_0$ and deformed element $\Omega^e$ is interpolated from the positions of control points $\mX_e$ and $\mx_e$, respectively, as
\eqb{lll}
\bX \is \mN\,\mX_e~, \quad
\bx = \mN\,\mx_e~, \\
\eqe
where $\mN(\bxi):= [N_1\bone,\, N_2\bone,\, ...,\, N_{n}\bone]$ is defined based on the NURBS shape functions of Eq.~\eqref{e:shpfct}. It follows that
\eqb{lll}
\delta\bx \is \mN\,\delta\mx_e ~, \\[1mm]
\ba_\alpha \is \mN_{,\alpha}\,\mx_e ~,\\[1mm]
\delta\ba_\alpha \is \mN_{,\alpha}\,\delta\mx_e ~,\\[1mm]
\ba_{\alpha,\beta} \is \mN_{,\alpha\beta}\,\mx_e ~,\\[1mm]
\ba_{\alpha;\beta} \is \tilde\mN_{;\alpha\beta}\,\mx_e ~,
\label{e:dxe}
\eqe
where $\mN_{,\alpha}(\bxi):= [N_{1,\alpha}\bone,\, N_{2,\alpha}\bone,\, ...,\, N_{{n},\alpha}\bone]$, $\mN_{,\alpha\beta}(\bxi):= [N_{1,\alpha\beta}\bone,\, N_{2,\alpha\beta}\bone,\, ...,\, N_{n,\alpha\beta}\bone]$, $N_{A,\alpha}=\partial N_A/\partial \xi^\alpha$, $N_{A,\alpha\beta}=\partial^2 N_A/(\partial \xi^\alpha\partial \xi^\beta)\;(A=1, ..., n)$ and\footnote{A tilde is included in $\tilde\mN_{;\alpha\beta}$ to distinguish it from $\mN_{,\alpha\beta} $.} 
\eqb{l}
\tilde\mN_{;\alpha\beta} := \mN_{,\alpha\beta} - \Gamma^\gamma_{\alpha\beta}\,\mN_{,\gamma}~.
\eqe
\begin{samepage}
Inserting those expressions into the formulas for $\delta a_{\alpha\beta}$ and $\delta b_{\alpha\beta}$ \citep{shelltheo} gives 
\eqb{lll}
\delta \auab 
\is \delta\mx_e^\mrT\, \big(\mN_{,\alpha}^\mrT\,\mN_{,\beta} + \mN_{,\beta}^\mrT\,\mN_{,\alpha}\big)\,\mx_e~,
\label{e:dd_aabh}
\eqe
and
\eqb{l}
\delta \buab =
\delta\mx_e^\mrT\,\tilde\mN_{;\alpha\beta}^\mrT\,\bn~.
\label{e:dd_babh}
\eqe
\subsection{FE force vectors}\label{s:fvectors}
\end{samepage}
Substituting Eqs.~\eqref{e:dd_aabh} and \eqref{e:dd_babh} into Eq.~\eqref{e:wfu} gives
\eqb{l}
\ds\sum_{e=1}^{n_\mathrm{el}} \left( G^e_\mathrm{in} + G^e_\mathrm{int} - G^e_\mathrm{ext}\right)  = 0 \quad\forall\,\delta\mx_e\in\sV~,
\eqe
where $n_\mathrm{el}$ is the number of elements and 
\eqb{l}
G^e_\mathrm{in} = \delta\mx_e^\mrT\,\mf^e_\mathrm{in} ~,\quad  $with$ \quad  \mf^e_\mathrm{in} :=  \ds\int_{\Omega^e_0} \mN^\mrT\,\rho_0\,\dot\bv\,\dif A  ~.
\label{e:Pin}
\eqe
Similarly,
\eqb{l}
G^e_\mathrm{int} = \delta\mx_e^\mrT\,\big(\mf^e_\mathrm{int\tau} + \mf^e_{\mathrm{int}M}\big)~,
\label{e:Pinth}
\eqe
with the internal FE force vectors due to the membrane stress $\tau^{\alpha\beta}$ and the bending moment $M_0^{\alpha\beta}$
\eqb{l}
\mf^e_\mathrm{int\tau} 
:=  \ds\int_{\Omega^e_0} \tau^{\alpha\beta} \,\mN_{,\alpha}^\mrT\,\vaub\,\dif A
\label{e:Fint}
\eqe
and
\eqb{l}
\mf^e_{\mathrm{int}M} 
:=  \ds\int_{\Omega^e_0} \, M_0^{\alpha\beta}\,\tilde\mN_{;\alpha\beta}^\mrT\,\bn\, \dif A~.
\label{e:Mint}
\eqe
Here the symmetry of $\tau^{\alpha\beta}$ has been exploited. The external virtual work follows as \citep{membrane,shelltheo}
\eqb{l}
G^e_\mathrm{ext} = \delta\mx_e^\mrT\,\big(\mf^e_{\mathrm{ext0}} + \mf^e_{\mathrm{ext}p}+\mf^e_{\mathrm{ext}t}+\mf^e_{\mathrm{ext}m}\big)~,
\label{e:Pexth}
\eqe
where the external FE force vectors are
\eqb{lll}
\mf^e_{\mathrm{ext0}} \dis \ds\int_{\Omega^e_0}\mN^\mrT\,\bff_0\,\dif A~, \\[5mm]
\mf^e_{\mathrm{ext}p} \dis \ds\int_{\Omega^e}\mN^\mrT\,p\,\bn\,\dif a~, \\[5mm]
\mf^e_{\mathrm{ext}t} \dis \ds\int_{\partial_t\Omega^e}\mN^\mrT\,\bt\,\dif s~, \\[5mm]
\mf^e_{\mathrm{ext}m} \dis -\ds\int_{\partial_m\Omega^e}\mN_{,\alpha}^\mrT\,\nu^\alpha\,m_\tau\,\bn\,\dif s ~.
\label{e:fext}
\eqe
Here $\bff_0$ is a constant surface force and $p$ is an external pressure acting always normal to $\sS$ \citep{membrane}. $\bt$ is the effective boundary traction of Eq.~\eqref{e:bt} and both $\nu^\alpha$ and $m_\tau$ are defined in Sec.~\ref{s:stress}. We note that, in the following sections, the inertia term is neglected, i.e.~$\rho_0\,\dot\bv = \boldsymbol{0}$. The corresponding tangent matrices can be found in Appendix~\ref{s:tangents}.

\subsection{Edge rotation conditions}\label{s:sym}
\begin{figure}[!ht]
\begin{center} \unitlength1cm
\unitlength1cm
\begin{picture}(0,2.5)
\put(-7.5,0.5){\includegraphics[height=20mm]{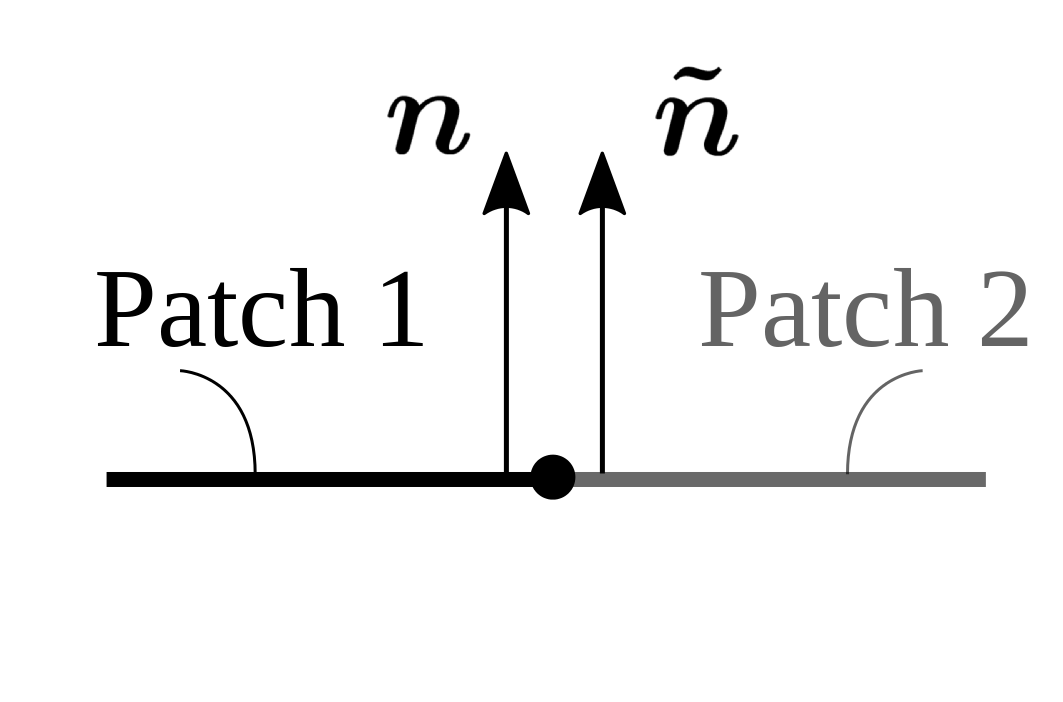}} 
\put(-4.5,0.5){\includegraphics[height=20mm]{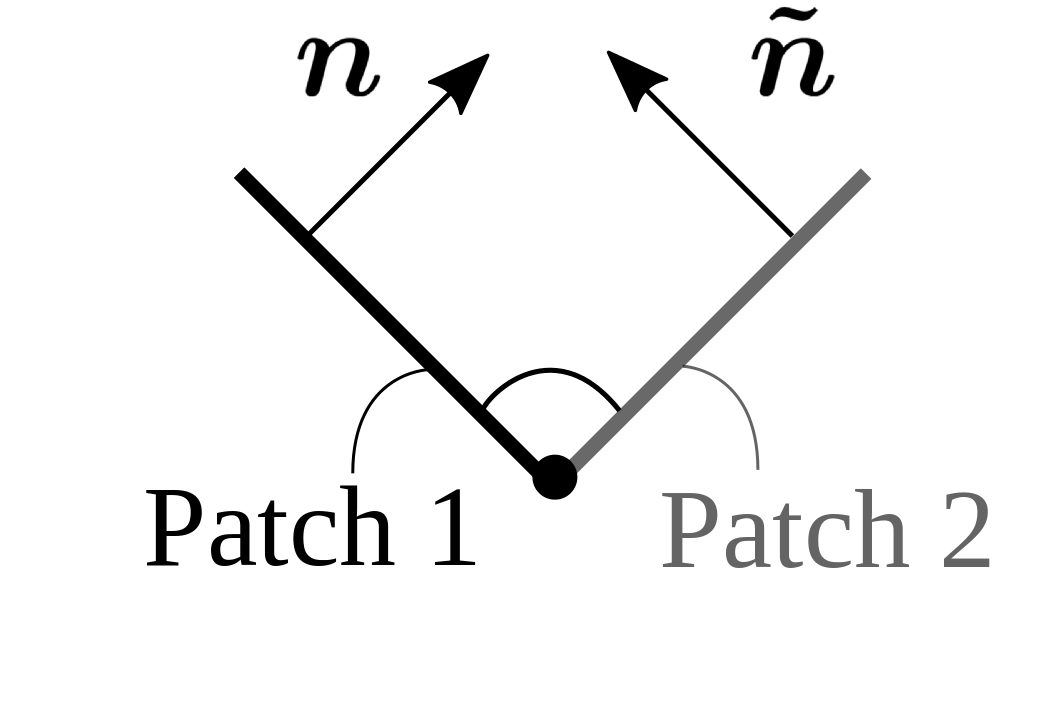}}
\put(-1.5,0.5){\includegraphics[height=20mm]{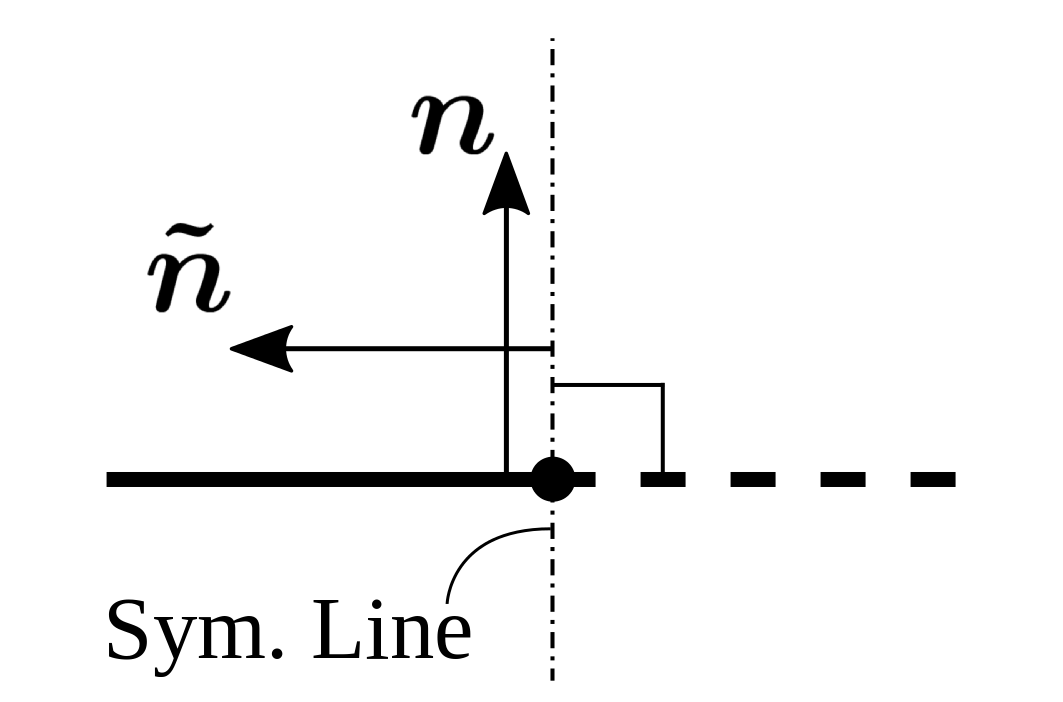}}
\put(1.5,0.5){\includegraphics[height=20mm]{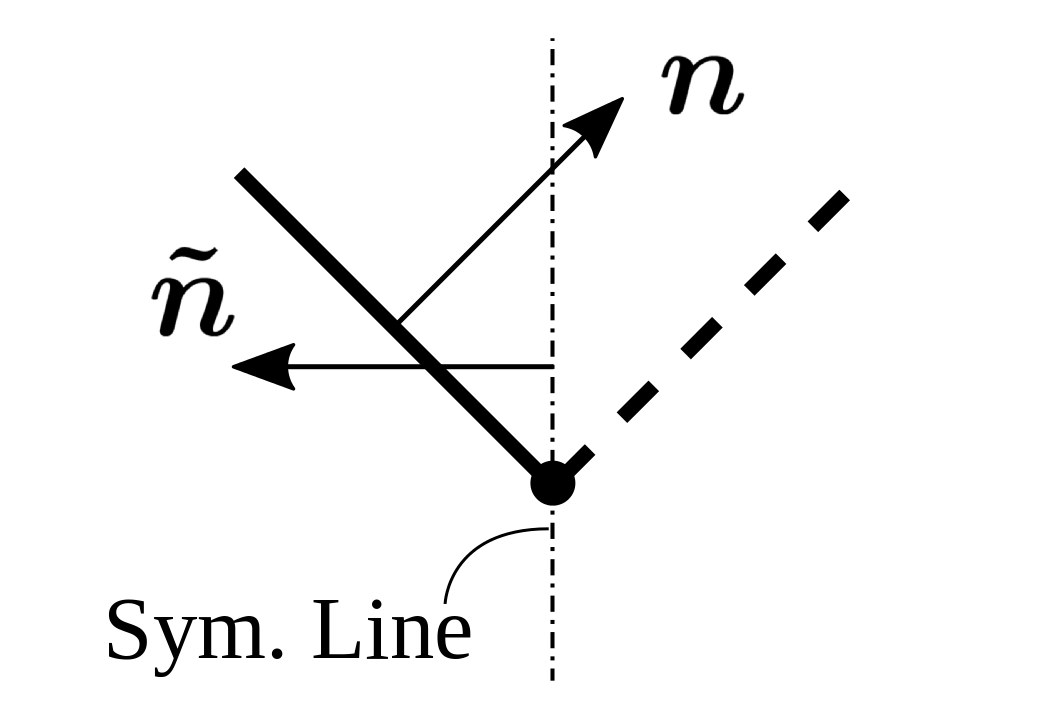}}
\put(4.5,0.5){\includegraphics[height=20mm]{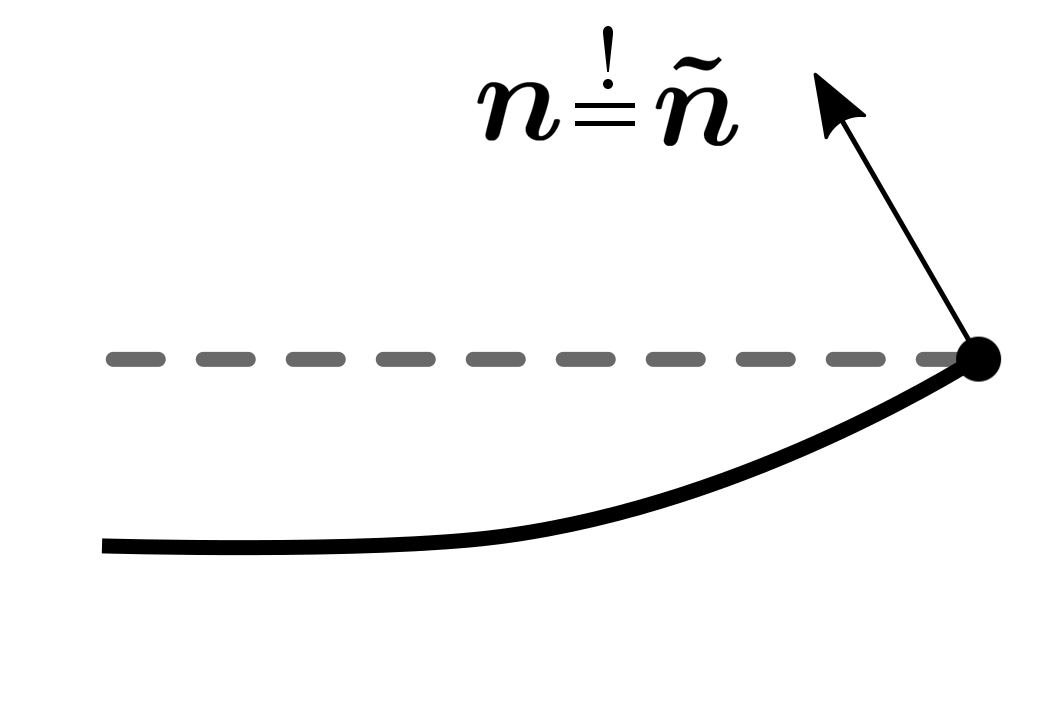}}

\put(-6.0,0){a.}
\put(-3.0,0){b.}
\put(0,0){c.}
\put(3,0){d.}
\put(6,0){e.}
\end{picture}
\caption{Edge rotation conditions: a.~$G^1$-continuity constraint,  b.~fixed surface folds (e.g.~V-shapes and L-shapes), c.~symmetry (or clamping) constraints, d.~symmetry constraint at a kink, e.~rotational Dirichlet boundary condition. Surface edge $\sL$, shown by a filled circle, is perpendicular to the plane and parallel to the inward pointing unit direction $\btau$.}
\label{f:gccase}
\end{center}
\end{figure}
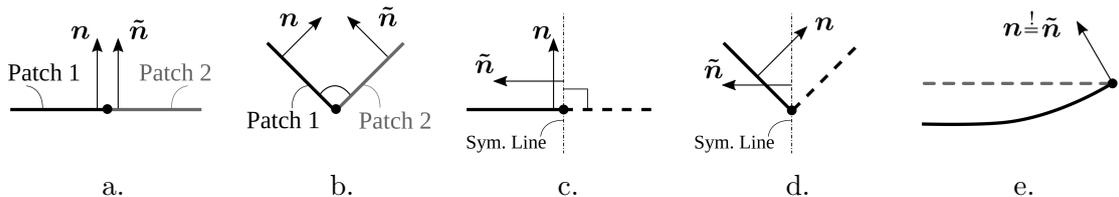

This section presents a general approach to describe different rotation conditions. These conditions are required due to the fact that the presented formulation has only displacement degrees of freedom as unknowns. As shown in Fig.~\ref{f:gccase}, the $G^1$-continuity constraint is required for multi-patch NURBS in order to transfer moments. Additionally, other rotation conditions may be needed such as fixed surface fold constraints, symmetry (or clamping) constraints, symmetry constraints at a kink and rotational Dirichlet boundary conditions.

There are various methods to enforce the continuity between patches, such as using T-Splines \citep{schill12}, the bending strip method \citep{kiendl10} and the Mortar method for non-conforming patches \citep{Dornisch-phd}. Alternatively, Nitsche's method as described in \citet{nguyen14nitsche,guo15} can also be used. Recently, \citet{lei15b} applied static condensation and a penalty method to enforce $G^1$-continuity of adjacent patches based on subdivision algorithms. 

Here, we introduce a general approach that can systematically enforce different edge rotation conditions, including $G^1$-continuity of adjacent patches, within the framework of a curvilinear coordinate system. The presented approach has new features compared to existing formulations. For instance, our approach is conceptually and technically different from the method of \citet{lei15b}. First and foremost, the presented method is not only restricted to $G^1$-continuity of adjacent patches of NURBS-based models but it also includes other constraints as mentioned above. Second, here the constraint is enforced by systematically adding a potential to the weak form (Eq.~\eqref{e:wfu}). In \citet{lei15b}, the $G^1$-continuity is enforced by constraining the position of control points in the vicinity of the shared edge so that the adjacent NURBS surfaces have the same tangent plane at the intersection points. This leads to a completely different constraint equation.\\
Like the bending strip method, we focus on conforming meshes, where the control points of adjacent patches coincide at their interface. For nonconforming meshes, the proposed method should be modified or alternative methods, like Nitsche's or Mortar methods, can be used. Compared to the bending strip method \citep{kiendl10}, the proposed method requires only line integration instead of surface integration. For nonlinear problems, the implementation of Nitsche's method (e.g.~\citep{guo15}) becomes more complex as it requires the tractions and their variations on the interface, which depend on the type of constitutive equations. Our formulation is independent of the choice of material model. Further, as shown in Secs.~\ref{s:conti} and \ref{s:edgeLM}, our constraint equation, enforced by both the penalty and Lagrange multiplier methods, gives an exact transmission of both traction and moment across the interface.

In order to model all five cases depicted in Fig.~\ref{f:gccase}, we first consider the constraint equation on the surface edge $\sL$
\eqb{l}
g_\mrc:= \cos\alpha_0 - \cos\alpha = 0~, \quad  \quad\forall \bx\in\sL~,
\label{e:gc}
\eqe
where
\eqb{l}
 \cos\alpha_0:=\bN\cdot\tilde\bN ~,  \quad \cos\alpha:= \bn\cdot \tilde{\bn}~.
\eqe
Here, $\bN$ is the surface normal of a considered patch and $\tilde\bN$ is the surface normal of a neighboring patch. They are defined in the initial configuration. Similarly, $\bn$ and $\tilde\bn$ are the corresponding quantities in the deformed configuration. As shown in Fig.~\ref{f:gccase}a, in the case of  $G^1$-continuity between two patches, we have $\bN\cdot\tilde\bN  = 1$. Similarly for fixed surface folds, $\bN\cdot\tilde\bN=\cos \alpha_0$, where $\alpha_0$ is the angle between the normal of the patches (see Fig.~\ref{f:gccase}b). In the case of symmetry, $\tilde{\bn} = \tilde{\bN}$ is the (fixed) normal vector of the symmetry plane (see Fig.~\ref{f:gccase}c-d). Furthermore, the constraint equation~\eqref{e:gc} can be used to apply rotational Dirichlet boundary conditions, by setting $\bN \cdot \tilde{\bN} = 1$, and $\tilde \bn$ is then the prescribed normal direction.

The constraint equation \eqref{e:gc} can uniquely realize angles within $ [0,~\pi]$. However, Eq.~\eqref{e:gc} becomes non-unique if $\alpha$ and $\alpha_0$ go beyond this range. Therefore, an additional constraint equation in the form of a $\mathrm{sine}$ function is considered, 
\eqb{l}
g_\mrs:= \sin\alpha_0 - \sin\alpha = 0~, \quad  \quad\forall \bx \in  {\sL}
\label{e:gs}
\eqe
with
\eqb{l}
 \sin\alpha_0:=\big(\bN\times\tilde\bN\big)\cdot\btau_0 ~,\quad \sin\alpha:= \big(\bn\times\tilde{\bn}\big)\cdot\btau~,
 \label{e:signedalpha}
\eqe
where $\btau$ and $\btau_0$ are the unit directions of interfaces $\sL$ and $\sL_0$ (see Sec.~\ref{s:stress}). Here, Eq.~\eqref{e:signedalpha} implies that $\alpha$ is measured from $\bn$ to $\tilde{\bn}$, and likewise for $\alpha_0$. Together, Eqs.~\eqref{e:gc} and \eqref{e:gs} uniquely define any physical angles of $\alpha_0$ and $\alpha$ $\in [0,~2\pi]$. One can easily show that they are equivalent to the system of equations 
\eqb{lll}
\bar{g_\mrc} \dis 1- \cos(\alpha-\alpha_0) = 0~, \\[2mm]
\bar{g_\mrs} \dis \sin(\alpha-\alpha_0) = 0~.
\label{e:gsALT}
\eqe
It is also worth noting that an initial NURBS mesh imported from CAD programs is usually given with $G^1$-continuity instead of $C^1$-continuity between patches. In this case, a strict $C^1$-continuity enforcement without a mesh modification will affect the FE solution. Thus, a $G^1$-continuity constraint is more practical in such a case. 
In the following, we present a penalty and a Lagrange multiplier method for enforcing constraints~\eqref{e:gc} and \eqref{e:gs}.

\subsubsection{Penalty method for edge rotations}\label{s:conti}
The system of constraints \eqref{e:gc} and \eqref{e:gs} can be enforced  by  the penalty formulation
\eqb{l}
\Pi_\sL =  \ds \int_{\sL_0} \frac{\epsilon}{2}\,\big( g_\mrc^2 + g_\mrs^2\,\big)\, \dif S = \ds \int_{\sL_0}\epsilon\,\big( 1 - c_0\,\cos\alpha - s_0\,\sin\alpha \,\big)\, \dif S~,
\label{e:pen}
\eqe
which is a locally convex function w.r.t.~$\alpha$ and $\alpha_0$. Thus, for the Newton-Raphson method, the existence of a unique solution is guaranteed, provided that $|\alpha - \alpha_0| < \pi$. Here $\epsilon$ is the penalty parameter and $s_0:=\sin\alpha_0$ and $c_0:=\cos\alpha_0$. Taking the variation of the above equation yields
\eqb{l}
\delta\Pi_\sL = - \ds \int_{\sL_0} \epsilon\, \big(\delta\btau\cdot \btheta + \delta\bn\cdot\tilde\bd + \delta\tilde\bn\cdot\bd\,\big) \,\dif S~,
\label{e:pen2}
\eqe
where we have defined
\eqb{l}
\btheta:=s_0\,\bn\times\tilde\bn~, 
\quad \tilde\bd:= c_0\,\tilde\bn + s_0\,\tilde\bnu ~,
\quad \bd:= c_0\,\bn + s_0\,\bnu 
\eqe
and $\tilde{\bnu}:=\tilde{\btau}\times\tilde{\bn}=-\btau\times\tilde{\bn}$.
The linearization of the above equation can be found in Appendix~\ref{s:linpen}. We note that, for some applications, the variation of $\btau$ and $\tilde{\bn}$ in Eq.~\eqref{e:pen2} vanishes since they are considered fixed.\\
From Eq.~\eqref{e:pen2}, the bending moments $m_\tau = m_{\tilde\tau} = \epsilon\,\sin(\textcolor{darkgreen}{\alpha - \alpha_0})$ can be identified. This implies that the bending moment is transmitted across the interface exactly.

\refstepcounter{remark}
\textbf{Remark \arabic{remark}:}
If $\alpha_0=0$ (for the cases in Figs.~\ref{f:gccase}a and \ref{f:gccase}e), the integrand in Eq.~\eqref{e:pen} reduces to $\epsilon\,(1 - \cos\alpha)$ such that $\Pi_\sL$ becomes
\eqb{l}
\Pi_{\sL} =  \ds \int_{\sL_0} \frac{\epsilon}{2} (\bn - \tilde\bn)\cdot(\bn - \tilde\bn)\, \dif S~,
\label{e:penC}
\eqe
enforcing the constraint
\eqb{l}
\bg_{\sL} := (\bn - \tilde\bn)= \boldsymbol{0}~.
\eqe
\subsubsection{Lagrange multiplier method for edge rotations} \label{s:edgeLM}
Alternatively, the Lagrange multiplier approach can be used to enforce the system of continuity constraints \eqref{e:gsALT}. In this case, we consider the constraint potential
\eqb{l}
\Pi_{\sL} =  \ds \int_{\sL_0} q\,\big(\bar{g}_c + \bar{g}_s \big)\, \dif S~,
\label{e:LMgcL}
\eqe
where $q$ is the Lagrange multiplier. This potential guarantees unique solutions of $\alpha$ and  $\alpha_0\in [0,~2\pi]$  provided that $|\alpha - \alpha_0| < \pi/4$.  Furthermore, within this range, the gradient of $g_{\sL}:=\bar{g}_c + \bar{g}_s$ w.r.t $\alpha$ and  $\alpha_0$ is non-zero, which is the necessary condition to have a solution for the Lagrange multiplier $q$ (see e.g.~\cite{dimitri82}). 

Taking the variation of Eq.~\eqref{e:LMgcL} yields
\eqb{l}
\delta\Pi_{\sL} = \ds \int_{\sL_0} g_\sL\, \delta q\, \dif S - \ds \int_{\sL_0} q\, \big(\delta\btau\cdot \,\btheta + \delta\bn\cdot\tilde\bd + \delta\tilde\bn\cdot\bd\,\big) \,\dif S~,
\label{e:LMgc}
\eqe  
where $\btheta$, $\bd$ and $\tilde\bd$ are now defined as
\eqb{llll}
\btheta \dis (s_0 - c_0)\,\bn\times\tilde\bn~,\\[2mm] 
\bd \dis(s_0 + c_0)\,\bn + (s_0 - c_0)\,\bnu~,\\[2mm]
\tilde\bd \dis (s_0 + c_0)\,\tilde\bn + (s_0 - c_0)\,\tilde\bnu~.
\label{e:defdLM}
\eqe
Here, we find  the bending moment $ m_\tau = m_{\tilde\tau} = -q $, which is also transmitted exactly across the interface. The linearization of Eq.~\eqref{e:LMgc} can be found in Appendix~\ref{s:linpen}~.
\section{Numerical examples}\label{s:examples}
In this section, the performance of the proposed shell formulation is illustrated by several benchmark examples, considering both linear and non-linear problems. The computational results are verified by available reference solutions. 

\subsection{Linear problems}
For the linear problems discussed below, we will consider two material models: the \textit{Koiter shell material model} of Eq.~\eqref{e:Koiter}, and the \textit{projected shell material model} obtained from the numerical integration of Eqs.~\eqref{e:M2Dab2} and \eqref{e:case2NH}, where $\tilde\Lambda = E\,\nu/\big[(1+\nu)(1-2\,\nu)\big]$ and $\tilde\mu = E/\big[2\,(1+\nu)\big]$. For the sake of simplicity, hereinafter we denote the former as the \textit{Koiter model} and the latter as the \textit{projected model.} Furthermore, all physical quantities are introduced in terms of unit length $L_0$ and unit stress $E_0$.
\subsubsection{Pinching of a hemisphere}
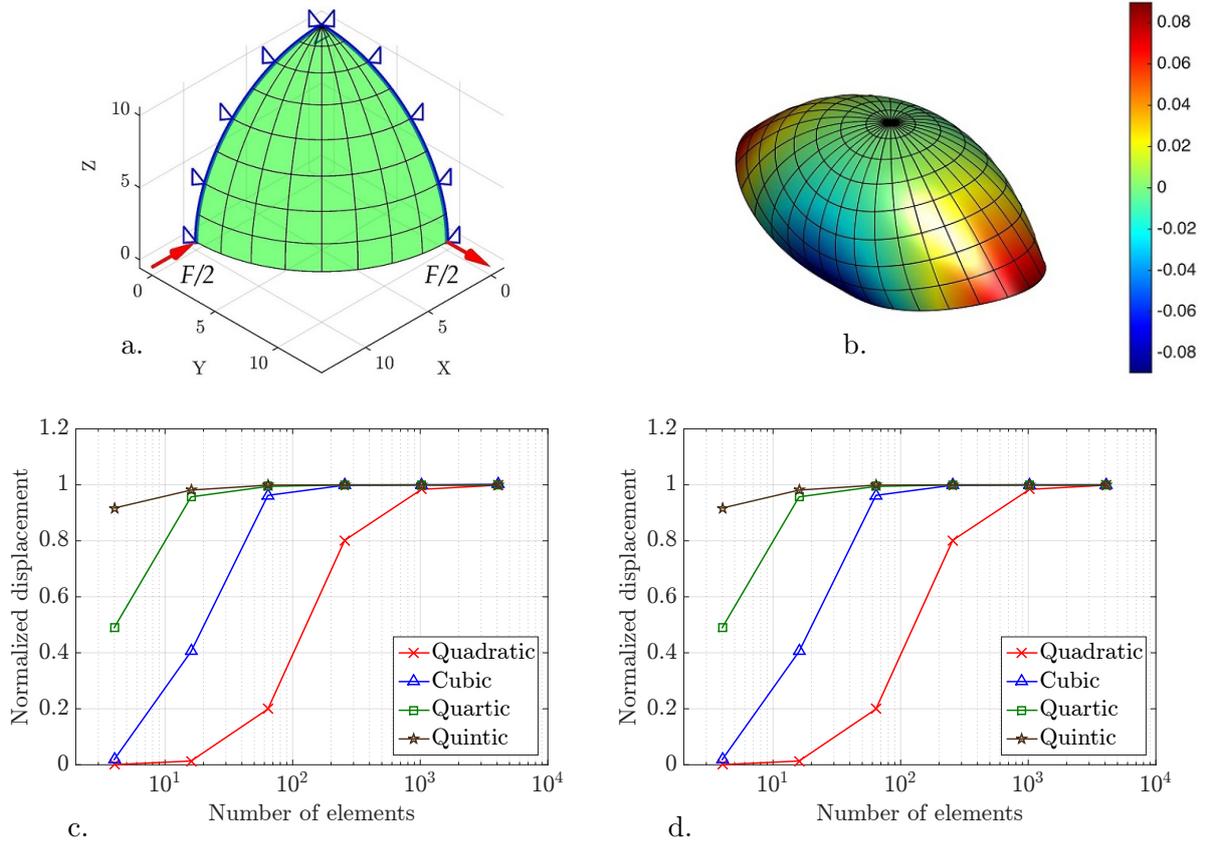
\begin{figure}[!ht]
\begin{center} \unitlength1cm
\unitlength1cm
\begin{picture}(0,10.7)
\put(-8.0,5.5){\includegraphics[height=60mm]{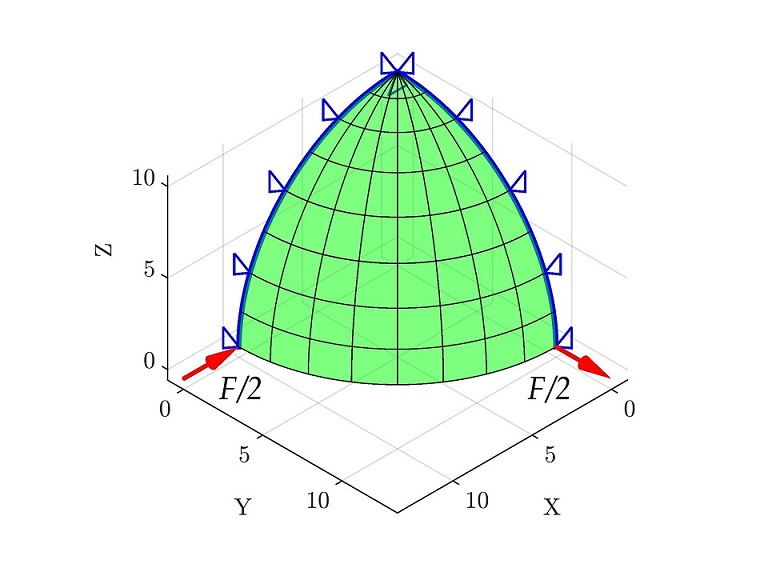}}
\put(-6.5,6.4){a.}
\put(0.0,5.5){\includegraphics[height=60mm]{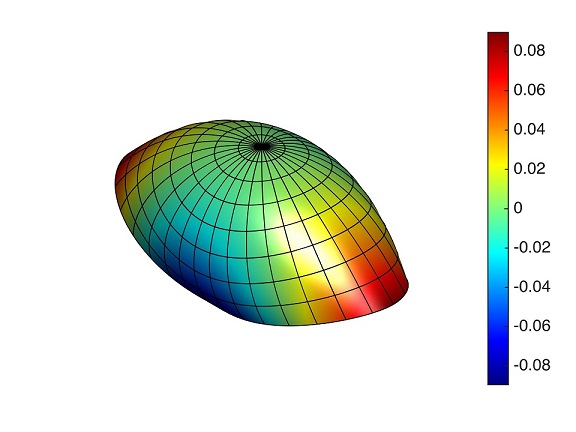}}
\put(3.0,6.4){b.}
\put(-8.0,0.1){\includegraphics[height=55mm]{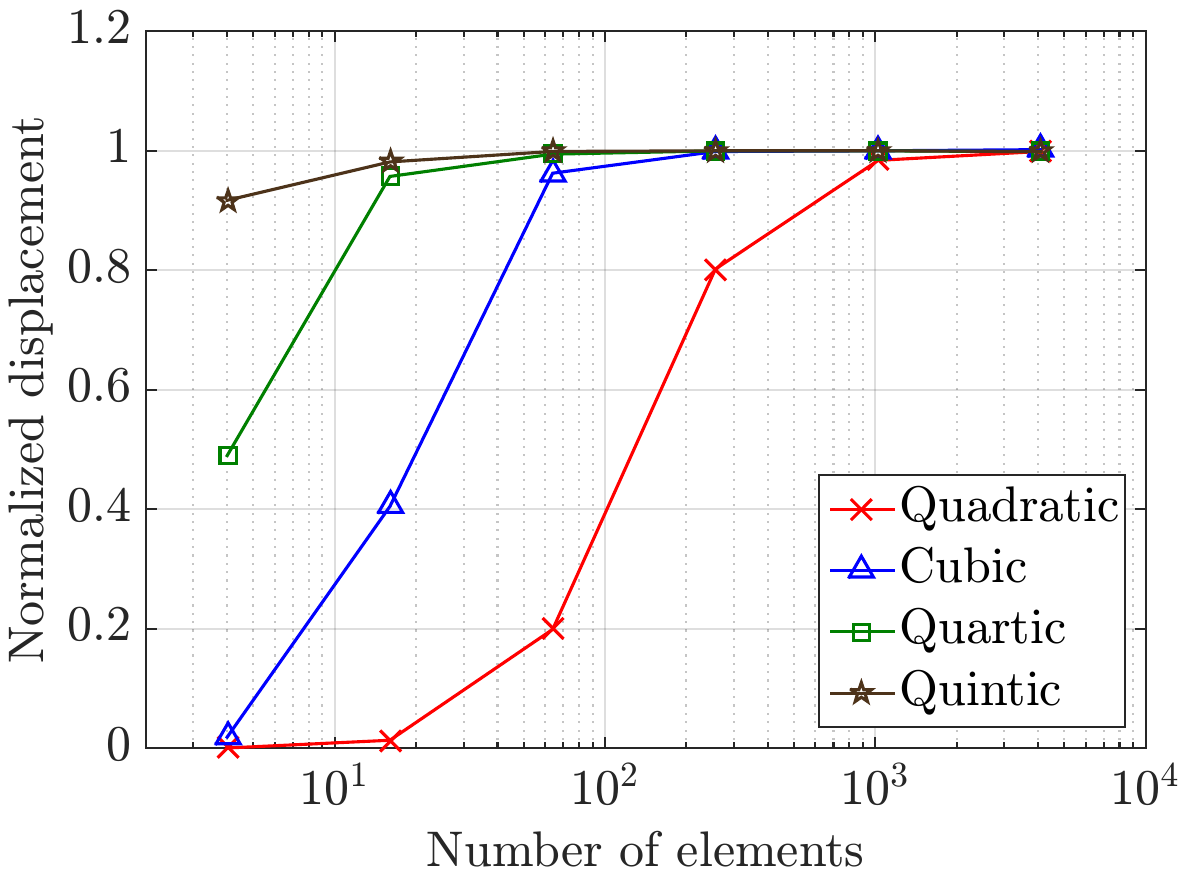}}
\put(-7.2,0.0){c.}
\put(0.0,0.1){\includegraphics[height=55mm]{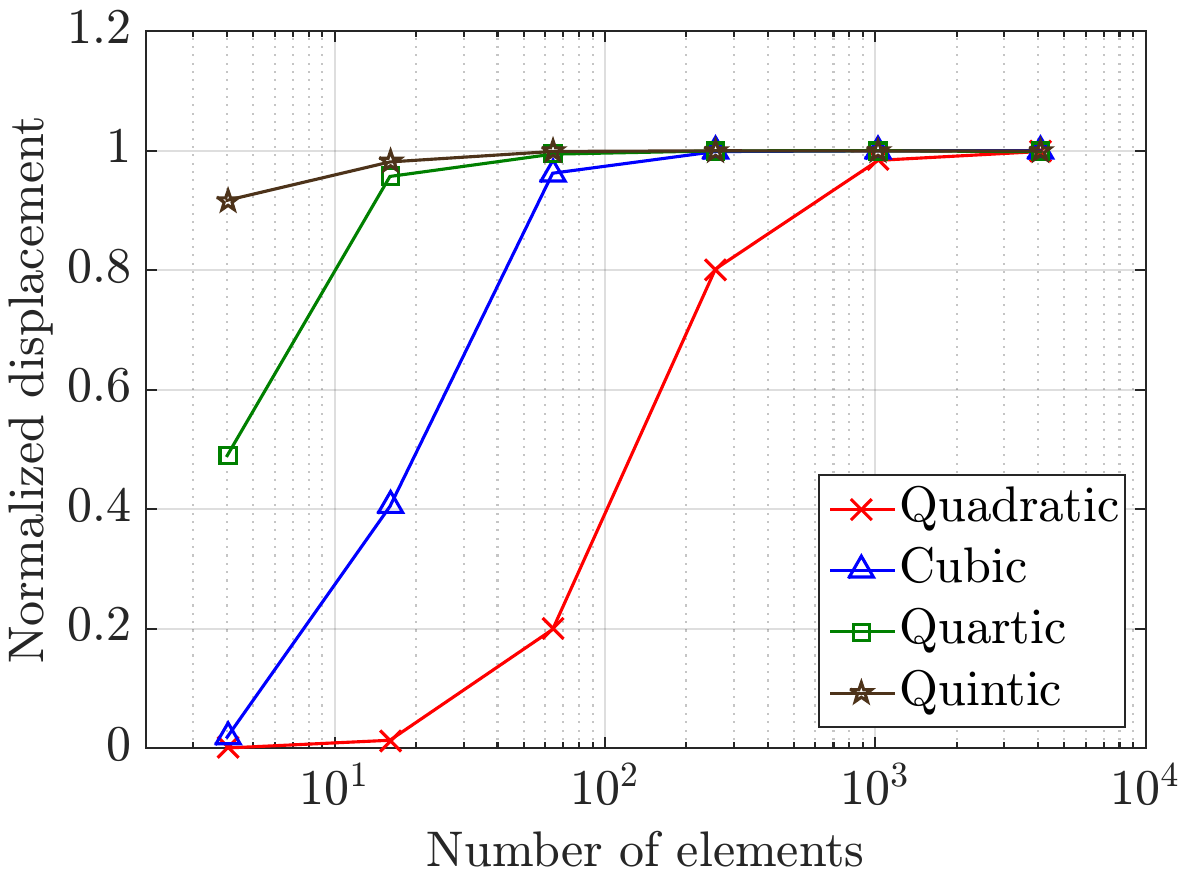}}
\put(0.7,0.0){d.}
\end{picture}
\caption{Pinching of a hemisphere: a. Undeformed configuration with boundary conditions. Here, the blue curves denote the symmetry lines. b. Deformed configuration (scaled $50$ times), colored by radial displacement. Radial displacement at the point load vs. mesh refinement for c.~the Koiter and d.~the projected shell model, considering various NURBS orders.}
\label{f:pinchHemF}
\end{center}
\vspace{-0.3cm}
\end{figure}

A hemisphere pinched by two pairs of diametrically opposite forces $F = 2\,E_0\,L_0^2 $ on the equator is examined in this example. The model parameters are adopted from \citet{belytschko85} as $ E = 6.825 \times 10^7 \,E_0$, $ \nu = 0.3 $, $ R = 10.0 \,L_0$ and $ T = 0.04 \,L_0$. Due to the symmetry, $ 1/4 $ of the hemisphere is modeled as shown in Fig.~\ref{f:pinchHemF}a. Here, the symmetry boundary conditions are applied on the $ Y = 0 $ and $ X = 0 $ planes by the penalty method of Eq.~\eqref{e:pen} with the penalty parameter $\epsilon=800\,E\,L_0^2 $ and the rigid body motions are restricted by fixing the top control point.
The benchmark reference solution for the radial displacement under the point loads is $0.0924 \,L_0$ \citep{belytschko85,macneal85, morley78}. As shown in Fig.~\ref{f:pinchHemF}, the numerical results converge to the reference solution as the mesh is refined and the NURBS order is increased. Here, the radial displacement of the force is normalized by the reference solution. As can be observed in Figs.~\ref{f:pinchHemF}c-d, for linear elastic deformations, both the Koiter and the projected model are identical. This is also the case for several other problems examined here. In the following, only the results of one of the models are reported if the differences are negligible.

\subsubsection{Simply supported plate under sinusoidal pressure loading}

\begin{figure}[!ht]
\begin{center} \unitlength1cm
\unitlength1cm
\begin{picture}(0,10.8)
\put(-8.0,5.5){\includegraphics[height=60mm]{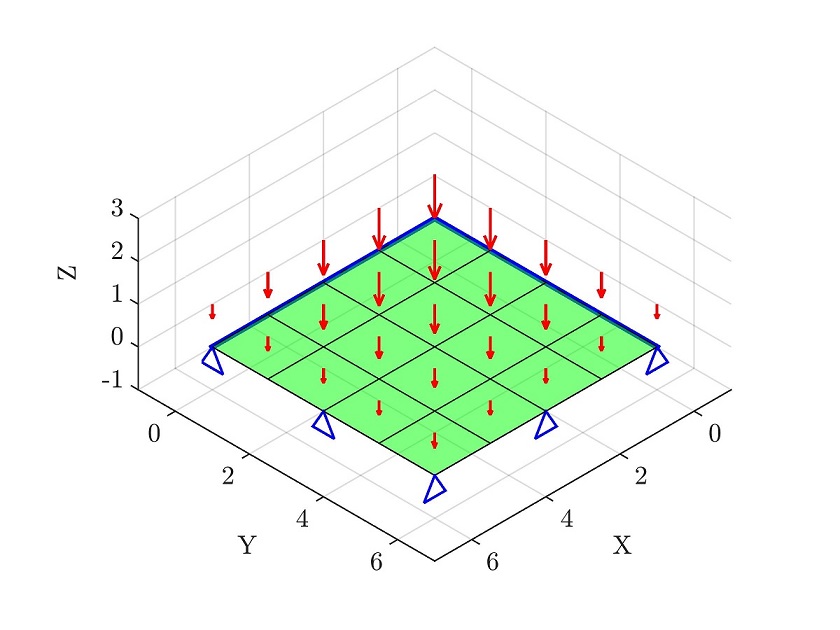}}
\put(0.0,5.5){\includegraphics[height=60mm]{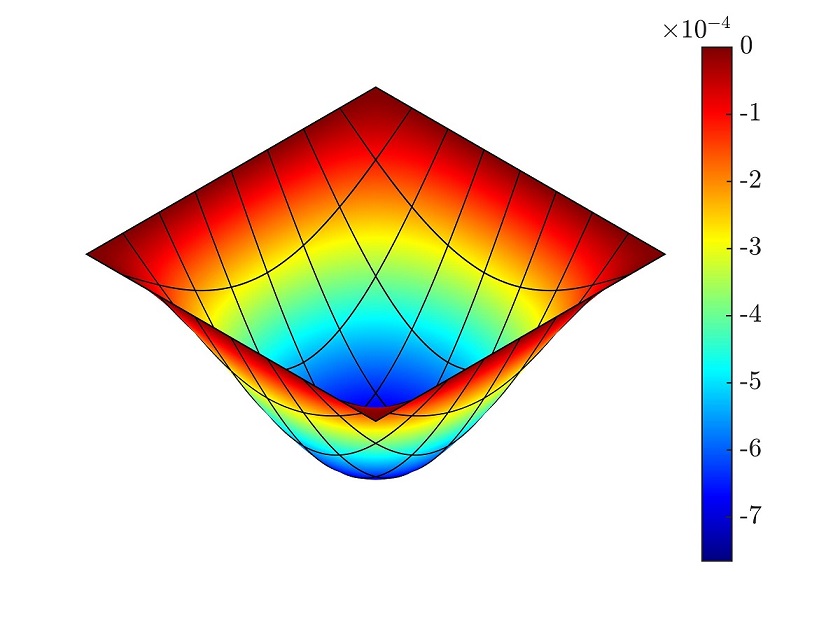}}
\put(-8.0,0.1){\includegraphics[height=55mm]{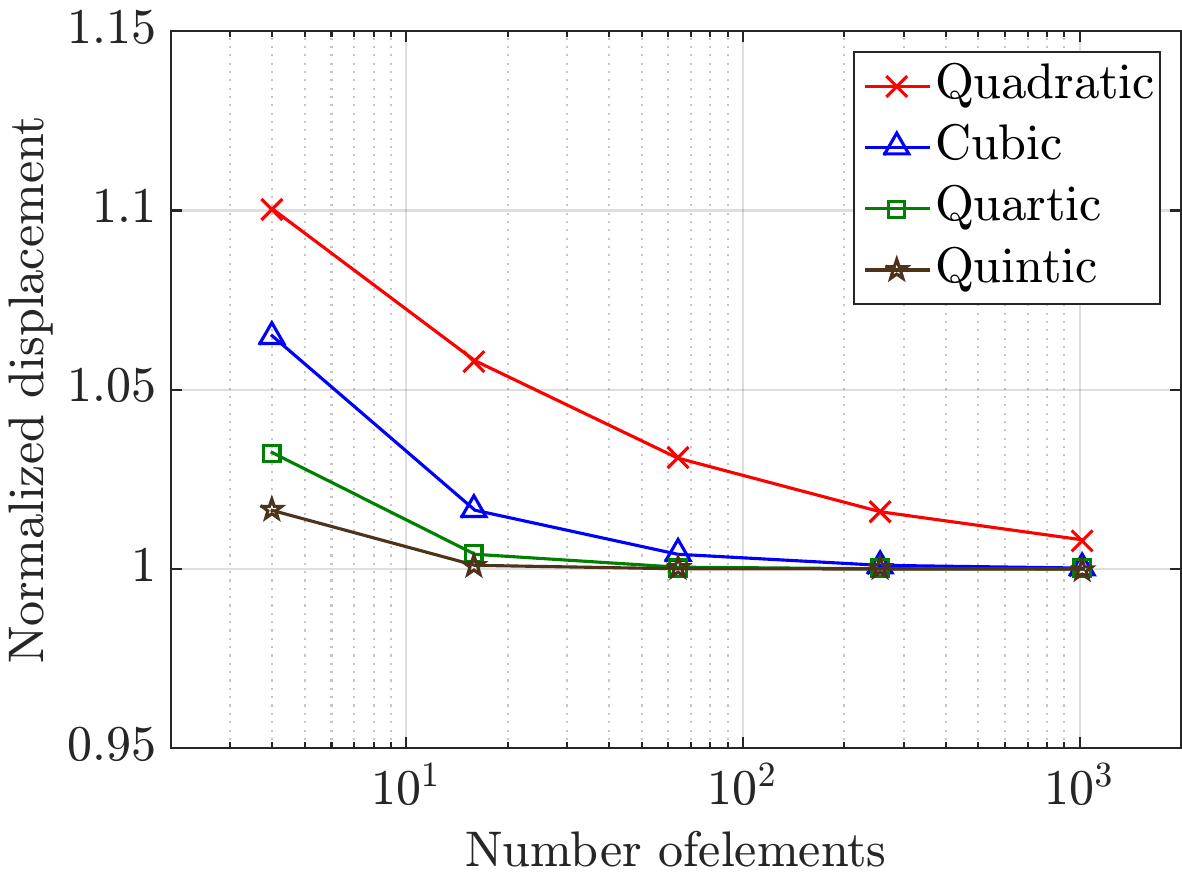}}
\put(0.0,0.1){\includegraphics[height=55mm]{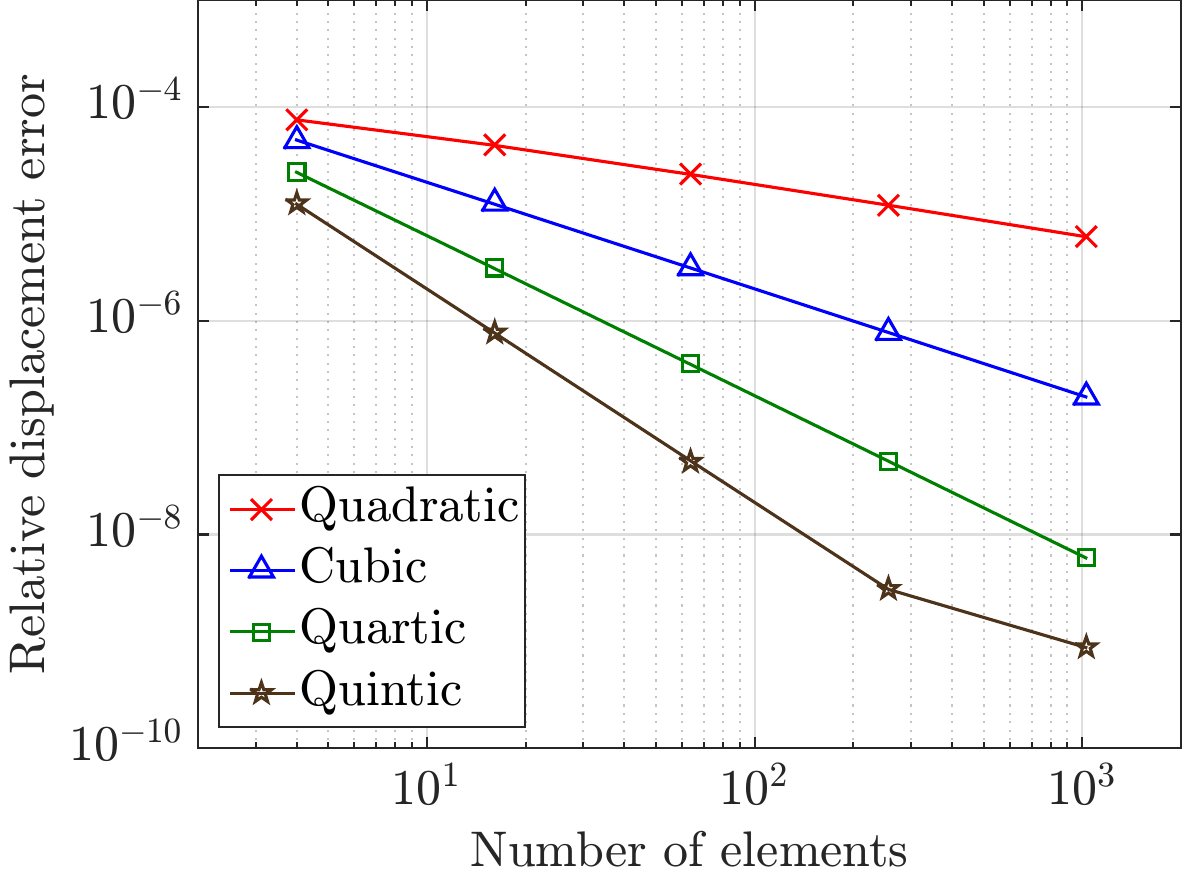}}

\put(-6.5,6.4){a.}
\put(3.0,6.4){b.}
\put(-7.2,0.0){c.}
\put(0.7,0.0){d.}
\end{picture}
\caption{Simply supported plate under sinusoidal pressure: a.~Initial configuration with boundary conditions, b.~deformed configuration (scaled $10^4$ times) colored by the vertical displacement, c.~displacement of the plate center normalized w.r.t.~the analytical solution and d.~relative error of the displacement.}
\label{f:plateSin1}
\end{center}
\end{figure}
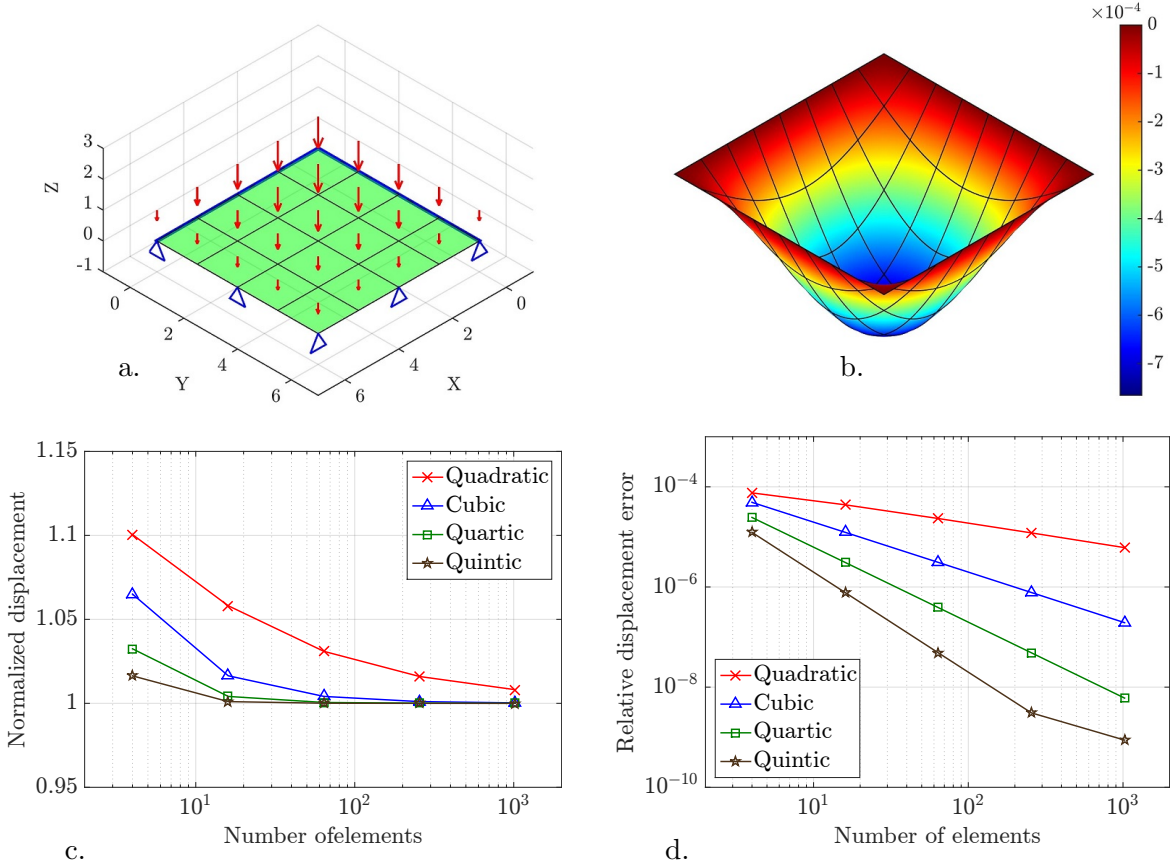

As a second example, we analyze a plate with size $L\times L = 12\times 12 \,L_0^2$, thickness $ T = 0.375 \,L_0$, Young's modulus $ E = 4.8 \times 10^5 \,E_0$, Poisson's ratio $ \nu = 0.38 $,  subjected to sinusoidal pressure $p(x,y) = p_0\,\sin(\pi\,x/L)\sin(\pi\,y/L) \,E_0$. According to \textit{Navier}'s solution \citep{ugural09}, the maximum deflection is at the middle point and given by
\eqb{lll}
w_\mathrm{max} =  \ds\frac{p_0\,L^{\textcolor{darkgreen}{4}}}{4\,\pi^4\,D}~,
\label{e:plate}
\eqe
where $D:=E\,T^3/12\,(1-\nu^2)$ is the flexural rigidity of the plate. The setup of the computational model is shown in Fig~\ref{f:plateSin1}a. Only $1/4$ of the plate is modeled using symmetry boundary conditions enforced by the penalty method of Sec.~\ref{s:conti} with the penalty parameter $\epsilon = \ds10^{-2}\,n^{p-1}\,E\,L_0^2$, where $p$ is the NURBS order and $n$ is the number of elements in each direction. Fig.~\ref{f:plateSin1}b shows the deformed plate with the Koiter model and Fig.~\ref{f:plateSin1}c shows the convergence of the computational solution to the analytical one as the mesh is refined. For the comparison, the vertical displacement at the center of the plate is normalized by the analytical solution given in Eq.~\eqref{e:plate}. The corresponding relative error is shown in Fig.~\ref{f:plateSin1}d. As expected, more accuracy is gained by increasing the NURBS order.

\subsubsection{Pinching of a cylinder}\label{s:LinCyl}
Next, we consider the pinched cylinder test with rigid diaphragms at its ends. It is designed to examine the performance of shell elements in inextensional bending modes and complex membrane states \citep{belytschko85}. The analytical solution for this problem was introduced by \citet{flugge62} based on a double Fourier series, see Appendix~\ref{s:fourier}. 

\begin{figure}[!ht]
\begin{center} \unitlength1cm
\begin{picture}(0,10.8)
\put(-8.0,5.5){\includegraphics[height=60mm]{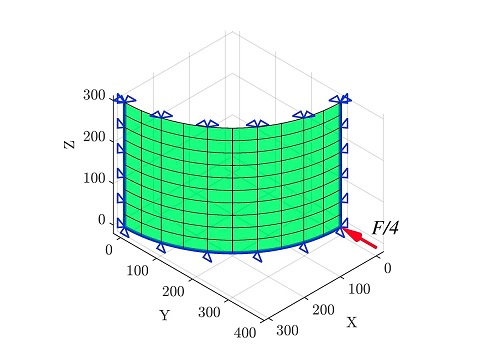}}
\put(0.0,5.5){\includegraphics[height=60mm]{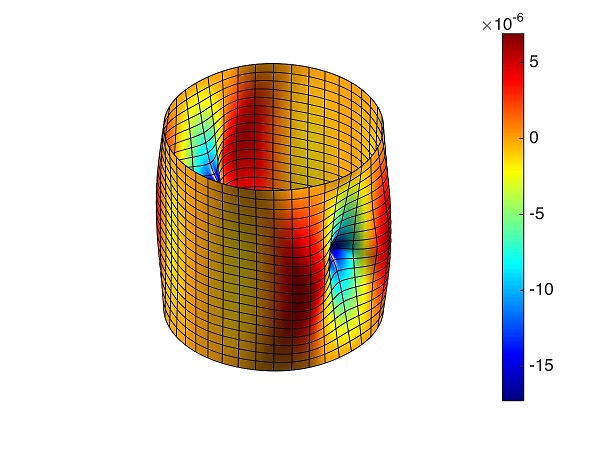}}
\put(-8.0,0.1){\includegraphics[height=54mm]{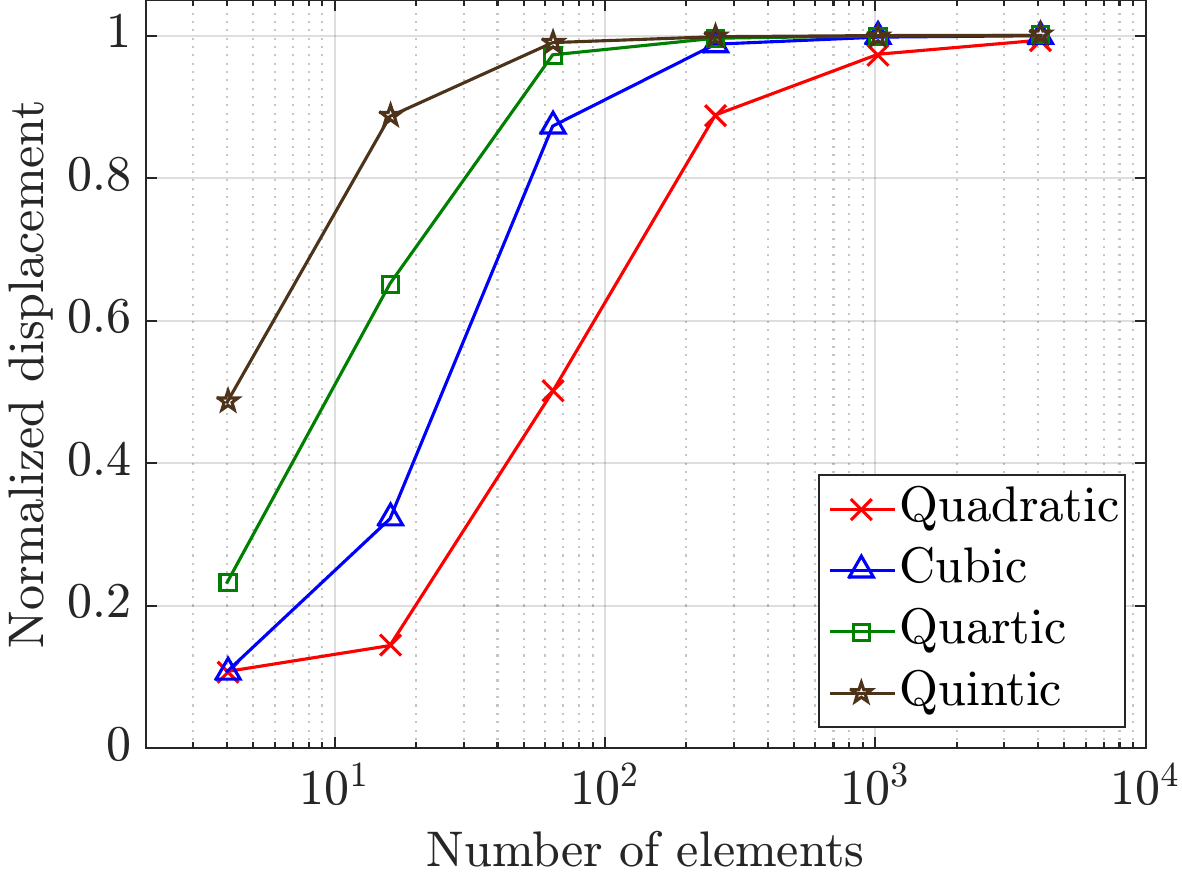}}
\put(-0.1,0.1){\includegraphics[height=55mm]{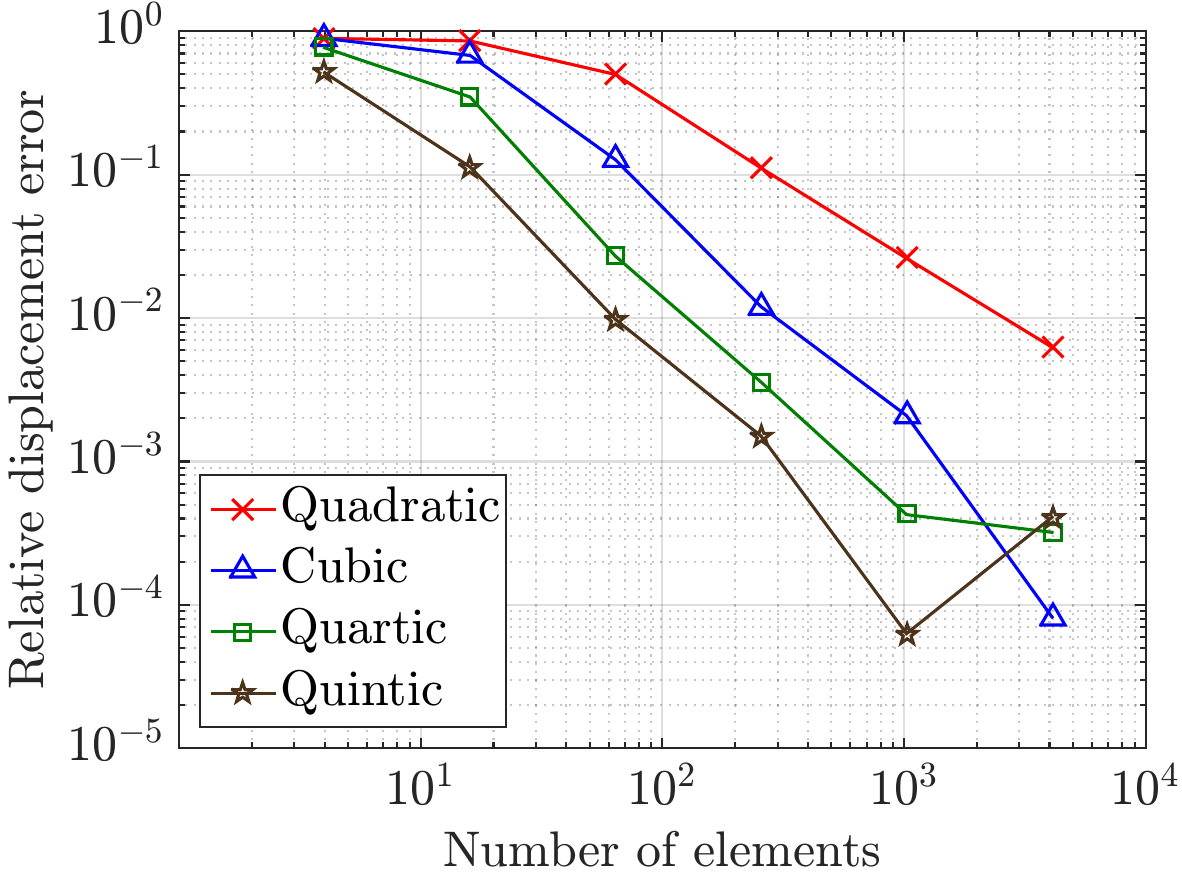}}

\put(-6.5,6.4){a.}
\put(2.0,6.4){b.}
\put(-7.2,0.0){c.}
\put(0.7,0.0){d.}
\end{picture}
\caption{Pinching of the cylinder with rigid end diaphragms:  
a.~Setup of the computation model, b.~deformed shell (scaled $10^6$ times) colored by the radial displacement, c.~normalized radial displacement at the point load and d.~error w.r.t.~the reference solution as the mesh is refined.}
\label{f:pinchCylR1}
\end{center}
\end{figure}
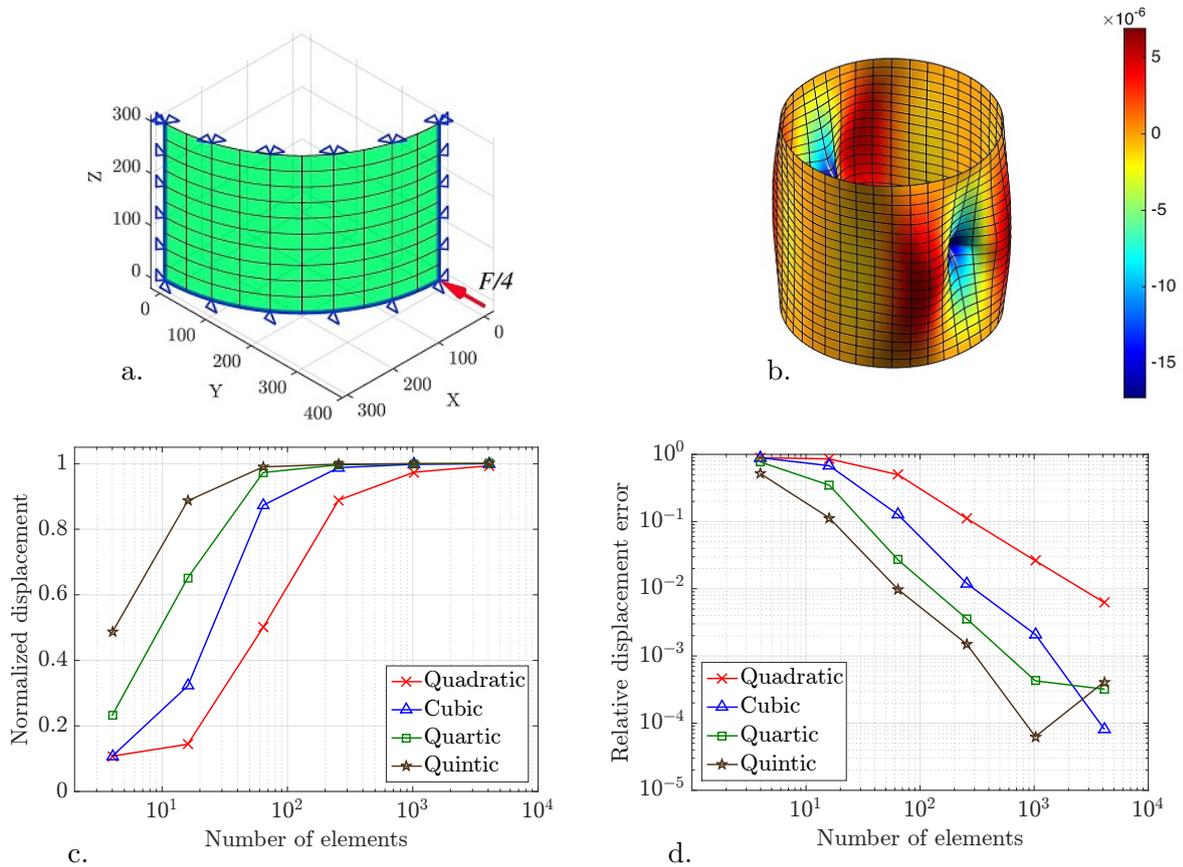

The parameters are adopted from \citet{belytschko85} as $ E = 3 \times 10^6 \,E_0 $, $ \nu = 0.3 $, $ R = 300 \,L_0 $, $ L = 600 \,L_0 $, $ T = 3 \,L_0 $, where $ E $ is Young's modulus, $\nu$ is Poisson's ratio; $R$, $L$ and $T$ are radius, length and thickness of the cylinder, respectively. For the FE computation, $ 1/8 $ of the cylinder is modeled  due to its symmetry as is shown in Fig.~\ref{f:pinchCylR1}a. The symmetry boundary conditions are enforced by the penalty method discussed in Sec.~\ref{s:conti}. The penalty parameters used are $\epsilon = \ds2\times10^2\,n_l^{p-1}\,E\,L_0^2$ for the axial symmetry, and $\epsilon = \ds2\times10^2\,n_t^{p-1}\,E\,L_0^2$ for the circumferential symmetry. Here, $n_t$ and $n_l$ are the number of elements in circumferential and axial directions, respectively.

Two opposing pinching forces $ F =1 \,E_0\,L_0^2$ are applied at the middle of cylinder. For the symmetric model, a quarter of the force is applied. The deflection due to the loads is measured for comparison with the reference solution. For $80 \times 80$ Fourier terms, the deflection under the point load is $ 1.82488 \times 10^{-5} $, which is the reference value commonly used in the literature. However, for $8192 \times 8192$ Fourier terms the converged solution is $ 1.82715781 \times 10^{-5} $ and  for $2^{14} \times 2^{14}$ Fourier terms, we obtain the solution of $1.82715797 \times 10^{-5}$.

Fig.~\ref{f:pinchCylR1}c shows the convergence behavior of the computed displacement at the loading point w.r.t.~the analytical solution as the mesh is refined. From Fig.~\ref{f:pinchCylR1}d, we observe that the relative error only converges up to a certain point and then seems to get stuck. This is due to several numerical and theoretical reasons. Firstly, although the NURBS order can be increased for the patches, the patch boundaries are only $G^1$-continuous according to Eq.~\eqref{e:pen}. This affects higher order NURBS, especially at the boundaries around the singular point load. Secondly, the analytical solution represents the concentrated load by a pressure distribution in the form of a Dirac delta function. Numerically, this is represented by a Fourier series and thus the Gibbs effect can lead to a loss in numerical accuracy for high order terms (see Appendix~\ref{s:fourier}). Therefore, the Fourier solution is truncated at some point and we cannot expect the error to go down to machine precision as the mesh is refined. We note that, although both the results of the \textit{Koiter shell material model} of Eq.~\eqref{e:Koiter}, and the \textit{projected shell material model} of Eqs.~\eqref{e:M2Dab2} and \eqref{e:case2NH} are identical, the computational time of the former is roughly half of the latter (considering three quadrature points through the thickness). Thus, the presented \textit{Koiter shell material model} is more efficient for the same accuracy.

\subsection{Nonlinear problems}
In the following, several nonlinear test cases are presented to illustrate the robustness and accuracy of the proposed shell formulation.

\subsubsection{Pure bending of a flat strip} \label{s:PB}
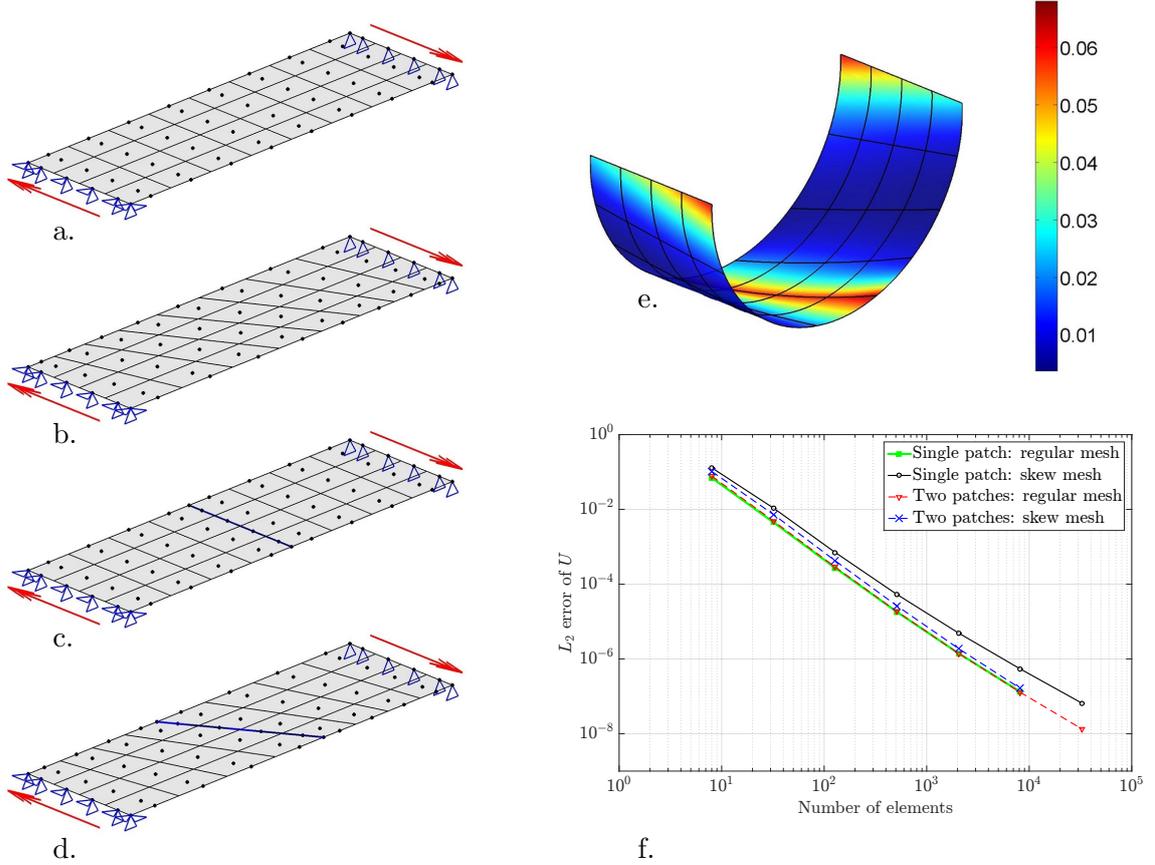
\begin{figure}[ht]
\begin{center} \unitlength1cm
\unitlength1cm
\begin{picture}(0,11.5)
\put(-7.5,0.5){\includegraphics[height=30mm]{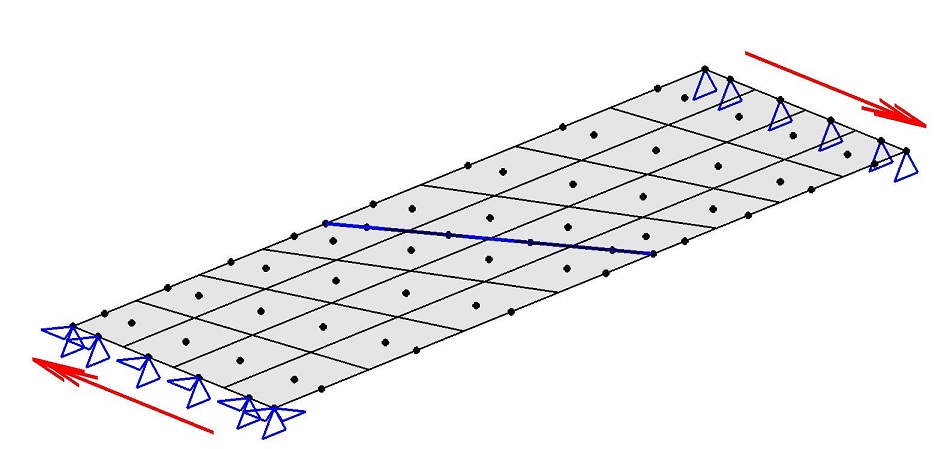}}
\put(-7.5,3.2){\includegraphics[height=30mm]{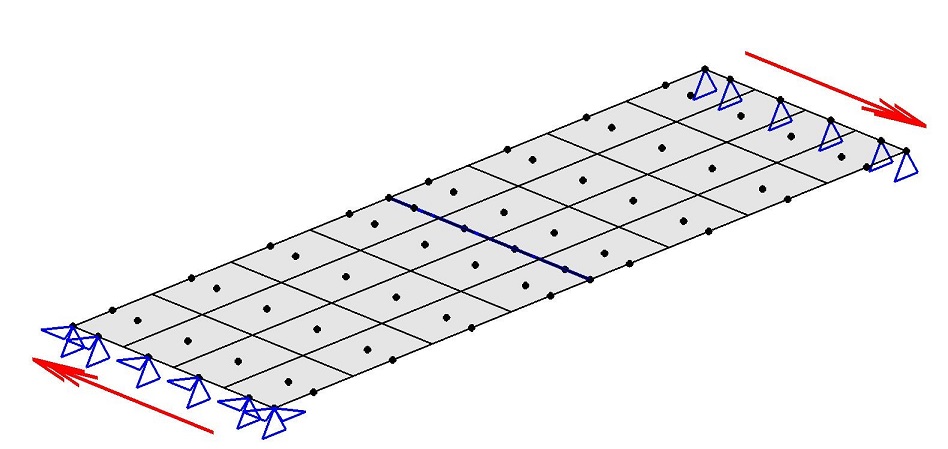}}
\put(-7.5,5.9){\includegraphics[height=30mm]{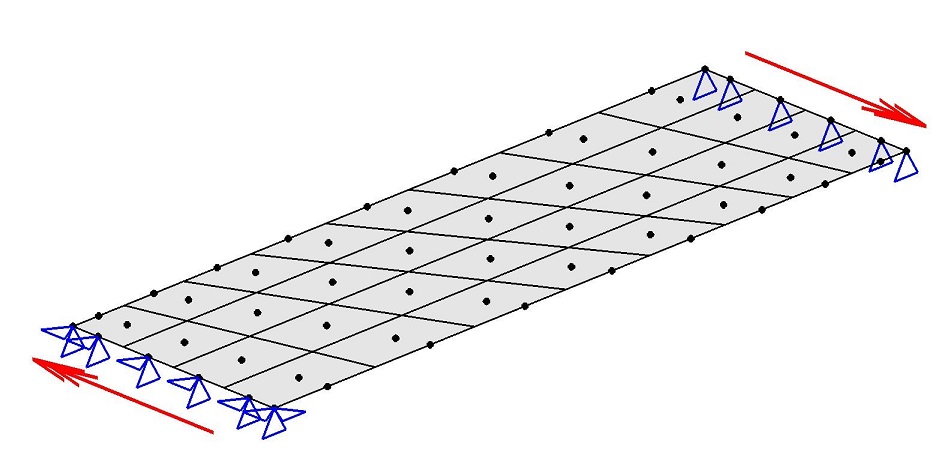}}
\put(-7.5,8.6){\includegraphics[height=30mm]{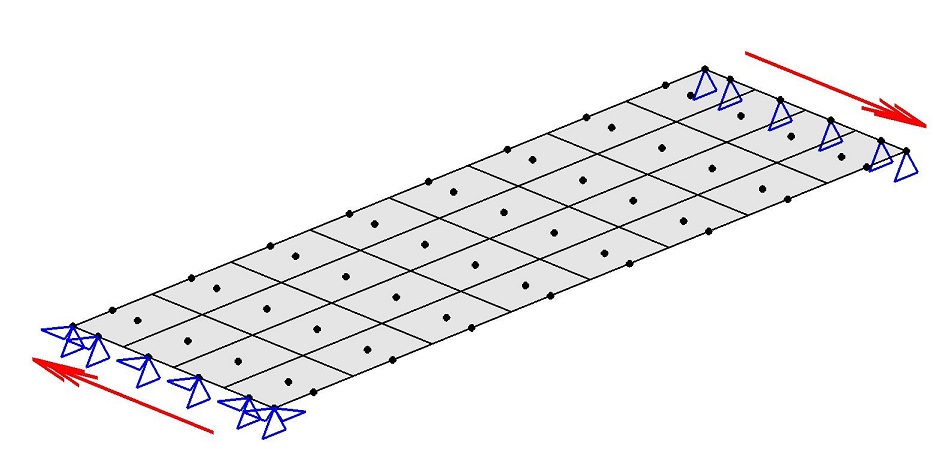}}
\put(-0.5,6.0){\includegraphics[height=60mm]{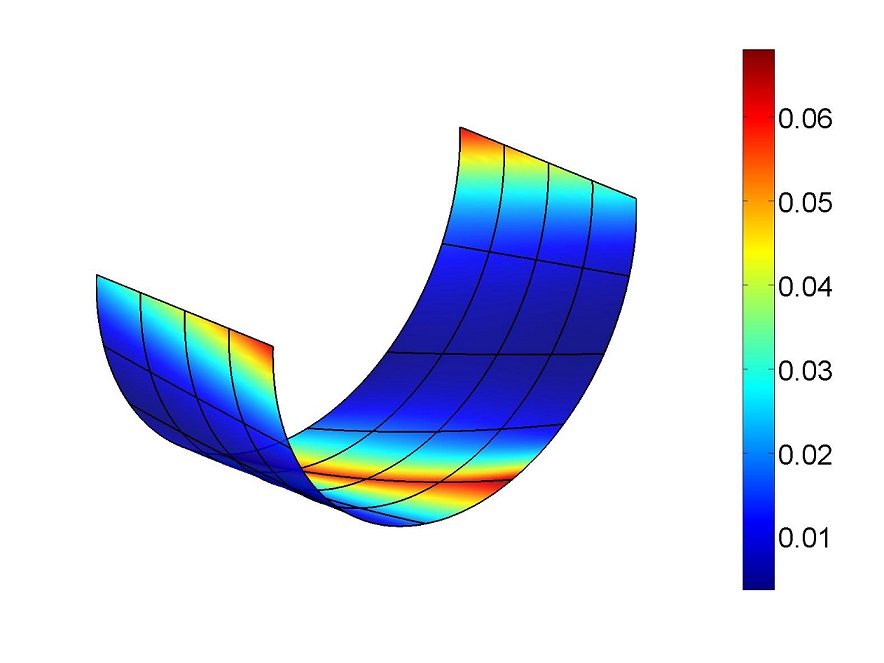}}
\put(-0.5,0.5){\includegraphics[height=55mm]{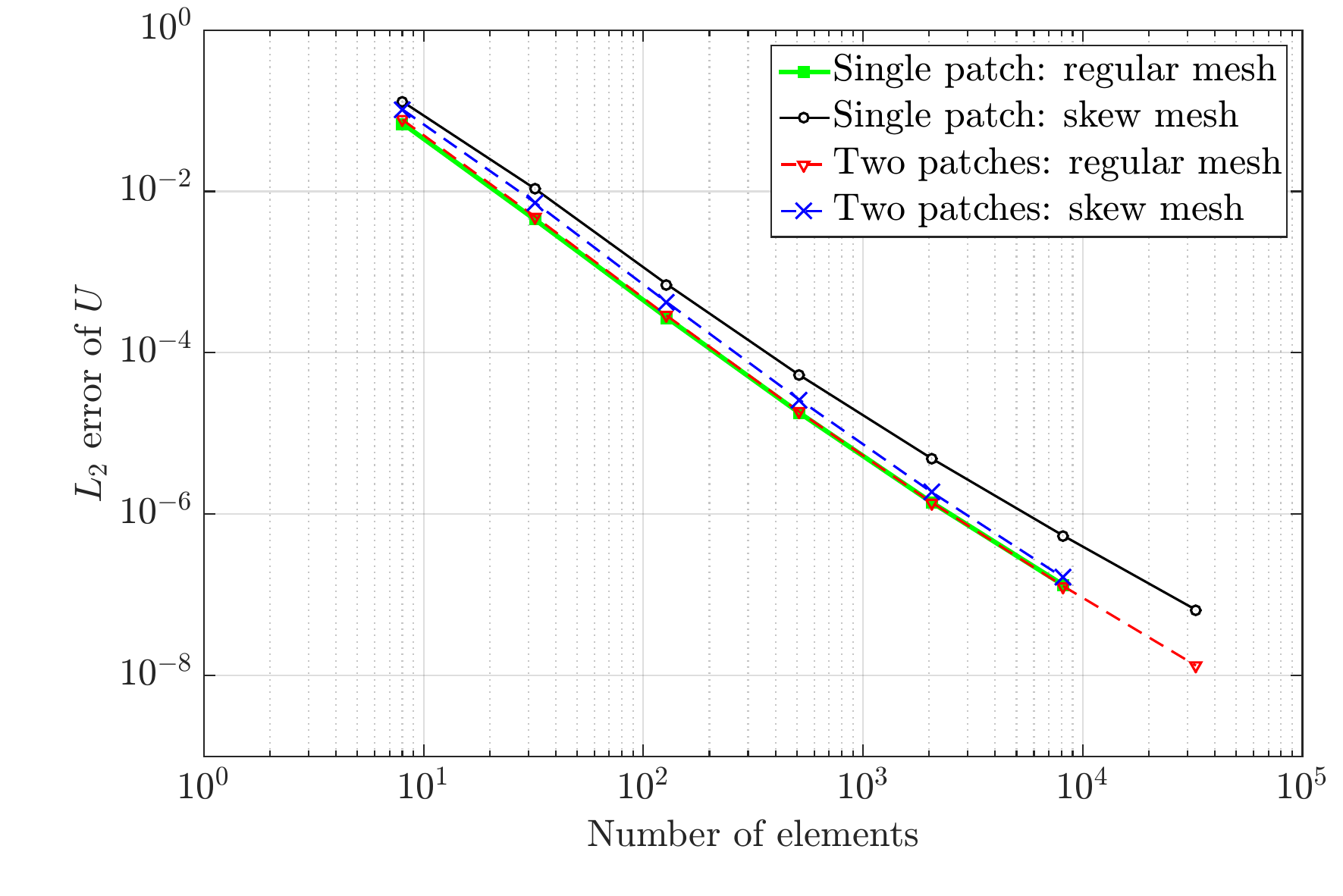}}
\put(-6.7,8.4){a.}
\put(-6.7,5.7){b.}
\put(-6.7,3.0){c.}
\put(-6.7,0.2){d.}
\put(1,7.5){e.}
\put(1,0.2){f.}

\end{picture}
\caption{Pure bending of a flat strip considering: a.~Single patch with regular mesh, b.~single patch with skew mesh, c.~two patches with regular mesh, and d.~two patches with skew mesh. e.~Deformed configuration for the mesh in d. The color here shows the relative error of mean curvature $H$. f.~$L_2$ error of $\bu$  (Eq.~\ref{e:pbErrU}) w.r.t.~mesh refinement.}
\label{f:convM_UCSL}
\end{center}
\end{figure}

We first examine the pure bending of a flat strip with the dimensions $S\times L$. The strip is fixed as is shown in Figs.~\ref{f:convM_UCSL}a-d in order to apply the moment $M$ at the two ends of the strip. The Canham material model, Eq.~\eqref{e:WmsF}, is used. For this problem, the analytical solution is given by \cite{shelltheo}. Accordingly, the strip is deformed into a curved sheet with dimensions $s\times l = \lambda_1\,S \times\lambda_2\,L$  with constant curvature $\kappa_1$ that is linearly related to $M$ as
\eqb{l}
M = c\,\kappa_1~.
\label{e:M_pb}
\eqe
Since all the boundaries are free to move, we expect that the in-plane stress components vanish. This condition leads to the stretches $\lambda_1$ and $\lambda_2$ in terms of the applied moment $M$ as
\eqb{l}
\lambda_2=\lambda_1/a_0~,\quad a_0:=\ds\frac{M^2}{2\mu c} + \sqrt{\left(\frac{M^2}{2\mu c}\right)^2 + 1}~,
\label{e:a0}\eqe
and
\eqb{l}
\lambda_1 = \ds\sqrt{-\bar\mu\,(a_0^2+1) + \sqrt{\bar\mu^2(a_0^2+1)^2 + a_0^2\,(4\,\bar\mu + 1)}}~,\quad \bar\mu :=\frac{\mu}{2\Lambda}~.
\eqe

To assess the FE computation, four different mesh schemes using quadratic NURBS are considered as is shown in Figs.~\ref{f:convM_UCSL}a-d: 1.~A single patch with regular mesh; 2.~A single patch with skew mesh with distortion ratio $r=1/5$;  3.~A double patch with regular mesh; and 4.~A double patch with skew mesh with the same distortion ratio. Here, the mesh refinement is carried out by the knot insertion algorithm (see e.g~\cite{hughes05}). In particular for the second scheme, the skew mesh is obtained by distortion of the knot vectors as explained in detail in Appendix~\ref{s:mesh}. 

For all mesh schemes, the material parameters $\mu = 10\,L^2/c$, $\Lambda = 5\,L^2/c$ and the bending moment $M = 1 \,c/L$ are applied (setting $m_\tau = M$ in Eq.~(\ref{e:fext}.4)).  For mesh schemes 3 and 4, the $G^1$-continuity constraint is used between the two patches (see Sec.~\ref{s:conti}) with $\epsilon = 10\;000\,(n_S\times n_L)\,L/c$, where $n_S$ and $n_L$ are the number of elements along the length and width of the strip, respectively.
 Fig.~\ref{f:convM_UCSL}e shows the deformed mesh and the mean curvature error for mesh $4\times 8$ of scheme 4. Fig.~\ref{f:convM_UCSL}f shows the $L_2 $ error of the displacement field for the four considered mesh schemes. Here, the $L_2 $ error is  defined as 
\eqb{l}
\| \bu^h - \bu \|_{L_2} := \sqrt{\ds\frac{1}{S\,L}\,\int_{\sS_0} \left(\bu^h - \bu\right)^2 \, \dif A},
\label{e:pbErrU}\eqe
where $ \bu^h $ is the displacement obtained from the FE analysis and $\bu $ is the corresponding analytical quantity calculated at each point on $\sS$. Here, $\bu$ can be computed for any applied moment by using the parametrization described in \cite{shelltheo}.

The convergence observed in Fig.~\ref{f:convM_UCSL}f verifies the presented finite element formulation. It also shows, that the accuracy of the double patch meshes is of the same order as the single patch results. This indicates the effectiveness of the penalty constraint presented in Sec.~\ref{s:sym}.  
\subsubsection{Pure bending of a folded strip} \label{s:PBfolded}
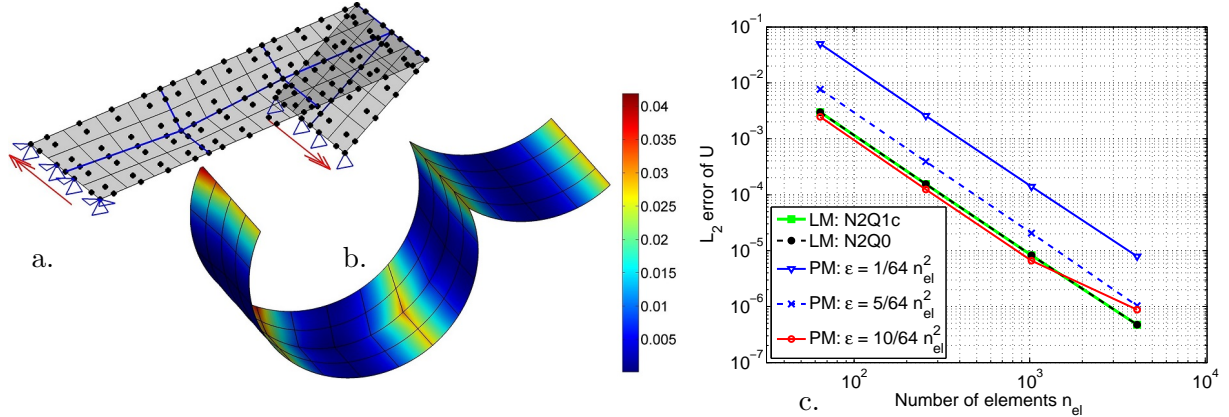
\begin{figure}[ht!]
\begin{center} \unitlength1cm
\begin{picture}(0,5.5)
\put(-7.9,0.5){\includegraphics[height=50mm]{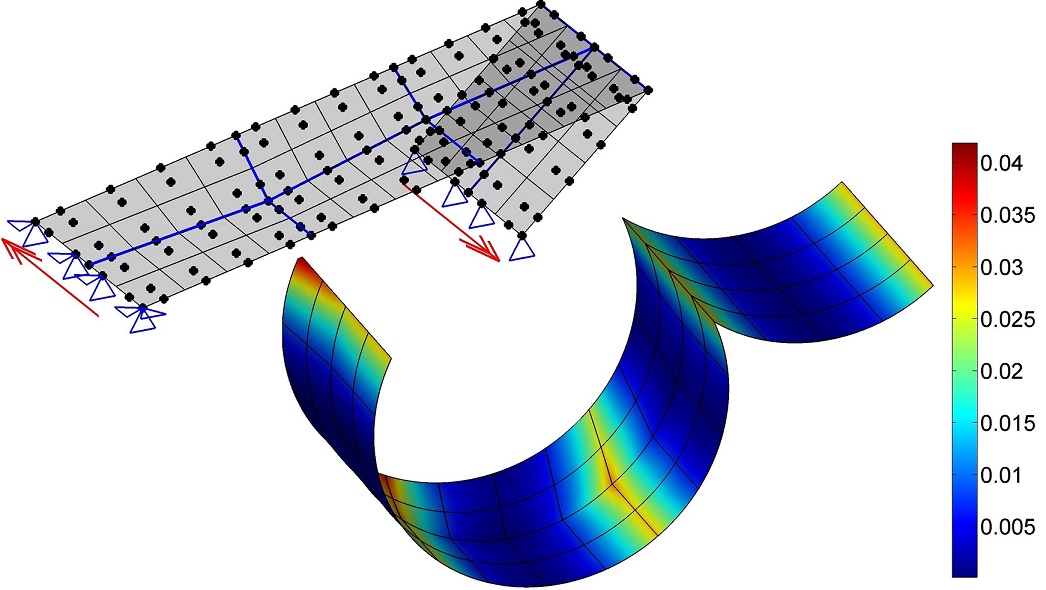}}
\put(1.1,0.0){\includegraphics[height=56mm]{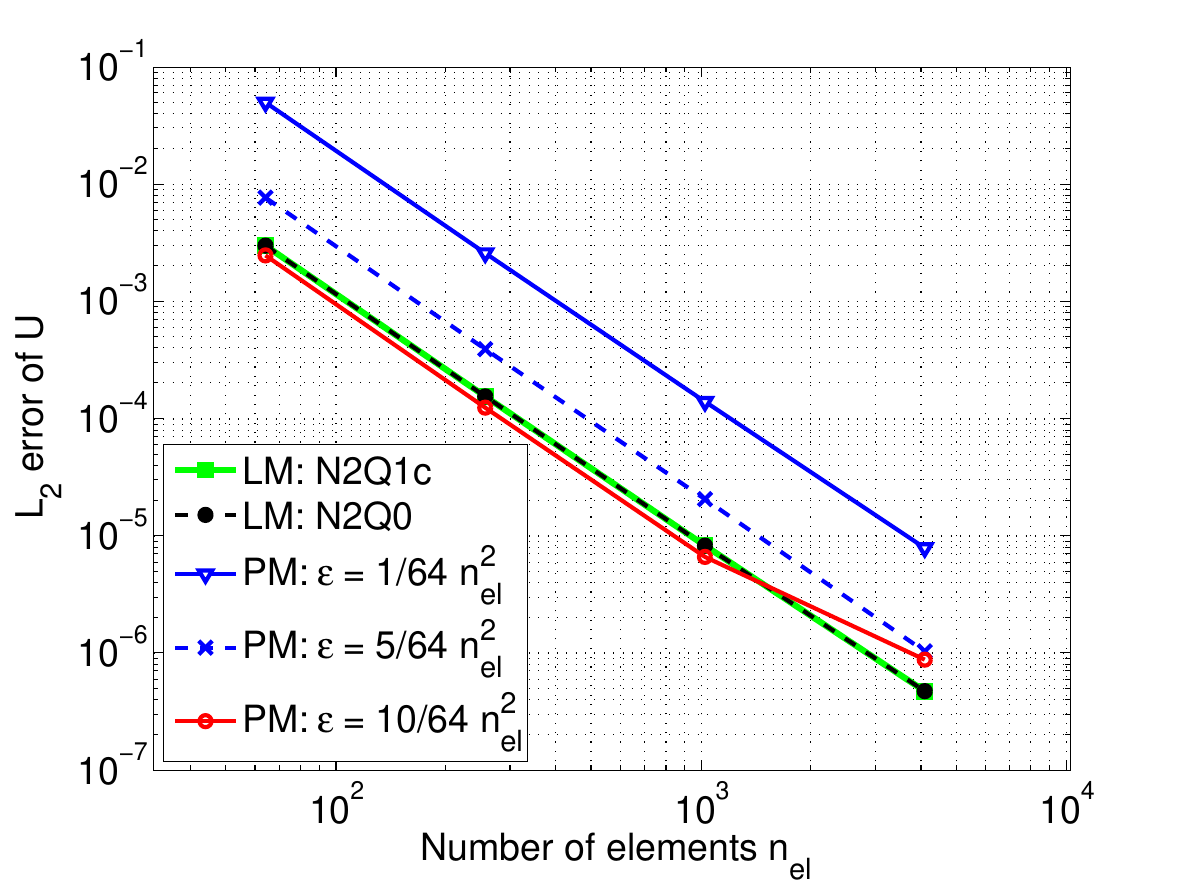}}
\put(-7.6,2){a.}
\put(-3.5,2){b.}
\put(2.5,0.1){c.}
\end{picture}
\caption{Pure bending of a folded strip: a.~Initial FE configuration and boundary conditions (for $S_1 + S_2=\pi L$, opening angle $\beta_0 = \pi/6$, discretized with quadratic NURBS elements); b.~current FE configuration and relative error of $H$ w.r.t $H_\mathrm{exact} = 0.8\,(1/L)$ for an imposed bending moment $M = 1.6~c/L$; c.~$L_2$ error of $\bu$ (Eq.~\eqref{e:pbErrU}) w.r.t. mesh refinement, considering the penalty method (PM) and the Lagrange multiplier method (LM).}
\label{f:bend_fold}
\end{center}
\end{figure}

Next, we demonstrate the robustness of the edge rotation constraint presented in Sec~\ref{s:sym} for multiple patch interfaces with kinks. For this purpose, we reconsider the pure bending test of the previous section, but here the mesh consists of 8 patches and has a kink at $3/4\,\pi\,L$ from the left end, as shown in Fig.~\ref{f:bend_fold}a. The material of the strip is the same as the example in Sec.~\ref{s:PB}. The deformed configuration is shown in Fig.~\ref{f:bend_fold}b. 

Here, the continuity between patch interfaces including both $G^1$-continuity and the kink constraints are enforced by the penalty method (PM) \eqref{e:pen} and the Lagrange multiplier method (LM) \eqref{e:LMgcL}. Further, for LM, we consider a constant  and a linear interpolation scheme of the Lagrange multiplier $q$ denoted as {N2Q0} and {N2Q1c}, respectively.  We note that at patch junctions, $q$ can jump across interfaces as there is a change of patch pairings. In this case, {N2Q0} can automatically capture the jump since $q$ is interpolated discontinuously over the elements, whereas for {N2Q1c}, the continuous interpolation of $q$ would result in over-constraining, i.e. is LBB-unstable. However, this problem can be simply solved by repeating the pressure degrees of freedom at patch junctions for each interface so that the jump can be captured. 

In order to assess the FE results, the $L_2$ error defined by Eq.~\eqref{e:pbErrU} is computed. As the analytical solution, given by Eqs.~\eqref{e:M_pb} and \eqref{e:a0}, is also valid for this problem, we can use the same procedure as in Sec~\ref{s:PB} to compute $\bu$ and $\bu^h$. It only needs to be adapted to include the kink.

Fig.~\ref{f:bend_fold}c shows the convergence of the computed $L_2$ error as the mesh is refined. With an appropriate choice of the penalty parameter $\epsilon$, the rate of convergence of PM can be achieved at the same order as for LM.  Furthermore, the rate of convergence here is also the same order as for bending of the flat strip (see Fig.~\ref{f:convM_UCSL}). Besides, the accuracy of PM approaches that of LM as $\epsilon$ is increased. However, note that the stiffness matrix becomes ill-conditioned if $\epsilon$  is too high. For LM, we observe that both interpolation schemes {N2Q0} and {N2Q1c} are robust and stable. It can be seen that {N2Q0} is as accurate as {N2Q1c} here.  
\subsubsection{Cantilever subjected to end shear forces}\label{s:canbeam}
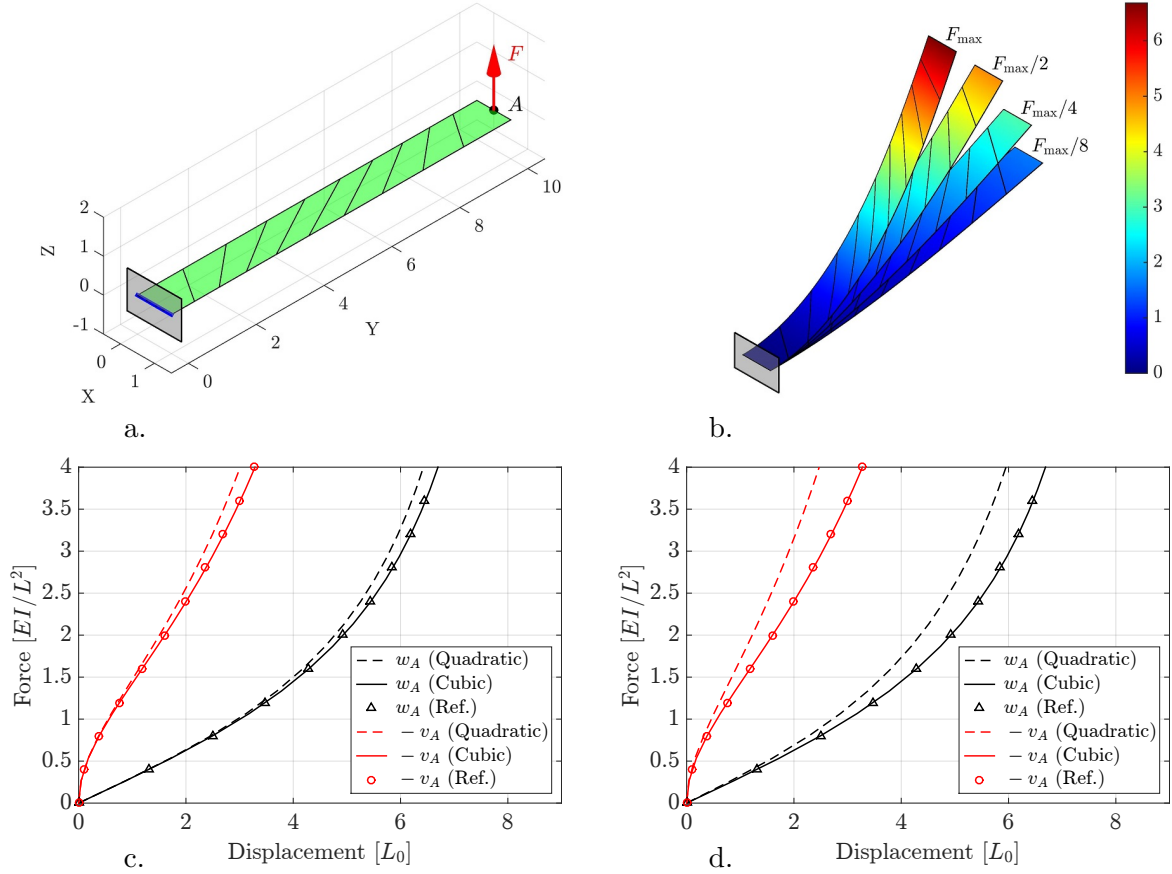
\begin{figure}[ht]
\begin{center} \unitlength1cm
\unitlength1cm
\begin{picture}(0,11.5)
\put(-8.2,6.0){\includegraphics[height=60mm]{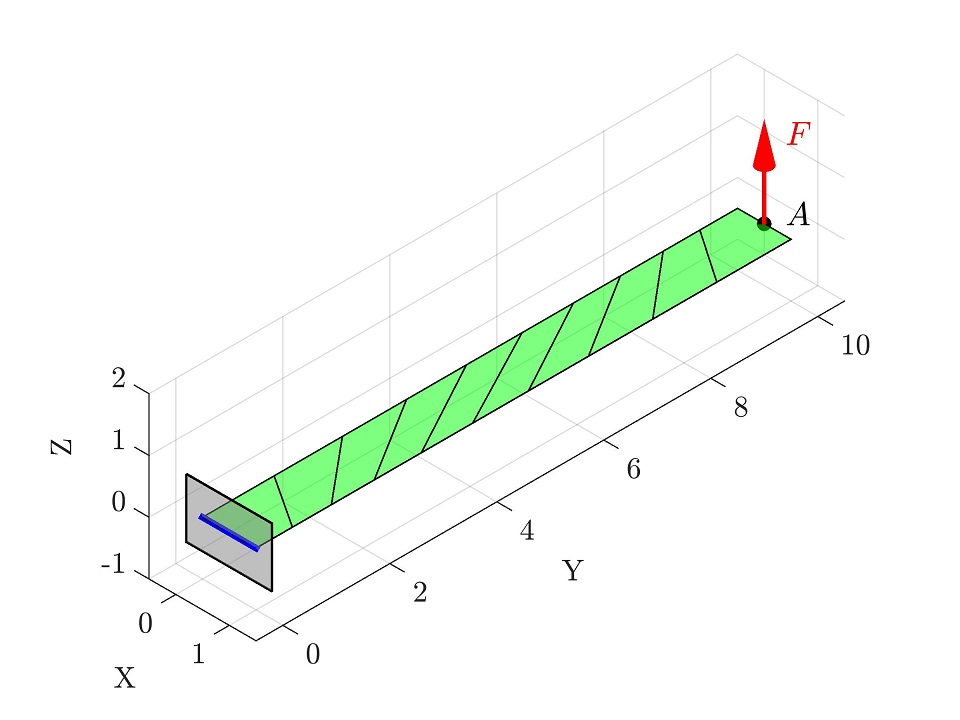}}
\put(-0.2,6.0){\includegraphics[height=60mm]{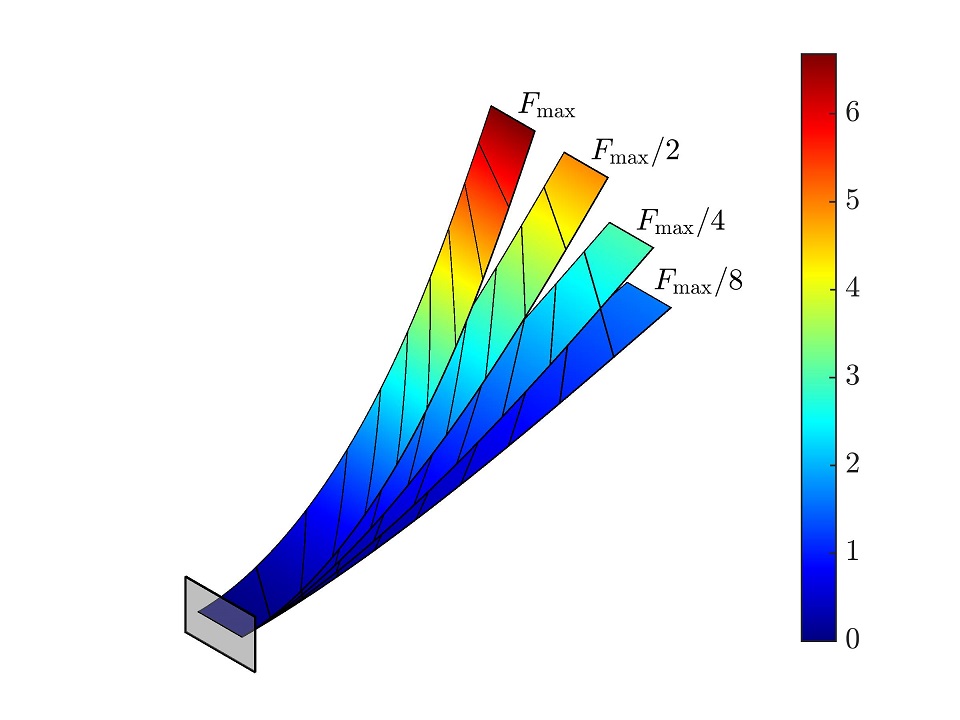}}
\put(-8.2,0.1){\includegraphics[height=55mm]{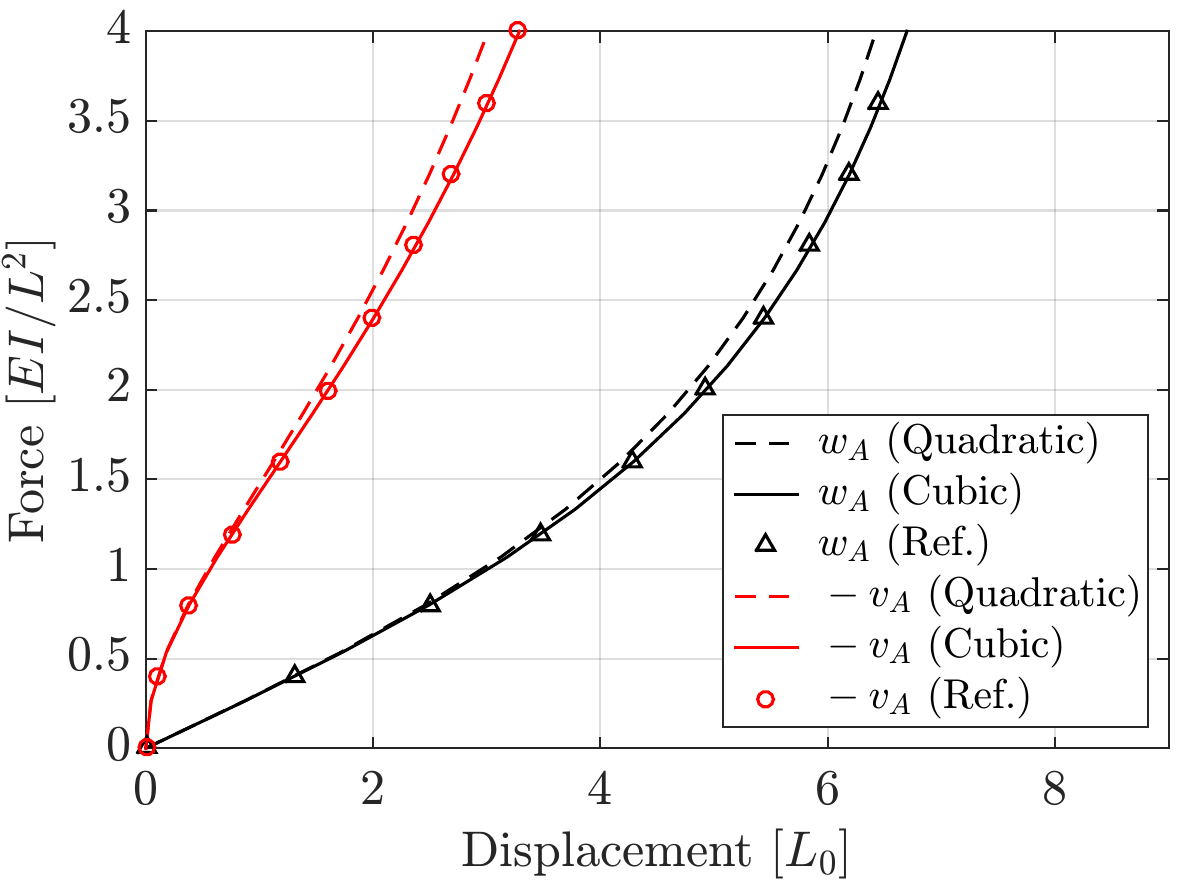}}
\put(-0.2,0.1){\includegraphics[height=55mm]{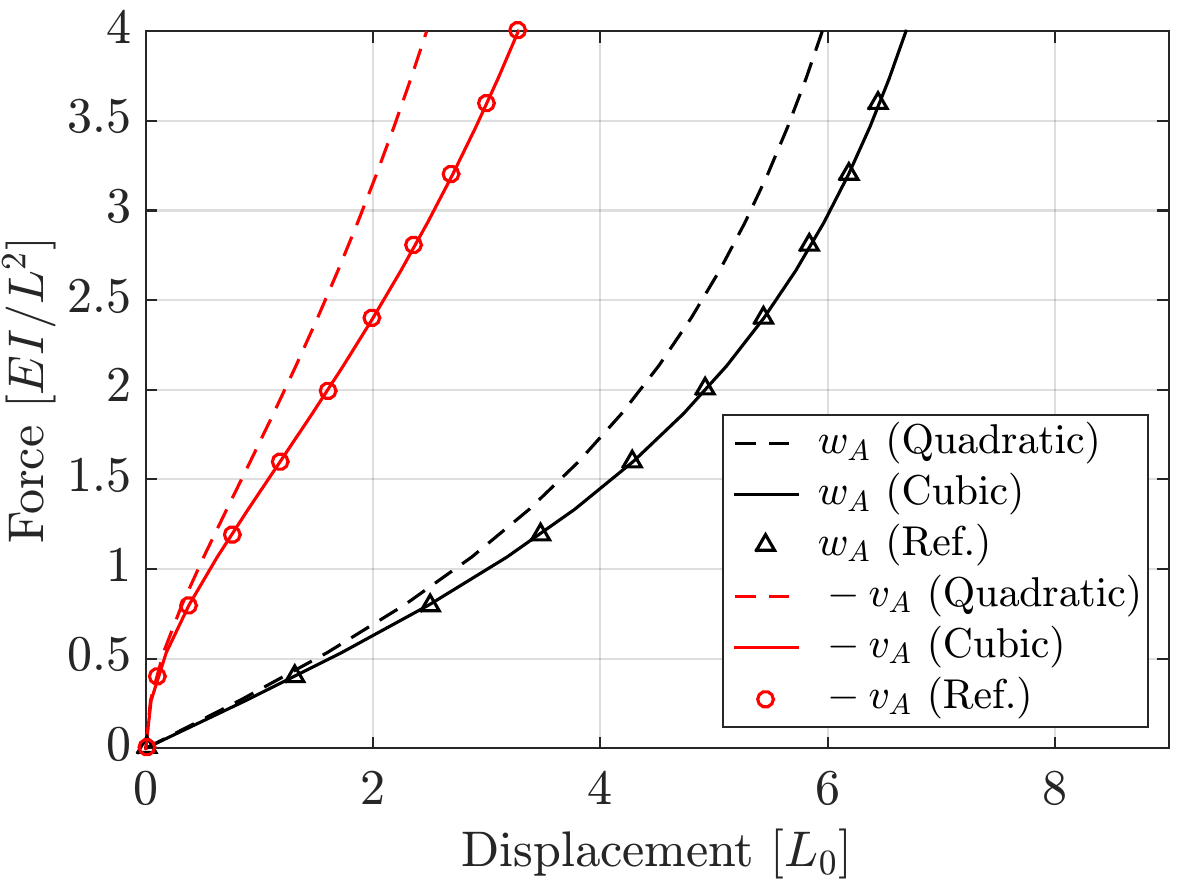}}
\put(-6.7,5.8){a.}
\put(1,5.8){b.}
\put(-6.7,0.1){c.}
\put(1,0.1){d.}
\end{picture}
\caption{Cantilever subjected to end shear force: a.~Undeformed configuration, b.~deformed configuration colored by the vertical displacement. Horizontal ($-v_A$) and vertical ($w_A$) displacement at tip $A$ for c.~a regular mesh with uniform element length and d.~a skew mesh as shown in b. The results are compared with \citet{sze04}.}
\label{f:cantilever}
\end{center}
\end{figure}

The large deflection of a cantilever beam, with dimensions $L\times W\times T = 10\times1\times0.1 \,L_0^3$, due to shear traction $\bar\bt = \bF/W$ applied to its free end is computed. The forces at the nodes located on the free end are derived from Eq.~(\ref{e:fext}.3). The fixed end is clamped by the penalty method  (Sec.~\ref{s:conti}) with $\epsilon = 1000\,E \,L_0^2$.  The material parameters are $ E = 1.2 \times 10^6 \,E_0$ and $ \nu = 0.0 $ \citep{sze04}. The beam is modeled with the Koiter shell material and is discretized by $1\times 10$ NURBS elements, considering both a regular mesh and a skew mesh with distortion ratio $r=0.2$ in $y$-direction. The applied maximum shear force is $F_\mathrm{max}=4\,F_0$ with $F_0 = EI/L^2$ and $I=1/12\,W\,T^3$. As shown in Fig.~\ref{f:cantilever}, considering 10 elements, cubic NURBS give very good results for both regular and skew meshes.   

\begin{figure}[ht]
\begin{center} \unitlength1cm
\unitlength1cm
\begin{picture}(0,5.5)
\put(-7.8,0.0){\includegraphics[height=55mm]{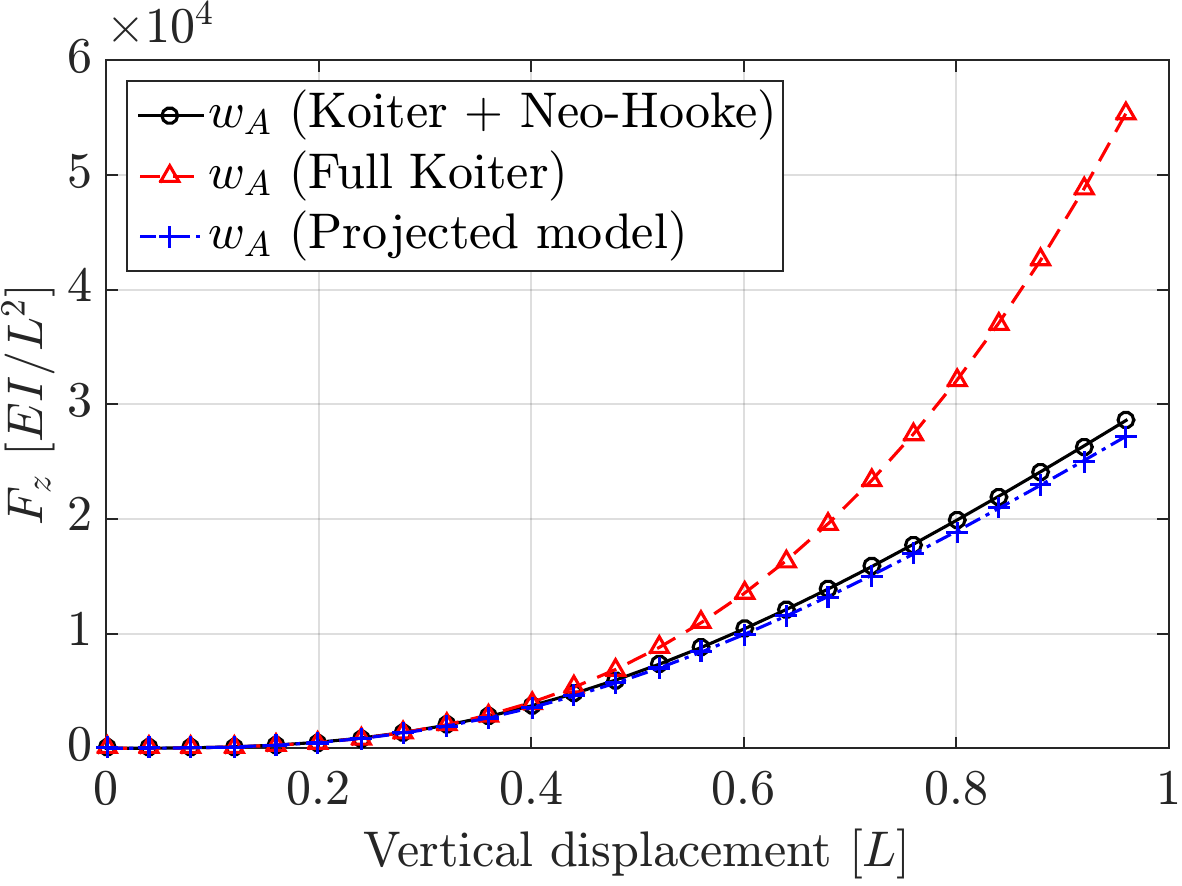}}
\put(0.0,0.0){\includegraphics[height=55mm]{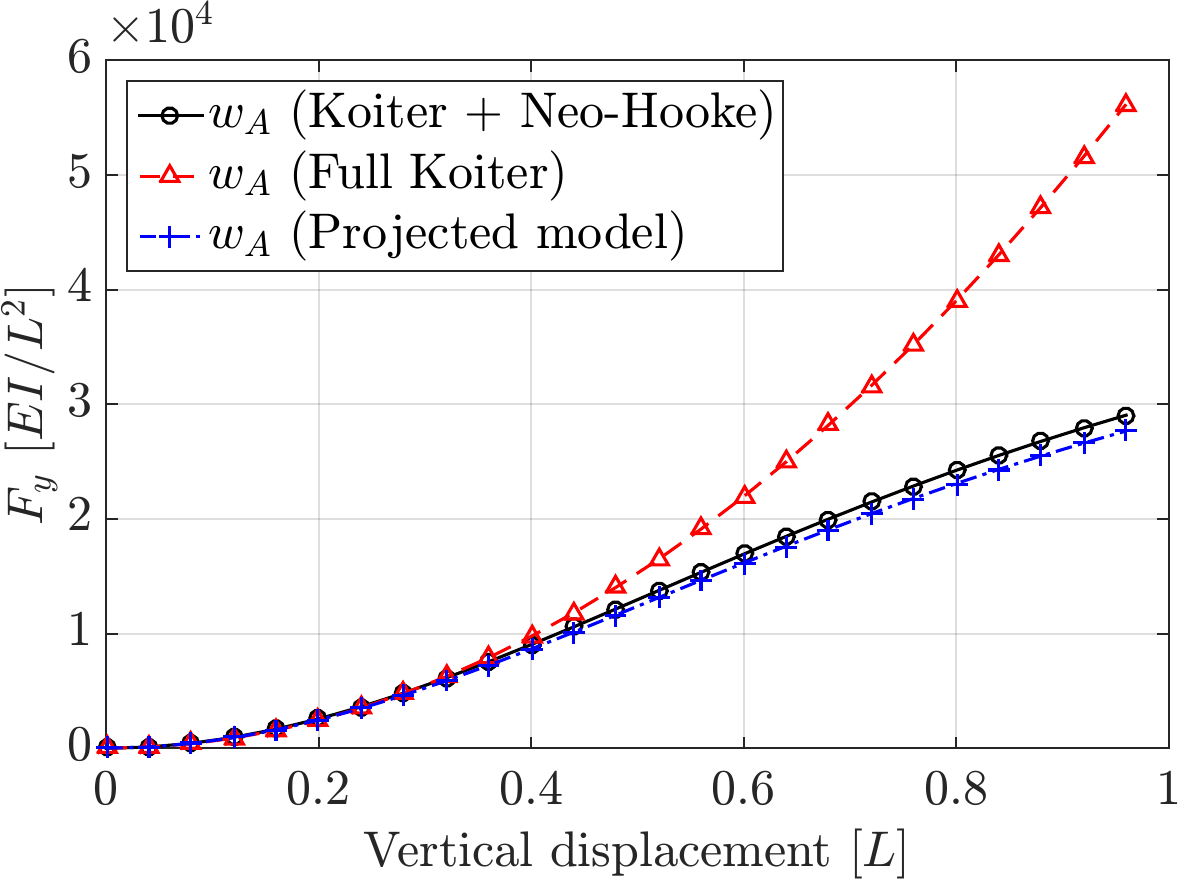}}
\put(-7.9,0.2){a.}
\put(0.1,0.2){b.}
\end{picture}
\caption{Cantilever with prescribed tip displacement: a.~Vertical reaction force $F_z$ vs. vertical displacement and b.~horizontal reaction force $F_y$ vs. vertical displacement.}
\label{f:cantilever2}
\end{center}
\end{figure}

To examine the effects of both bending and large membrane strains, the same problem is studied with an extra Dirichlet boundary condition applied at the beam tip imposing zero displacements in the $y$-direction. Here, since the beam is considerably stretched, the in-plane membrane strains are significant. Instead of a shear force, a vertical displacement is applied at the tip and the reaction forces are measured. In addition to the Koiter model and the projected shell model, a mixed model that combines the bending part of a Koiter shell material with a Neo-Hookean membrane formulation is considered (see Eq.~\eqref{e:KoiterNH}). As shown in Fig.~\ref{f:cantilever2}, all three models predict similar results as long as the bending effects are dominant. By increasing the membrane strains, the mixed formulation is very close to the projected model while the full Koiter model has a different trend. This shows that the simple, more efficient mixed Koiter model can accurately capture the full 3D model behavior. 
\subsubsection{Pinching of a hemisphere with a hole}
\begin{figure}[ht!]
\begin{center} \unitlength1cm
\unitlength1cm
\begin{picture}(0,10.5)
\put(-8.0,5.1){\includegraphics[height=55mm]{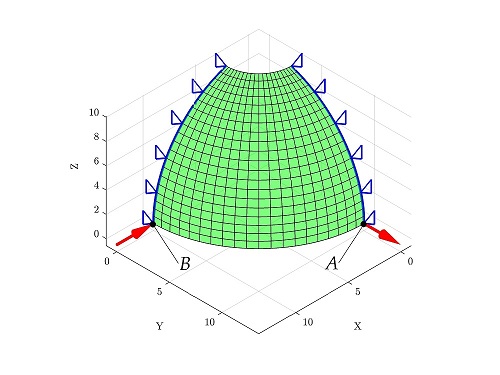}}
\put(-6.5,6.1){a.}
\put(0,5.0){\includegraphics[height=55mm]{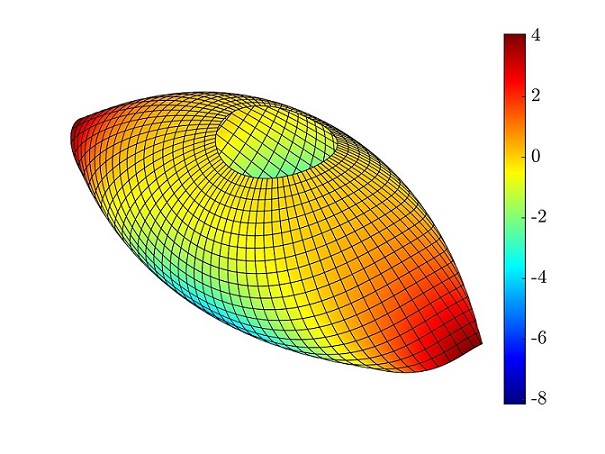}}
\put(2.0,6.1){b.}
\put(-8.5,0.0){\includegraphics[height=55mm]{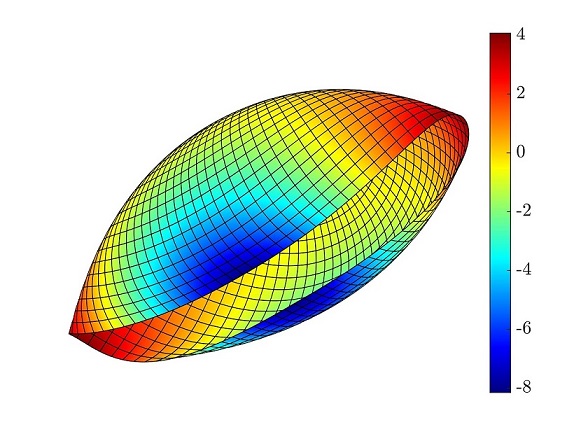}}
\put(-6.5,0.1){c.}
\put(-0.7,0.0){\includegraphics[height=55mm]{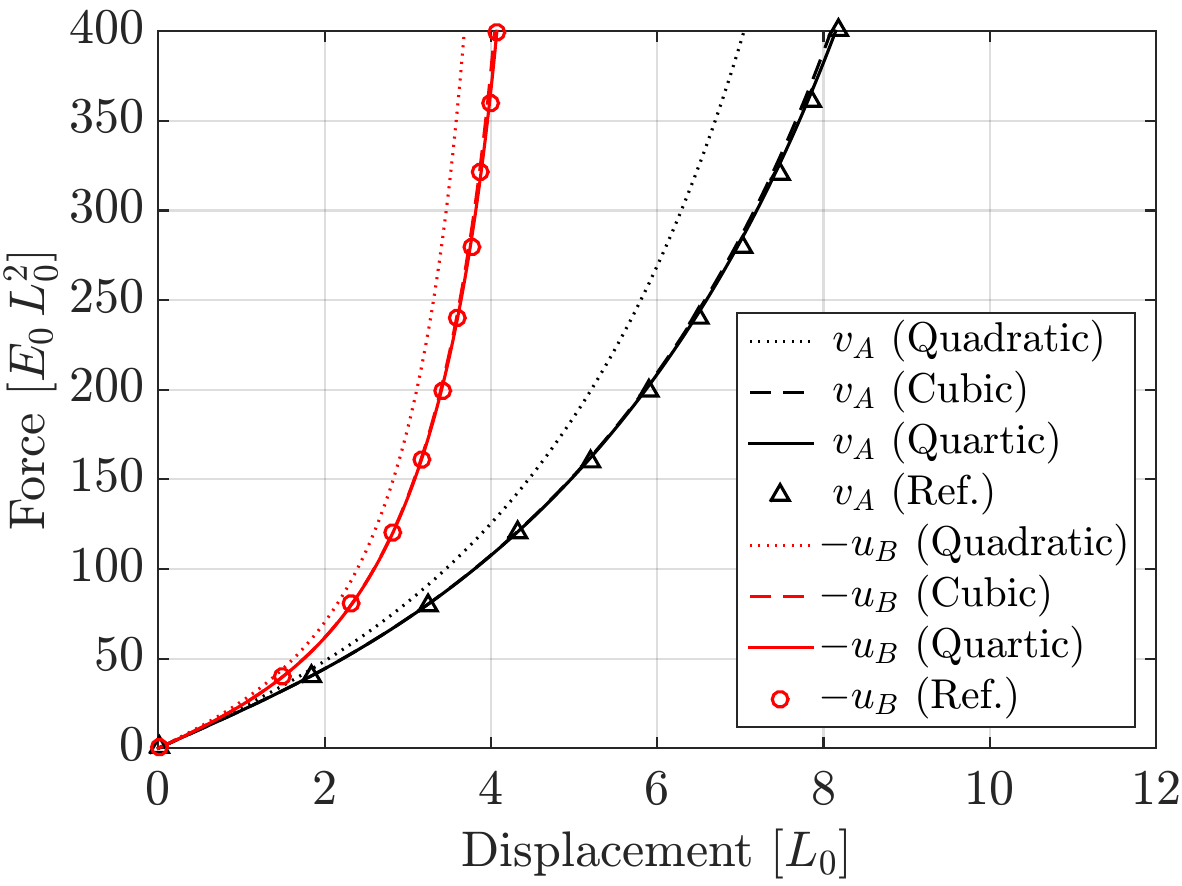}}
\put(-0.5,0.1){d.}
\end{picture}
\caption{Pinching of a hemisphere with a hole at $ 18^\circ $:~a.~Undeformed configuration with boundary conditions, b-c. deformed configuration colored by the radial displacement and d. force vs. displacement of points $A$ and $B$ compared with the reference solution of \citet{sze04}.}
\label{f:pNonHem}
\end{center}
\end{figure}
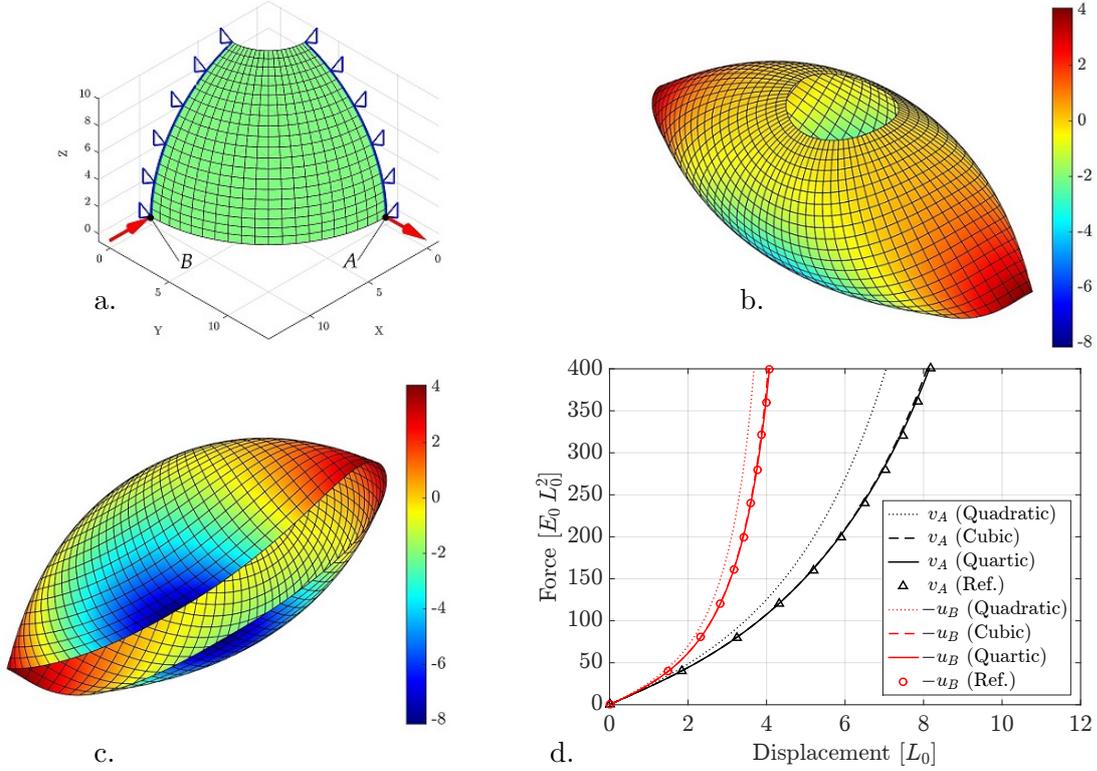

In this example, we compute the large deformation of a hemisphere with a hole at the top, under two pairs of equally large but opposing forces that are applied on the equator of the hemisphere as shown in Fig.~\ref{f:pNonHem}. 
The parameters are extracted from \citet{sze04}: $ R = 10\,L_0 $, $ T= 0.04\,L_0 $, $ E = 6.825 \times 10^7\,E_0 $, $ \nu = 0.3 $ and the point load $ F_\mathrm{max} = 400 \,E_0\,L_0^2$. The symmetry of the hemisphere is modeled by the penalty method  with $ \epsilon = 6\times 10^3\,E\,L_0^2 $ (see Fig.~\ref{f:pNonHem}a). The Koiter shell model is used and the surface is meshed with $20\times20$ quadratic, cubic and quartic NURBS based finite elements. As observed in Fig.~\ref{f:pNonHem}d, the computed results approach the reference solution as the NURBS order is increased.
\subsubsection{Pinching of a cylinder with end rigid diaphragms}
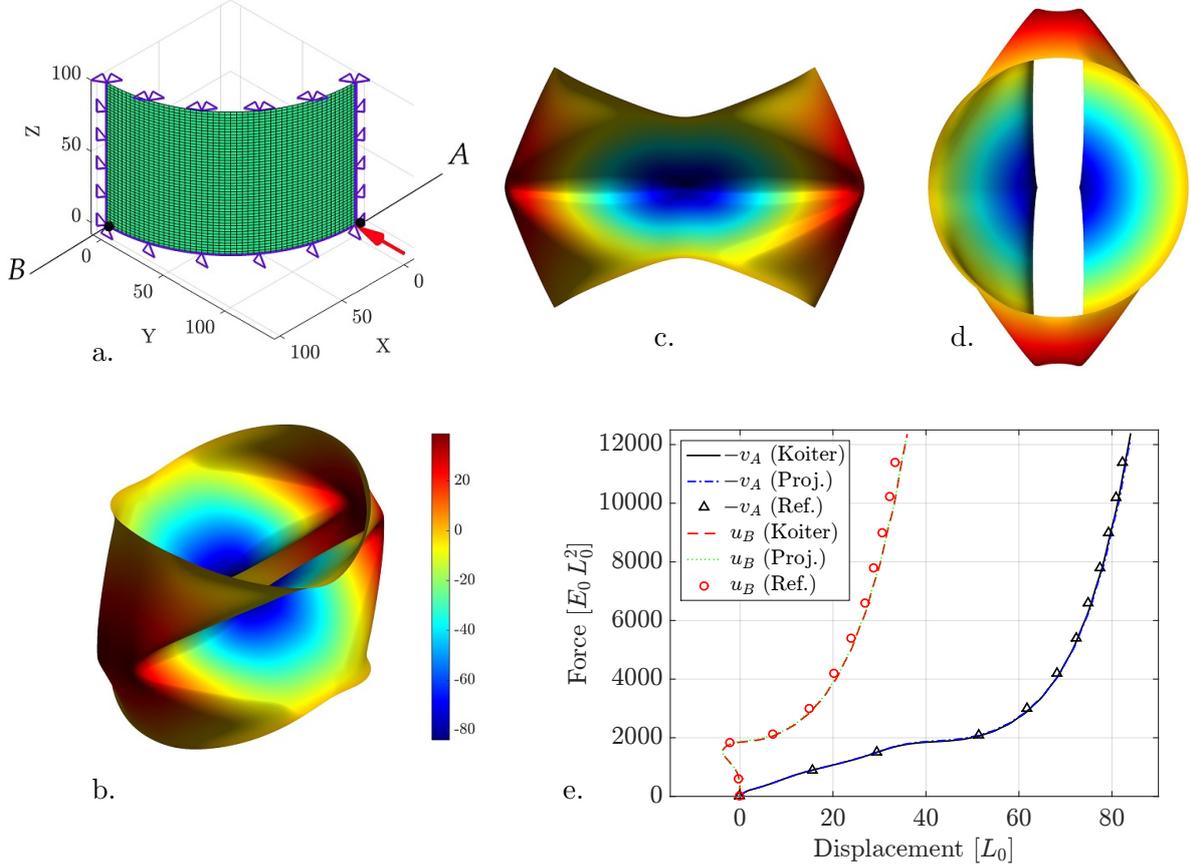
\begin{figure}[ht]
\begin{center} \unitlength1cm
\unitlength1cm
\begin{picture}(0,12)

\put(-8.0,6.5){\includegraphics[height=55mm]{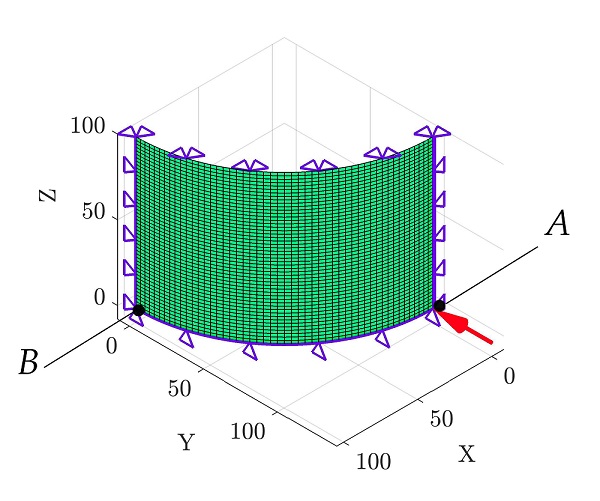}}
\put(-1.5,6.5){\includegraphics[height=50mm]{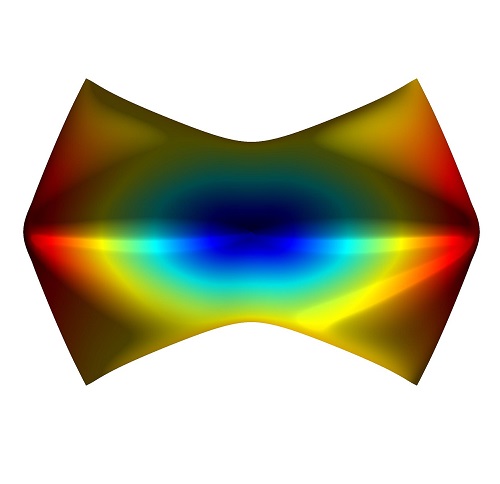}}
\put(4.0,6.5){\includegraphics[height=50mm]{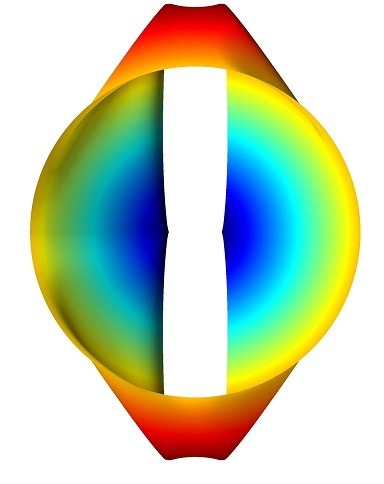}}
\put(-6.8,1.2){\includegraphics[height=50mm]{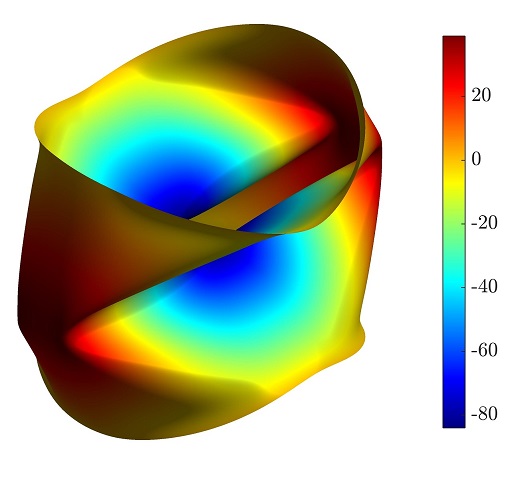}}
\put(-0.5,0.1){\includegraphics[height=60mm]{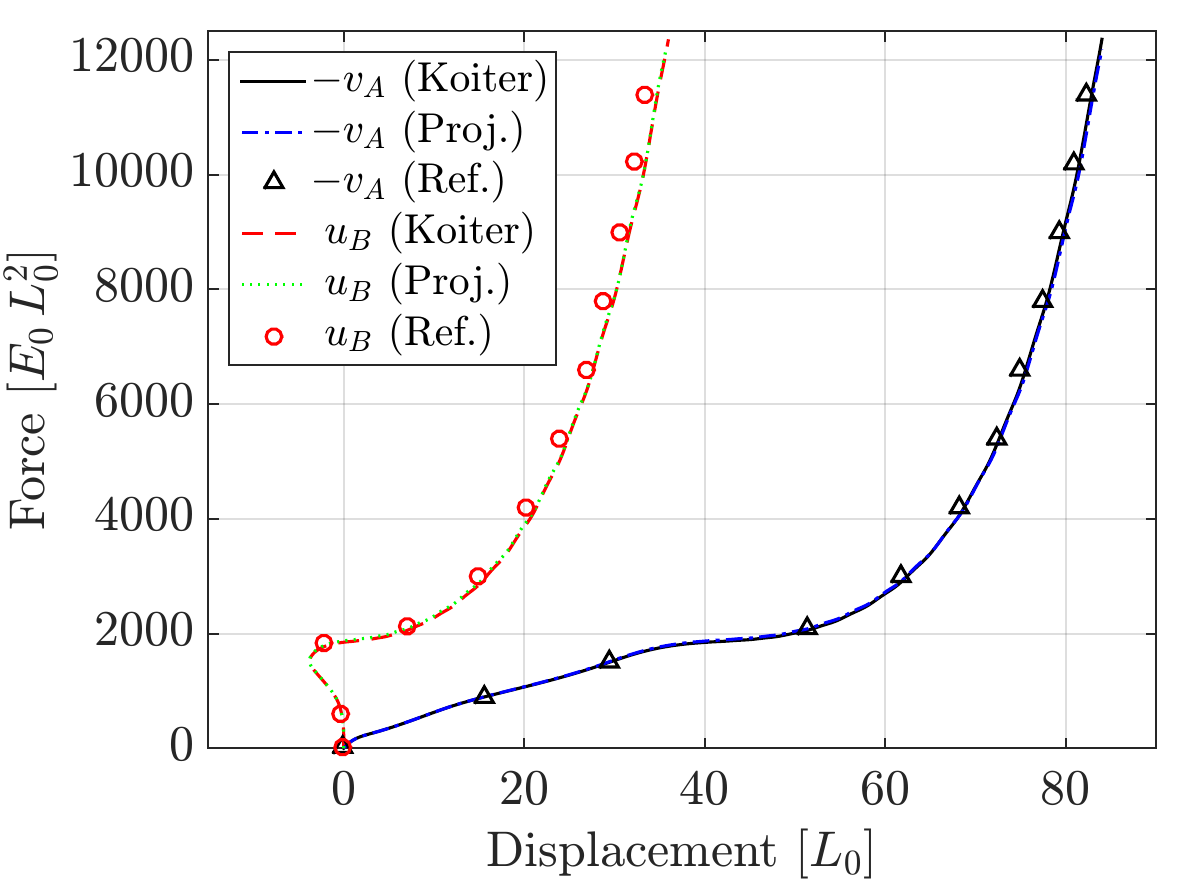}}

\put(-6.7,6.8){a.}
\put(-6.7,1.0){b.}
\put(0.7,7.0){c.}
\put(4.6,7.0){d.}
\put(-0.5,1.0){e.}
\end{picture}
\caption{Pinching of a cylinder with rigid end diaphragm: a.~Undeformed configuration with boundary conditions. Deformed configurations in b.~3D view, c.~y-axis view, d.~z-axis view. The color here denotes the radial displacement. f.~Force vs. displacement at points A and B compared to the results of \citet{sze04}.}
\label{f:NonCylRD}
\end{center}
\end{figure}

We reconsider the pinched cylinder test of Sec.~\ref{s:LinCyl}, but here with large deformations as shown in Fig.~\ref{f:NonCylRD}. Here, the length, thickness and radius of the cylinder are $ L = 200\,L_0 $, $ T = 1.0 \,L_0$ and $ R = 100\,L_0$, respectively. Both the Koiter and the projected shell material models are used with $ E = 30 \times 10^3 \,E_0$ and $ \nu = 0.3$ and they give indistinguishable results. The point load $ F_\mathrm{max} = 12\times10^3 \,E_0\,L_0^2$ is applied. Due to the symmetry, only $1/8$ of the cylinder is modeled. The symmetry boundary conditions are enforced by the Lagrange multiplier method (see Sec.~\ref{s:sym}). The cylinder is discretized by $ 50 \times 50 $ quadratic NURBS finite elements.  Fig.~\ref{f:NonCylRD}f shows good agreement with the reference result of \citet{sze04}.
\subsubsection{Spreading of a cylinder with free ends}
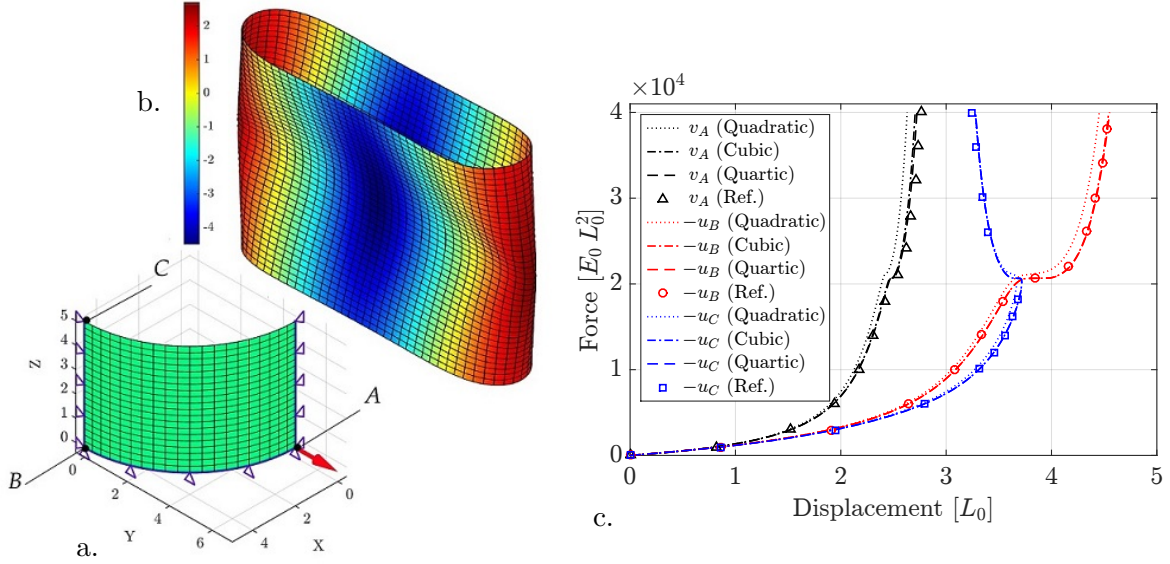
\begin{figure}[ht]
\begin{center} \unitlength1cm
\unitlength1cm
\begin{picture}(0,8)
\put(-7.8,0.5){\includegraphics[height=75mm]{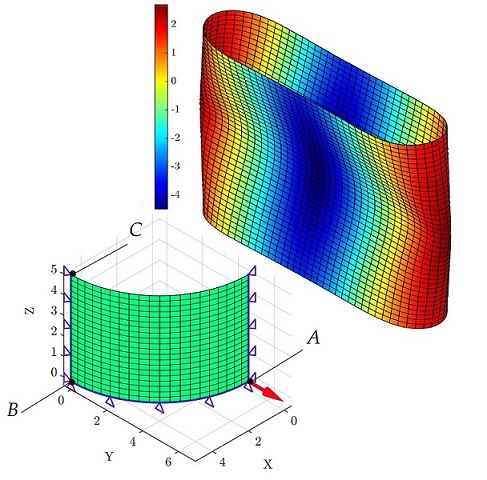}}
\put(-0.5,1.0){\includegraphics[height=60mm]{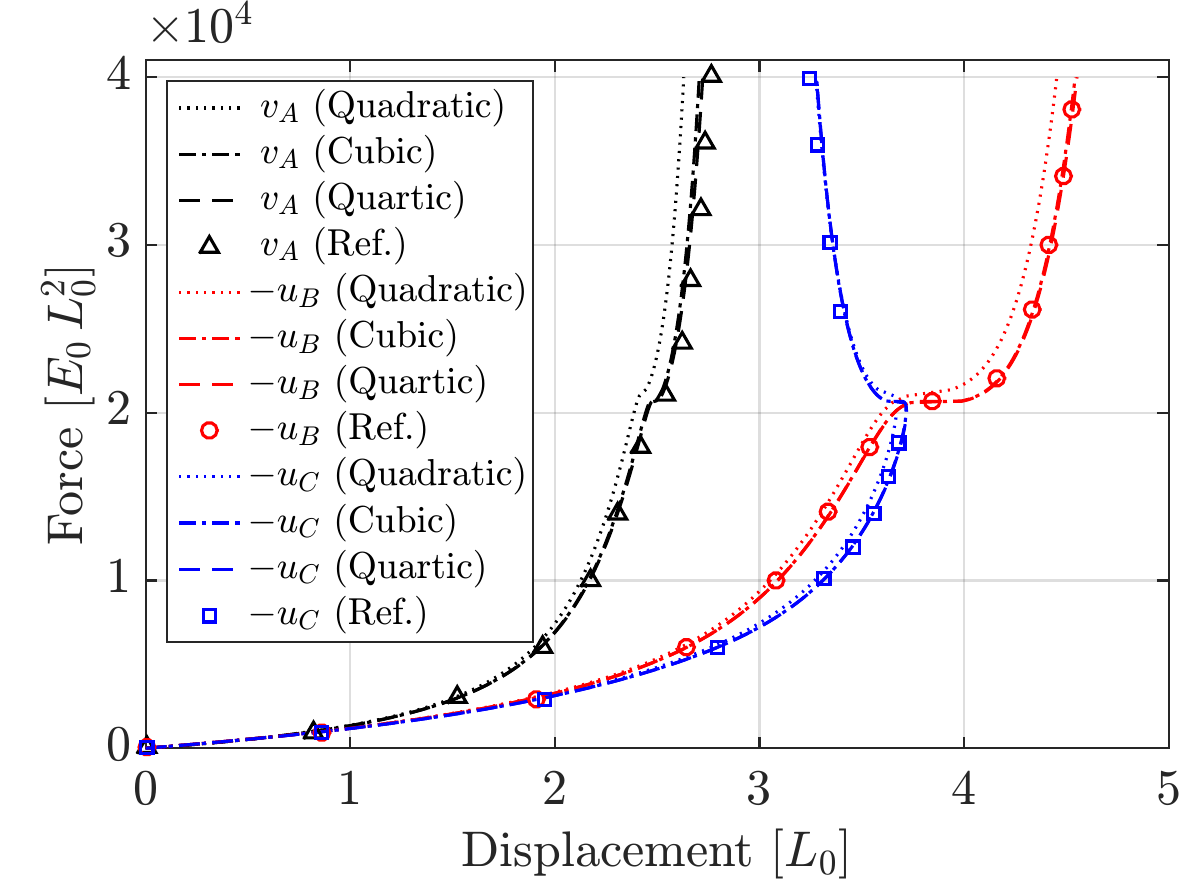}}
\put(-6.8,0.6){a.}
\put(-6.0,6.5){b.}
\put(-0.0,1){c.}
\end{picture}
\caption{Spreading of a cylinder with free ends: a. Undeformed configuration with boundary conditions, b. deformed configuration colored with radial displacement, c.~force vs.~displacement of points A, B and C compared to the results of \citet{sze04}.}
\label{f:NonCylFree}
\end{center}
\end{figure}

In this test, a cylinder, with dimensions $L\times R \times T =  10.35 \times 4.953 \times 0.094\,L_0^3$,  with open ends is pulled apart by a pair of opposite forces up to  $ F_\mathrm{max} = 40\times10^3\,E_0\,L_0^2$, which are applied at the middle of the cylinder (see Fig.~\ref{f:NonCylFree}). The Koiter material is used with $ E = 10.5 \times 10^6\,E_0 $ and $ \nu = 0.3125 $. Here, the results of \citet{sze04} are used as a reference for comparison. Due to the symmetry, only $1/8$ of the cylinder is modeled as shown in Fig.~\ref{f:NonCylFree}a. The cylinder is discretized by $ 20 \times 20 $ NURBS  finite elements. The symmetry boundary conditions are enforced by the penalty method (see Sec.~\ref{s:sym}) with $ \epsilon = 20\,E\,L_0^3/L$ along the axial edge and $ \epsilon = 20\,E\,L_0^3/\pi\,R$ along the circumferential edge.  A good agreement with the reference results is also observed in Fig.~\ref{f:NonCylFree}c.
%
%
\section{Conclusion}\label{s:con}
A new unified FE formulation for rotation-free thin shells is presented. The formulation uses isogeometric analysis to benefit from its high order continuity. We have also introduced a penalty and a Lagrange multiplier approach to enforce continuity at patch boundaries. The approach can be used to model $G^1$-continuity required for multi-patch NURBS, fixed surface folds, symmetry (or clamping) constraints, symmetry constraints at a kink and rotational Dirichlet boundary conditions. It is required in order to transfer moments at patch boundaries. We observed that the penalty method provides a simple and very efficient implementation, yet still maintains sufficient accuracy. The proposed formulation is tested by several numerical examples considering both linear as well as non-linear regimes. The numerical results are verified either by available analytical solutions or reliable reference solutions. The results demonstrate the robustness as well as good performance of the formulation. 

Further, we have presented a detailed and systematic procedure to obtain shell material models based on through-thickness integration of existing 3D constitutive laws. Besides that, our formulation is also designed to accept material models given directly in surface energy form. Since there is no need for numerical integration over the thickness, such material models are much more efficient. For the considered numerical examples, it turns out that these models, particularly the mixed Koiter formulation, are equally accurate as the expensive integration models, even for large deformations and stretches. Further, our formulation is presented fully in curvilinear coordinates. This allows for a direct and efficient implementation, since no local coordinate transformation is needed. The formulation also allows for a straightforward and efficient inclusion of material anisotropy \citep{biomembrane} and stabilization schemes for liquid shells \citep{liquidshell}.

%
\vspace{8mm}
{\bf Acknowledgment}\\
Financial support from the German Research Foundation (DFG) through grant GSC 111, is gratefully acknowledged. The authors also wish to thank {Callum J. Corbett} and {Yannick A.D. Omar} for their help.
%
%
\appendix
\section{FE tangent matrices}\label{s:tangents}
Using Eqs.~\eqref{e:dd_aabh} and \eqref{e:dd_babh}, the terms in Eq.~\eqref{e:dxdxWS} become
\eqb{lll}
c^{\alpha\beta\gamma\delta}\,\frac{1}{2}\delta\auab\,\frac{1}{2}\Delta\augd
\is c^{\alpha\beta\gamma\delta}\,\delta\mx_e^\mrT\,\mN^\mrT_{,\alpha}\,(\vaub\otimes\vaug)\, \mN_{,\delta}\,\Delta\mx_e~,
\\[2.5mm]
d^{\alpha\beta\gamma\delta}\,\frac{1}{2}\delta\auab\,\Delta\bugd
\is  d^{\alpha\beta\gamma\delta}\,\delta\mx_e^\mrT\,\mN^\mrT_{,\alpha}\,(\vaub\otimes\bn)
\,\tilde\mN_{;\gamma\delta}\,\Delta\mx_e~, 
\\[2.5mm]
e^{\alpha\beta\gamma\delta}\,\delta\buab\,\frac{1}{2}\Delta\augd
\is e^{\alpha\beta\gamma\delta}\,\delta\mx^\mrT_e\,\tilde\mN^\mrT_{;\alpha\beta}\,(\bn\otimes\vaug)\, \mN_{,\delta}\,\Delta\mx_e~,
\\[2mm]
f^{\alpha\beta\gamma\delta}\,\delta\buab\,\Delta\bugd
\is f^{\alpha\beta\gamma\delta}\,\delta\mx^\mrT_e \,\tilde\mN^\mrT_{;\alpha\beta}\,(\bn\otimes\bn)
\,\tilde\mN_{;\gamma\delta}\,\Delta\mx_e~,
\eqe
and
\eqb{l}
\tau^{\alpha\beta}\,\frac{1}{2}\Delta\delta a_{\alpha\beta}
= \delta\mx_e^\mrT\,\mN^\mrT_{,\alpha}\,\tau^{\alpha\beta}\,\mN_{,\beta}\,\Delta\mx_e~,
\eqe
\eqb{lll}
\Delta\delta b_{\alpha\beta} \is
-\,\delta\mx_e^\mrT\, \Big[ \mN^\mrT_{,\gamma}\,(\bn\otimes\ba^\gamma)\,\tilde\mN_{;\alpha\beta}
+ \tilde\mN^\mrT_{;\alpha\beta}\,(\ba^\gamma\otimes\bn)\,\mN_{,\gamma} + \mN^\mrT_{,\gamma}\, a^{\gamma\delta}\,b_{\alpha\beta}\,(\bn\otimes\bn) 
\,\mN_{,\delta}\Big]\,\Delta\mx_e~.
\eqe
Thus, the linearization of $G_\mathrm{int}^e$ yields
\eqb{l}
\Delta G^e_\mathrm{int} = \delta\mx_e^\mrT\,\big(\mk^e_{\tau\tau} + \mk^e_{\tau M} + \mk^e_{M\tau} + \mk^e_{MM} + \mk^e_{\tau} + \mk^e_{M}  \big)\, \Delta\mx_e~,
\eqe
with the material stiffness matrices
\eqb{lll}
\mk^e_{\tau\tau} \dis \ds\int_{\Omega_0^e} c^{\alpha\beta\gamma\delta}\,\mN^\mrT_{,\alpha}\,(\vaub\otimes\vaug)\, \mN_{,\delta}\,\dif A~, \\[4mm]
\mk^e_{\tau M} \dis \ds\int_{\Omega_0^e} d^{\alpha\beta\gamma\delta}\,\mN^\mrT_{,\alpha}\,(\vaub\otimes\bn)\,\tilde\mN_{;\gamma\delta}\,\dif A~,  \\[4mm]
\mk^e_{M \tau} \dis \ds\int_{\Omega_0^e} e^{\alpha\beta\gamma\delta}\,\tilde\mN^\mrT_{;\alpha\beta}(\bn\otimes\vaug)\, \mN_{,\delta}\, \dif A~, \\[4mm]
\mk^e_{MM} \dis \ds\int_{\Omega_0^e}
f^{\alpha\beta\gamma\delta}\,\delta\mx^\mrT_e\,\tilde\mN^\mrT_{;\alpha\beta}\,(\bn\otimes\bn)
\,\tilde\mN_{;\gamma\delta}\,\dif A
\label{e:matKij}
\eqe
and the geometric stiffness matrices
\eqb{lll}
\mk^e_{M}=\mk^e_{M1}+\mk^e_{M2}+(\mk^e_{M2})^\mrT~,
\label{e:matGeo}
\eqe
 with
\eqb{lll}
\mk^e_{M1} \dis - \ds\int_{\Omega_0^e} b_{\alpha\beta}\,M_0^{\alpha\beta}\,a^{\gamma\delta}\,\mN^\mrT_{,\gamma}\,(\bn\otimes\bn)\,\mN_{,\delta}\,\dif A~, \\[4mm]
\mk^e_{M2} \dis - \ds\int_{\Omega_0^e} M_0^{\alpha\beta}\,\mN_{,\gamma}^\mrT\,(\bn\otimes\ba^\gamma)\,\tilde\mN_{;\alpha\beta}\,\dif A~.
\label{e:matGeosim}
\eqe

For an efficient implementation of these matrices see Appendix~\ref{s:efficient}. Similarly, the linearization of $G_\mathrm{ext}^e$ yields
\eqb{l}
\Delta G^e_\mathrm{ext} = \delta\mx_e^\mrT\,\big(\mk^e_{\mathrm{ext}p} + \mk^e_{\mathrm{ext}m} \big)\, \Delta\mx_e~,
\label{e:DGextM}
\eqe
with the external FE tangent matrix\footnote{The unsymmetric rear term in Eq.~\eqref{e:kextM} disappears if $m_\tau \dif s$ is constant, which corresponds to dead loading.}
\eqb{lll}
\mk^e_{\mathrm{ext}m} \is \ds\int_{\partial_m\Omega^e} m_\tau\,\mN^\mrT_{,\alpha}\,\big(\nu^\beta\,\bn\otimes\ba^\alpha + \nu^\alpha\,\ba^\beta\otimes\bn\big)\,\mN_{,\beta}\,\dif s~ \\[5mm]

\mi \ds\int_{\partial_m\Omega^e} m_\tau\, \nu^\alpha\,\mN^\mrT_{,\alpha}\,(\bn\otimes\ba^\xi) 
 \, \mN_{,\xi}\,\dif s~
\textcolor{darkgreen}{ + \ds\int_{\partial_m\Omega^e} \mN^\mrT_{,\alpha}\, \tau^\alpha\,\bn\otimes\bnu\, \mN_{,\xi}\, \frac{m_\tau}{\norm{\ba_\xi}} \,\dif s}~,
 
\label{e:kextM}
\eqe
which corresponds to $\mf^e_{\mathrm{ext}m}$, in Eq.~(\ref{e:fext}.4). Here, $\xi$ denotes the convective coordinate of  the curve $\partial_m\Omega^e$. For the application of surface pressure through Eq.~(\ref{e:fext}.2), the corresponding $\mk^e_{\mathrm{ext}p}$ can be found e.g.~in \citet{membrane}.

\section{Efficient FE implementation}\label{s:efficient}
This section presents an efficient implementation of the above equations. First a general algorithm is suggested, which can be used for a variety of models like for example the Koiter model \eqref{e:Koiter} and the projected shell model \eqref{e:NeoH3D}. Then a special implementation for models like the Canham model \eqref{e:WmsF} is presented.

\subsection{For general cases}\label{s:eff_gen}
\begin{samepage}
For an efficient implementation, we arrange the tangent components $c^{\alpha\beta\gamma\delta}$  as 
\eqb{l}
\mC := \left[\begin{matrix}
c^{1111} & c^{1122} & c^{1112} \\ 
c^{2211} & c^{2222} & c^{2212}  \\ 
c^{1211} & c^{1222} & c^{1212}\\ 
\end{matrix} \right] , ~
\eqe
and likewise $d^{\alpha\beta\gamma\delta}$, $e^{\alpha\beta\gamma\delta}$ and $f^{\alpha\beta\gamma\delta}$
are arranged as $\mD$, $\mE$ and $\mF$. Note that for model \eqref{e:Koiter}, $\mC=\mC^\mrT$, $\mE = \mD^\mrT$ and $\mF=\mF^\mrT$, leading to $\mk^e_{M\tau} = \mk^{e\mrT}_{\tau M}$ . The stress, moment, and curvature tensors are written in Voigt notation as
 \eqb{lll}
 \hat{\btau} \dis [\tau^{11}~,~ \tau^{22}~,~  \,\tau^{12}]^T ~, \\[2mm]
 \hat{\bM}_0 \dis [M_0^{11}~,~ M_0^{22}~,~  \,M_0^{12}]^T ~, \\[2mm]
 \hat{\bb} \dis [b_{11}~,~ b_{22}~,~ 2\,b_{12}]^T ~.
\eqe
Taking $n$ as the number of control points per element,  we define the ($3n\times 1$) arrays 
\end{samepage}
\eqb{lll}
\mL_{\alpha\beta}^a \dis \mN^\mrT_{,\alpha}\,\ba_\beta~, \\[2mm]
\mL_{\alpha}^n \dis \mN^\mrT_{,\alpha}\,\bn~, \\[2mm]
\mG_{\alpha\beta}^n \dis \tilde\mN_{;\alpha\beta}^\mrT\,\bn \\[2mm]
\eqe
and organize them into
 \eqb{lll}
 \hat{\mL}_a \is [\mL^a_{11}~,~ \mL^a_{22}~,~  \mL^a_{12}+\mL^a_{21}]~, \\[2mm]
 \hat{\mG}_n \is [\mG^n_{11}~,~ \mG^n_{22}~,~  \mG^n_{12}+\mG^n_{21}]~.
\eqe

We thus obtain
\eqb{lll}
\mf^e_\mathrm{int\tau} 
\is  \ds\int_{\Omega^e_0} \hat{\mL}_a\, \hat{\btau}  \,\dif A~, \\[4mm]
\mf^e_{\mathrm{int}M}  
\is  \ds\int_{\Omega^e_0} \hat{\mG}_n\, \hat{\bM}_0  \, \dif A~,
\eqe
for the FE forces in Eq.~\eqref{e:Fint} and \eqref{e:Mint} introduced in Sec.~\ref{s:fvectors}, and
\eqb{llll}
\mk^e_\mathrm{\tau\tau} 
\is \ds\int_{\Omega^e_0} \hat{\mL}_a\,\mC\,\hat{\mL}_a^{\mrT}\,\dif A~, \\[4mm]
\mk^e_{\tau M}  
\is  \ds\int_{\Omega^e_0}  \hat{\mL}_a \,\mD\, \hat{\mG}_n^\mrT\, \dif A~, \\[4mm] 
\mk^e_{M\tau}
\is  \ds\int_{\Omega^e_0}  \hat{\mG}_n \,\mE\, \hat{\mL}_a^\mrT\, \dif A~, \\[4mm] 
\mk^e_{MM} \is  \ds\int_{\Omega^e_0}  \hat{\mG}_n\,\mF\,\hat{\mG}_n^{\mrT} \, \dif A~,
\eqe
for the material stiffness matrices in Eq.~\eqref{e:matKij}. Similarly, for the geometric stiffness matrices in Eq.~\eqref{e:matGeosim}, we get 
\eqb{llll}
\mk^e_\mathrm{M1} 
\is -\ds\int_{\Omega^e_0} b_M\, \Big(a^{11}\, \mL^n_1\,\mL^{n\mrT}_1 +   a^{22}\, \mL^n_2\,\mL^{n\mrT}_{2} +  \textcolor{darkgreen}{a^{12}\,\big[\mL^n_1\,\mL^{n\mrT}_{2} +  \mL^n_2\,\mL^{n\mrT}_{1}\big]} \Big)\,\dif A~, \quad b_M:= \hat{\bb}^\mrT\hat{\bM}_0~, \\[4mm]
\mk^e_\mathrm{M2} 
\is -\ds\int_{\Omega^e_0}  \Big(\mL^n_1\, \ba^{1\mrT} + \mL^n_2\, \ba^{2\mrT} \Big)\,\Big(M_0^{11}\, \tilde\mN_{;11} + M_0^{22}\, \tilde\mN_{;22} + 2\,M_0^{12}\, \tilde\mN_{;12}\Big)\,\dif A~. 
\eqe

\subsection{For particular cases}
For some material models, the material tangents fit into the format
\eqb{l}
\hat c^{\alpha\beta\gamma\delta}
= \hat c_{aa}\,a^{\alpha\beta}\,a^{\gamma\delta} 
+ \hat c_{a}\,a^{\alpha\beta\gamma\delta} 
+ \hat c_{ab}\,a^{\alpha\beta}\,b^{\gamma\delta}
+ \hat c_{ba}\,b^{\alpha\beta}\,a^{\gamma\delta}
+ \hat c_{bb}\,b^{\alpha\beta}\,b^{\gamma\delta}~,
~\hat c = c,\,d,\,e,\,f~,
\label{e:hat_c}
\eqe
with suitable definitions of coefficients $\hat c_{aa}$, $\hat c_a$, $\hat c_{ab}$, $\hat c_{ba}$ and $\hat c_{bb}$. We can then obtain a more efficient implementation of the possible contractions within Eqs.~\eqref{e:matKij}~and~\eqref{e:matGeo}. Examples for this case are  the Canham model (Sec.~\ref{s:canham}) and Helfrich model \citep{liquidshell}. The sequential computation is as follows.

Precompute the ($3\times3n$) arrays
\eqb{lll}
\tilde\mN^\alpha_\beta \dis a^{\alpha\gamma}\,\tilde\mN_{;\beta\gamma}~, \\[2mm]
\tilde\mN_\mra \dis \tilde\mN^\alpha_\alpha~, \\[2mm]
\tilde\mN_\mrb \dis b^{\alpha\beta}\,\tilde\mN_{;\alpha\beta}~
\eqe
and the ($3n\times1$) vectors
\eqb{llllll}
\mL^\alpha_\beta \dis \mN^\mrT_{,\beta}\,\ba^\alpha~, \quad\quad \tilde\mL_\beta^\alpha\dis \tilde{\mN}^{\alpha\,\mrT}_\beta\,\bn ~,\\[2mm]
\mL_\mra \dis \mL^\alpha_\alpha~, \hfill \tilde\mL_\mra \dis \tilde\mN^\mrT_\mra\,\bn~, \\[2mm]
\mL_\mrb \dis \mN^\mrT_{,\alpha}\,\bb^\alpha~, \hfill \tilde\mL_\mrb \dis \tilde\mN^\mrT_\mrb\,\bn~,\\[2mm]
\tilde\mL_\mra^\alpha\dis \tilde\mN^\mrT_\mra\,\ba^\alpha~, \hfill \tilde\mL_\mrb^\alpha\dis \tilde\mN^\mrT_\mrb\,\ba^\alpha~,\\[2mm]
\mL_\alpha \dis \mN^\mrT_{,\alpha}\,\bn~, \hfill \mL_\mathrm{A} \dis \mN^\mrT_{,\alpha}\,\tilde\bA^\alpha ~(*)~,
\eqe
with $\bb^\alpha:=b^{\alpha\beta}\,\ba_\beta$ and $\tilde\bA^\alpha:= \tilde{A}^{\alpha\beta} \ba_\beta$. The equation marked by (*) is only needed for the stabilization schemes for liquid shells \citep{liquidshell} and can be skipped for solid shells. (Depending on the type of stabilization scheme used, either $\tilde{A}^{\alpha\beta}:=A^{\alpha\beta}$ or $\tilde{A}^{\alpha\beta}:=a^{\alpha\beta}_\mathrm{pre}$, where $a^{\alpha\beta}_\mathrm{pre}$ is the corresponding quantity at the previous load step.)

We then compute the ($3n\times3n$) arrays
\eqb{lllll}
\mN_\mra^2 \dis a^{\alpha\beta}\,\mN^\mrT_{,\alpha}\,\mN_{,\beta}~,\\[2mm]
\mN_\mathrm{A}^2 \dis {\tilde{A}^{\alpha\beta}\,\mN^\mrT_{,\alpha}\,\mN_{,\beta}}~,\\[2mm]
\mN_\mrb^2 \dis b^{\alpha\beta}\,\mN^\mrT_{,\alpha}\,\mN_{,\beta}~,\\[2mm]
\mL^2 \dis a^{\alpha\beta}\,\mL_{\alpha}\,\mL_{\beta}^\mrT~.\\[2mm]
\eqe
With
\eqb{lll}
\tau^{\alpha\beta} \is \tau_\mra\,a^{\alpha\beta} + \tau_\mrb\,{b}^{\alpha\beta} + \tau_\mathrm{A}\,\tilde{A}^{\alpha\beta}~, \\[2mm]
M_0^{\alpha\beta} \is M^0_\mra\,a^{\alpha\beta} + M^0_\mrb\,{b}^{\alpha\beta} ~,
\eqe
 and
\eqb{lll}
c^{\alpha\beta\gamma\delta} \is c_{\mra\mra}\,a^{\alpha\beta}\,a^{\gamma\delta}
+ c_\mra\,a^{\alpha\beta\gamma\delta} + c_{\mra\mrb}\,\big(a^{\alpha\beta}\,b^{\gamma\delta}+ b^{\alpha\beta}\,a^{\gamma\delta}\big)
+ c_{\mrb\mrb}\,b^{\alpha\beta}\,b^{\gamma\delta} ~,\\[2mm]
d^{\alpha\beta\gamma\delta} \is d_{\mra\mra}\,a^{\alpha\beta}\,a^{\gamma\delta}
+ d_\mra\,a^{\alpha\beta\gamma\delta} + d_{\mra\mrb}\,a^{\alpha\beta}\,b^{\gamma\delta}+ d_{\mrb\mra}\,b^{\alpha\beta}\,a^{\gamma\delta}
+ d_{\mrb\mrb}\,b^{\alpha\beta}\,b^{\gamma\delta}~,\\[2mm]
f^{\alpha\beta\gamma\delta} \is f_{\mra\mra}\,a^{\alpha\beta}\,a^{\gamma\delta}
+ f_\mra\,a^{\alpha\beta\gamma\delta}~,
\label{e:unifcdef}
\eqe
we thus get
\eqb{lll}
\mf^e_\mathrm{int\tau} 
\is  \ds\int_{\Omega^e_0} \big(\tau_\mra\,\mL_\mra+\tau_\mrb\,\mL_\mrb + \tau_\mathrm{A}\,\mL_{\textcolor{darkgreen}{\mrA}}\big)\,\dif A~, \\[4mm]
\mf^e_{\mathrm{int}M}  
\is  \ds\int_{\Omega^e_0} \big(M^0_\mra\,\tilde\mL_\mra + M^0_\mrb\,\tilde\mL_\mrb 
\big)\, \dif A
\eqe
and
\eqb{l}
\mk^e_\mathrm{\tau\tau} 
=  \ds\int_{\Omega^e_0} \mathbf{c}\,\dif A~, \quad
\mk^e_{\tau M}  
=  \ds\int_{\Omega^e_0} \md\, \dif A = (\mk^e_{M\tau})^\mrT~, \quad
\mk^e_{MM}  
=  \ds\int_{\Omega^e_0} \mf\, \dif A~,
\eqe
with
\eqb{lll}
\mathbf{c} \dis c_{\mra\mra}\,\mL_\mra\,\mL^\mrT_\mra + c_{\mra\mrb}\,\big(\mL_\mra\,\mL^\mrT_\mrb + \mL_\mrb\,\mL^\mrT_\mra\big) + c_{\mrb\mrb}\,\mL_\mrb\,\mL^\mrT_\mrb - \ds\frac{c_\mra}{2}\big(\mL^\beta_\alpha\,{\mL^\alpha_\beta}^\mrT+\mN_\mra^2-\mL^2\big) \\[2mm]
\md \dis d_{\mra\mra}\,\mL_\mra\,\tilde\mL^\mrT_\mra + d_{\mra\mrb}\,\mL_\mra\,\tilde\mL^\mrT_\mrb + d_{\mrb\mra}\,\mL_\mrb\,\tilde\mL^\mrT_\mra + d_{\mrb\mrb}\,\mL_\mrb\,\tilde\mL^\mrT_\mrb - d_\mra\,\mL^\beta_\alpha\,\tilde\mL^{\alpha\,\mrT}_\beta~, \\[2mm]
\mf \dis f_{\mra\mra}\,\tilde\mL_\mra\,\tilde\mL^\mrT_\mra
- f_\mra\,\tilde\mL^\beta_\alpha\,\tilde\mL^{\alpha\,\mrT}_\beta~,
\eqe
and
\eqb{l}
\mk^e_\mathrm{\tau} =  \ds\int_{\Omega^e_0} \mathbf{t}\,\dif A~, \quad
\mk^e_{M1} =  \ds\int_{\Omega^e_0} \mm_1\, \dif A ~, \quad
\mk^e_{M2}  =  \ds\int_{\Omega^e_0} \mm_2\, \dif A
\eqe
with
\eqb{lll}
\mt \dis \tau_\mra\,\mN^2_\mra + {\tau_\mathrm{A}\,\mN^2_\mathrm{A}}+ \tau_\mrb\,\mN^2_\mrb~, \\[2mm]
\mm_1 \dis {-\big(2H}\,M^0_\mra + (4H^2-2\kappa)\,M^0_\mrb\big)\,\mL^2~, \\[2mm]
\mm_2 \dis -M^0_\mra\,\mL_\alpha\,\tilde\mL^{\alpha\,\mrT}_\mra -M^0_\mrb\,\mL_\alpha\,\tilde\mL^{\alpha\,\mrT}_\mrb ~.
\eqe

For further efficiency, the symmetric matrices $\mathbf{c}$, $\mf$, $\mt$ and $\mm_1$ should only be computed in upper triangular form, integrated and then reflected after integration (i.e.~after the quadrature loop). Likewise $\mk^e_{M\tau}$ and $\mk^e_M$ should be determined from $\mk^e_{\tau M}$, $\mk^e_{M1}$ and $\mk^e_{M2}$ after quadrature. 
\section{Linearization of the projected shell model}\label{s:linproj}
In this section, the material tangents $c^{\alpha\beta\gamma\delta}$, $d^{\alpha\beta\gamma\delta}$, $e^{\alpha\beta\gamma\delta}$ and $f^{\alpha\beta\gamma\delta}$ that are needed for the linearization of the projected shell model, i.e.~Eqs.~\eqref{e:M2Dab2} and \eqref{e:case2NH}, are derived considering $\xi\approx\xi_0$.
\subsection{Linearization of $g^{\alpha\beta}$}
The linearization of $g^{\alpha\beta}$ gives 
\eqb{l}
\Delta g^{\alpha\beta} = g^{\alpha\beta\gamma\delta}\,\Delta g_{\gamma\delta} ~, 
\eqe
where (see \cite{shelltheo})
\eqb{lllllll}
g^{\alpha\beta\gamma\delta} := -\ds\frac{1}{2}\,\big(g^{\alpha\gamma}\,g^{\beta\delta} + g^{\alpha\delta}\,g^{\beta\delta}\big)~. 
\eqe
Further, from Eq.~\eqref{e:deltgab} we have
\eqb{lll}
\Delta g_{\gamma\delta} = a^{\epsilon\eta}_{\gamma\delta}~\Delta a_{\epsilon\eta} + b^{\epsilon\eta}_{\gamma\delta}~\Delta b_{\epsilon\eta}~, 
\label{e:Dgab_2}
\eqe
with
\eqb{lll}
a^{\epsilon\eta}_{\gamma\delta} \dis g_a\,\delta^\epsilon_\gamma\,\delta^\eta_\delta + \kappa\,\xi^2\, a_{\gamma\delta}\,a^{\epsilon\eta} - \xi^2\,b_{\gamma\delta}\,b^{\epsilon\eta}~, \\[2mm]
b^{\epsilon\eta}_{\gamma\delta} \dis g_b\,\delta^\epsilon_\gamma\,\delta^\eta_\delta - \xi^2\, a_{\gamma\delta}\,\tilde{b}^{\epsilon\eta} + \xi^2\,b_{\gamma\delta}\,a^{\epsilon\eta}~.
\eqe
We thus obtain
\eqb{l}
\Delta g^{\alpha\beta} = g^{\alpha\beta\gamma\delta}_a\,\Delta a_{\gamma\delta} + g^{\alpha\beta\gamma\delta}_b\,\Delta b_{\gamma\delta}~, 
\eqe
with
\eqb{l}
g^{\alpha\beta\gamma\delta}_a : = g^{\alpha\beta\epsilon\eta}\,a_{\epsilon\eta}^{\gamma\delta}~, \quad g^{\alpha\beta\gamma\delta}_b : = g^{\alpha\beta\epsilon\eta}\,b_{\epsilon\eta}^{\gamma\delta}~.
\eqe
\subsection{Linearization of $\tilde\tau^{\alpha\beta}$}
Considering that $\tilde\tau^{\alpha\beta}$  has the form (see e.g.~Eq.~\eqref{e:case2NH})
\eqb{l}
\tilde\tau^{\alpha\beta} = \tilde\mu\,G^{\alpha\beta} + f(J^*)\,g^{\alpha\beta},
\eqe
we have
\eqb{l}
\Delta\tilde\tau^{\alpha\beta} = \left( \ds\frac{f'J^*}{2}\,g^{\alpha\beta}\,g^{\gamma\delta} + f\,g^{\alpha\beta\gamma\delta} \right)\,\Delta g_{\gamma\delta}~.
\eqe
Taking into account Eq.~\eqref{e:Dgab_2}, we obtain
\eqb{l}
\Delta\tilde\tau^{\alpha\beta} = \ds\frac{1}{2}\,\tilde{c}^{\alpha\beta\gamma\delta} \,\Delta a_{\gamma\delta} + \tilde{d}^{\alpha\beta\gamma\delta} \,\Delta b_{\gamma\delta}~,
\eqe
with
\eqb{lll}
\tilde{c}^{\alpha\beta\gamma\delta} \dis 2\,\ds\pa{\tilde{\tau}^{\alpha\beta}}{a_{\gamma\delta}} = J^*f'\,g^{\alpha\beta}\,g^{\epsilon\eta}\,a^{\gamma\delta}_{\epsilon\eta} + 2\,f\,g^{\alpha\beta\gamma\delta}_a ~,\\[5mm]
\tilde{d}^{\alpha\beta\gamma\delta} \dis \ds\pa{\tilde{\tau}^{\alpha\beta}}{b_{\gamma\delta}}  = \ds\frac{J^*f'}{2}\,g^{\alpha\beta}\,g^{\epsilon\eta}\,b^{\gamma\delta}_{\epsilon\eta} + 2\,f\,g^{\alpha\beta\gamma\delta}_b ~.
\eqe
\subsection{Linearization of $\tau^{\alpha\beta}$ and $M_0^{\alpha\beta}$}
From Eq.~\eqref{e:M2Dab2}, we have\footnote{Note that due to the approximation assumed in Eq.~\eqref{e:M2Dab2}, the tangents in Eq.~\eqref{e:cabgd} lose major symmetry, but they still have minor symmetry. Therefore $\mC\neq\mC^T$, $\mE\neq\mD^T$ and $\mF\neq\mF^T$ in Appendix~\ref{s:eff_gen}. If Eq.~\eqref{e:M2Dab3} is used the symmetry is retained.}
\eqb{llcll}
c^{\alpha\beta\gamma\delta} \dis 2\,\ds\pa{\tau^{\alpha\beta}}{a_{\gamma\delta}} \is \ds\int_{-\frac{T}{2}}^{\frac{T}{2}} s_0\,\left[2\,\kappa\,\xi^2\,\tilde{\tau}^{\alpha\beta}\,a^{\gamma\delta} + (1 - \xi^2\,\kappa)\,\tilde{c}^{\alpha\beta\gamma\delta}\right] \dif\xi ~,\\[5mm]
d^{\alpha\beta\gamma\delta} \dis \ds\pa{\tau^{\alpha\beta}}{b_{\gamma\delta}} \is \ds\int_{-\frac{T}{2}}^{\frac{T}{2}} s_0\,\left[-\xi^2\,\tilde{\tau}^{\alpha\beta}\,\tilde{b}^{\gamma\delta} + (1 - \xi^2\,\kappa)\,\tilde{d}^{\alpha\beta\gamma\delta}\right] \dif\xi ~,\\[5mm]
e^{\alpha\beta\gamma\delta} \dis 2\,\ds\pa{M_0^{\alpha\beta}}{a_{\gamma\delta}} \is \ds\int_{-\frac{T}{2}}^{\frac{T}{2}} s_0\,\left[-\xi^2\,\tilde{\tau}^{\alpha\beta}\,b^{\gamma\delta} - (\xi - H\, \xi^2)\,\tilde{c}^{\alpha\beta\gamma\delta}\right] \dif\xi ~,\\[5mm]
f^{\alpha\beta\gamma\delta} \dis \ds\pa{M_0^{\alpha\beta}}{b_{\gamma\delta}} \is \ds\int_{-\frac{T}{2}}^{\frac{T}{2}} s_0\,\left[\ds\frac{1}{2}\,\xi^2\,\tilde{\tau}^{\alpha\beta}\,a^{\gamma\delta} - (\xi - H\, \xi^2)\,\tilde{d}^{\alpha\beta\gamma\delta}\right] \dif\xi ~.
 \label{e:cabgd}
\eqe
\section{Linearization of edge rotation conditions}\label{s:linpen}
In this section, the linearization of the constraint equations, introduced in Sec.~\ref{s:sym} for the edge rotation conditions, is presented for both the penalty method and the Lagrange multiplier method. Simplifications are provided in Remarks \ref{r:alpha0} and \ref{r:fix}.
\subsection{Linearization for the penalty method} 
For the penalty method Eq.~\eqref{e:pen2}, one has (e.g. see~ \cite{shelltheo,droplet})
\eqb{l}
 \delta \bn=\bR^\alpha\,\delta\ba_\alpha,~ \quad \delta\tilde\bn=\tilde\bR^\alpha\,\delta\tilde\ba_\alpha~, 
\eqe
where
\eqb{ll}
 \bR^\alpha := -\ba^\alpha\otimes\bn ~, \quad  \tilde\bR^\alpha := -\tilde\ba^\alpha\otimes\tilde\bn~.
\eqe
Along the patch interface, one can define the tangents $\btau$ and $\btau_0$ as
\eqb{l}
 \btau:= \ds\frac{\ba_\xi}{\norm{\ba_\xi}} ~,\quad\quad  \btau_0:=\ds\frac{\bA_\xi}{\norm{\bA_\xi}} ,
\eqe 
where $\ba_\xi :  = \mN_{,\xi}\,\mx_e$, $\bA_\xi :  = \mN_{,\xi}\,\mX_e$, and  $\xi$ denotes the convective coordinate along the interface.  From this we find
\eqb{l}
\delta\btau = \ds\bM^\xi\,\delta\ba_\xi~,
\eqe
where 
\eqb{l}
 \ds\bM^\xi := \ds\frac{1}{\norm{\ba_\xi}}\,\big(\bone - \btau\otimes\btau\big) ~.
\eqe
We thus have from Eq.~\eqref{e:pen2}
\eqb{rll}
\delta\Pi_{\sL}^e \is \delta\mx_\mre^\mrT\, \mf^e_\mathrm{n}
+\delta\tilde\mx_\mre^\mrT\, \mf^e_\mathrm{\tilde n}~, \\[2mm]
\Delta\delta\Pi_{\sL}^e 
\is \delta\mx_\mre^\mrT\, \mk^e_\mathrm{nn}\,\Delta\mx_e 
+ \delta\mx_\mre^\mrT\, \mk^e_\mathrm{n\tilde n}\,\Delta\tilde\mx_e
+ \delta\tilde\mx_\mre^\mrT\,(\mk^e_\mathrm{n\tilde n})^\mrT\,\Delta\mx_e
+ \delta\tilde\mx_\mre^\mrT\, \mk^e_\mathrm{\tilde n\tilde n}\,\Delta\tilde\mx_e~,
\label{e:ftn}\eqe
with
\eqb{lll}
\mf_\mrn^e := -  \ds \int_{\Gamma_0^e}\epsilon\,\left(\mN_{,\xi}^\mrT\,\bM^\xi\,\btheta  - \mN_{,\alpha}^\mrT\,\tilde{d}_\mra^\alpha\,\bn  \,\right)\,\dif S~, \quad
\mf_{\tilde\mrn}^e := \ds \int_{\Gamma_0^e} \epsilon\, \tilde\mN^\mrT_{,\alpha}\,d_{\tilde\mra}^\alpha\,\tilde{\bn}\,\dif S~,
\label{e:fg1}
\eqe
and
\eqb{lll}
\mk_\mathrm{nn}^e \dis \ds \int_{\Gamma_0^e} \epsilon\, \left[ \mN^\mrT_{,\alpha}\,\bQ^{\alpha\beta}\,\mN_{,\beta}\, + \mN_{,\xi}^\mrT\,\bQ^{\xi\xi}\,\mN_{,\xi}  - 2\,s_0\,\mathrm{sym}\big( \mN^\mrT_{,\alpha}\,\bn\otimes\tilde{\bn}_\mra^\alpha\,\,\bM^\xi\,\mN_{,\xi} \big)\,\right]\dif S~,\\[5mm]
\mk_\mathrm{\tilde n\tilde n}^e \dis \ds \int_{\Gamma_0^e}  \epsilon\, \tilde\mN^\mrT_{,\alpha}\,\tilde\bQ^{\alpha\beta}\,\tilde\mN_{,\beta}\,\dif S~, \\[5mm]
\mk_\mathrm{n\tilde n}^e \dis  \ds \int_{\Gamma_0^e} \epsilon\, \left(s_0 \,\mN_{,\xi}^\mrT\,\bM^\xi\, \bn_{\tilde{\mra}}^\alpha\otimes\tilde\bn\, \tilde\mN_{,\beta} - \mN^\mrT_{,\alpha}\,\hat{a}^{\alpha\beta}\,\bn\otimes\tilde\bn\,\tilde\mN_{,\beta}  \right)  \,\dif S~ .
\label{e:kn}
\eqe
Here, we have defined
\eqb{lll}
\bQ^{\xi\xi} \dis    \ds \Big[ (\btau\cdot\btheta)\,(\bone - 3\,\btau\otimes\btau) + \btheta\otimes\btau + \btau\otimes\btheta\,\Big]/\norm{\ba_\xi}^2~,\\[2mm]
\bQ^{\alpha\beta} \dis  \tilde{d}_\mrn\,a^{\alpha\beta}\,(\bn\otimes\bn) +  \tilde{d}_\mra^\alpha\,\bR^\beta + (\tilde{d}_\mra^{\textcolor{darkgreen}{\beta}}\,\bR^{\textcolor{darkgreen}{\alpha}})^\mrT ~,\\[2mm]
\tilde\bQ^{\alpha\beta} \dis  d_{\tilde{\mrn}}\,\textcolor{darkgreen}{\tilde{a}}^{\alpha\beta}\,(\textcolor{darkgreen}{\tilde{\bn}}\otimes\textcolor{darkgreen}{\tilde{\bn}}) +  d_{\tilde{\mra}}^\alpha\,\textcolor{darkgreen}{\tilde{\bR}}^\beta + (d_{\tilde{\mra}}^{\textcolor{darkgreen}{\beta}}\,\textcolor{darkgreen}{\tilde{\bR}^{\alpha}})^\mrT ~,
\eqe
with $ \tilde{d}_\mra^\alpha:= \tilde\bd\cdot\ba^\alpha$, $\tilde{d}_\mrn:= \tilde\bd\cdot\bn$, $ d_{\tilde{\mra}}^\alpha:= \bd\cdot\tilde\ba^\alpha$, $d_{\tilde{\mrn}}= \bd\cdot\tilde\bn$, $\bn_{\tilde{\mra}}^\alpha:= \bn\times\tilde\ba^\alpha$, $\tilde{\bn}_\mra^\alpha:= \tilde\bn\times\ba^\alpha$ and $\hat{a}^{\alpha\beta}:= \ba^\alpha\cdot(c_0\,\tilde\ba^\beta - s_0\,\btau\times\tilde\ba^\beta)$. 

\refstepcounter{remark}
\textbf{Remark \arabic{remark}:} Note that here $\mN$ and $\tilde{\mN}$ denote the shape function arrays and should not be confused with $\bN$ and $\tilde{\bN}$. In Eqs.~\eqref{e:fg1} and \eqref{e:kn}, the shape functions and their derivatives with the indices $\alpha$ and $\beta$ affect all control points of the elements.  Here the shape function array $\mN_{,\xi}$ is understood to have the same dimension as $\mN_{,\alpha}$ -- i.e. appropriate zeros have to be added to $\mN_{,\xi}$.

\refstepcounter{remark}
\label{r:alpha0}
\textbf{Remark \arabic{remark}:} In case of $\alpha_0 = 0$, for Figs.~\ref{f:gccase}a and \ref{f:gccase}e, Eqs.~\eqref{e:fg1} and \eqref{e:kn} are further reduced to
\eqb{lll}
\mf_\mrn^e :=   \ds \int_{\Gamma_0^e}\epsilon\,\mN_{,\alpha}^\mrT\, (\bn\otimes\tilde\bn)\,\ba^\alpha\,\dif S~, \quad
\mf_{\tilde\mrn}^e := \ds \int_{\Gamma_0^e} \epsilon\, \tilde\mN^\mrT_{,\alpha}\, (\tilde\bn\otimes\bn)\,\tilde\ba^\alpha\,\dif S~,
\label{e:fg1G1}
\eqe
and
\eqb{lll}
\mk_\mathrm{nn}^e \dis \ds \int_{\Gamma_0^e} \epsilon\,  \mN^\mrT_{,\alpha}\,\left[ \tilde{d}_\mrn\,a^{\alpha\beta}\,(\bn\otimes\bn) -  \tilde{d}_\mra^\alpha\,(\ba^\beta\otimes\bn) - \tilde{d}_\mra^\beta\,(\bn\otimes\ba^\alpha)  \right]\,\mN_{,\beta}\,\dif S~,\\[5mm]
\mk_\mathrm{\tilde n\tilde n}^e \dis \ds \int_{\Gamma_0^e}  \epsilon\, \tilde\mN^\mrT_{,\alpha}\, \left[ d_{\tilde{\mrn}}\,\textcolor{darkgreen}{\tilde{a}}^{\alpha\beta}\,(\textcolor{darkgreen}{\tilde\bn}\otimes\textcolor{darkgreen}{\tilde\bn}) -  d_{\tilde{\mra}}^\alpha\, (\textcolor{darkgreen}{\tilde{\ba}}^\beta\otimes\textcolor{darkgreen}{\tilde{\bn}}) -  d_{\tilde{\mra}}^\beta\,(\textcolor{darkgreen}{\tilde{\bn}}\otimes\textcolor{darkgreen}{\tilde{\ba}}^\alpha)  \right]\,\tilde\mN_{,\beta}\,\dif S~, \\[5mm]
\mk_\mathrm{n\tilde n}^e \dis  \ds -\int_{\Gamma_0^e} \epsilon\, \mN^\mrT_{,\alpha}\,(\ba^\alpha\cdot\tilde\ba^\beta)\,(\bn\otimes\tilde\bn)\,\tilde\mN_{,\beta}    \,\dif S~ ,
\label{e:knG1}
\eqe
with $\tilde{d}_\mrn = d_{\tilde{\mrn}}:= \tilde\bn\cdot\bn$, $ \tilde{d}_\mra^\alpha:= \tilde\bn\cdot\ba^\alpha$ and $ d_{\tilde{\mra}}^\alpha:= \bn\cdot\tilde\ba^\alpha$. In the case of Fig.~\ref{f:gccase}\textcolor{darkgreen}{.e}, $\mf^e_{\tilde{n}}$ and the related tangents are not required.
%
%
\subsection{Linearization for the Lagrange multiplier method} 
For the Lagrange multiplier method, using the interpolation $q = \mN_\mrq\,\mq_\mre$, from Eq.~\eqref{e:LMgc} we find
\eqb{rll}
\delta\Pi_{\sL}^e \is \delta\mx_\mre^\mrT\, \mf^e_\mathrm{n}
+\delta\tilde\mx_\mre^\mrT\, \mf^e_\mathrm{\tilde n} +   \delta\mq^\mrT_\mre\, \mf^e_\mrq ~, \\[2mm]
\Delta\delta\Pi_{\sL}^e 
\is \delta\mx_\mre^\mrT\, \mk^e_\mathrm{nn}\,\Delta\mx_e 
+ \delta\tilde\mx_\mre^\mrT\, \mk^e_\mathrm{\tilde n\tilde n}\,\Delta\tilde\mx_e + \delta\mx_\mre^\mrT\, \mk^e_\mathrm{n\tilde n}\,\Delta\tilde\mx_e
+ \delta\tilde\mx_\mre^\mrT\,(\mk^e_\mathrm{n\tilde n})^\mrT\,\Delta\mx_e\\[2mm]
\plus \delta\mx_\mre^\mrT\, \mk^e_\mathrm{nq}\,\Delta\mq_\mre
+ \delta\mq_\mre^\mrT\, (\mk^e_\mathrm{nq})^\mrT\,\Delta\mx_\mre
+ \delta\tilde\mx_\mre^\mrT\, \mk^e_\mathrm{\tilde n q}\,\Delta\mq_\mre
+ \delta\mq_\mre^\mrT\, (\mk^e_\mathrm{\tilde n q})^\mrT\,\tilde\Delta\mx_\mre
~.
\label{e:ftnL}\eqe
As seen in Eq.~\eqref{e:LMgc}, the last integral has the same form as for the penalty method, Eq.~\eqref{e:pen2}. Therefore, the quantities $\mf^e_\mathrm{n}$, $\mf^e_\mathrm{\tilde n}$, $\mk^e_\mathrm{nn}$, $\mk_\mathrm{\tilde n\tilde n}^e$, and $\mk_\mathrm{n\tilde n}^e$ are already given in Eqs.~\eqref{e:fg1} and \eqref{e:kn} with the substitution $\epsilon=1$, $c_0~\hat{=}~q\,(s_0 + c_0)$, $s_0~\hat{=}~q\,(s_0 - c_0)$; and $\bd$, $\tilde\bd$, $\btheta$ are now given in Eq.~\eqref{e:defdLM}. Additionally, we find
\eqb{lll}
\quad \mf_\mrq^e := \ds \int_{\Gamma_0^e}  \mN^\mrT_\mrq\, g_\sL\,\dif S~,
\label{e:fg1L}
\eqe
and
\eqb{lll}
\mk_\mathrm{n q}^e := \ds   -\int_{\Gamma_0^e} \left[ \mN^\mrT_{,\alpha}\,(\bR^\alpha)^\mrT\,  \tilde\bd  + \mN^\mrT_{,\xi}\,\bM^\xi\,\btheta\,  \right]\,\mN_\mrq \,\dif S~ , 
\quad \mk_\mathrm{\tilde n q}^e := -\ds \int_{\Gamma_0^e} \tilde\mN^\mrT_{,\alpha}\,(\tilde\bR^\alpha)^\mrT\, \bd\,\,\mN_\mrq\,\dif S~ .
\label{e:knL}\eqe

\refstepcounter{remark}
\label{r:fix}
\textbf{Remark \arabic{remark}:} Note that in the cases of symmetry constraints and rotational Dirichlet boundary conditions, $\tilde{\bn}$ is given and therefore Eqs.~(\ref{e:fg1}.2),~(\ref{e:kn}.2-\ref{e:kn}.3)~and~(\ref{e:knL}.2) vanish.
\section{Analytical solution of the pinched cylinder with rigid end diaphragms}\label{s:fourier}
This appendix provides the analytical solution following the approach of \citet{flugge62} for the pinched cylinder with rigid diaphragms. It is based on the double Fourier representation of the displacement field and the applied loading. 

Accordingly, if the shell deformation is described by $u$, $v$ and $w$ in the axial, circumferential and radial directions, respectively, the equilibrium of the system satisfies the following system of partial differential equations \citep{flugge62}
\eqb{lll}
\begin{cases}
u'' + \ds\frac{1-\nu}{2}\,u^{**} + \frac{1+\nu}{2}\,v'^{*} + v\,w' + k\,\left[\frac{1-\nu}{2}\,u^{**} - w''' + \frac{1-\nu}{2}\,w'^{**}\right] = - \ds\frac{p_x\,R^2}{d}~,\\[4mm]\\
\ds\frac{1+\nu}{2}\,u'^* + v^{**} + \frac{1-\nu}{2}\,v'' + w^* + k\,\left[\frac{3}{2}\,(1-\nu)\,v'' - \frac{3-\nu}{2}\,w''^*\right] = -\ds\frac{p_\phi\,R^2}{d}~,\\[6mm]
\nu\,u' + v^* + w + k\,\left[\ds\frac{1-\nu}{2}\,u'^{**} - u''' - \frac{3-\nu}{2}\,v''^* + w^{''''} + 2\,w''^{**}\right. +\\[4mm]
 ~~~~~~~~~~~~~~~~~~~~~~~~~~~~~~~~~~~~~~~~~~~~~~~~~~~~~+ ~\ds w^{****} + 2\,w^{**} + w\Big] = \ds\frac{p_r\,R^2}{d}~,
\label{e:pcylPDE}
\end{cases}
\eqe
with $k:= T^2/(12\,R^2)$, $d:= (E\,T)/(1-\nu^2)$, where $R$ and $T$ are the radius and thickness of the cylinder, respectively, $\nu$ is Poisson's ratio, $E$ is Young's modulus and the partial derivatives are denoted by $ (\bullet)':=R\,\partial(\bullet)/\partial x $ and $ (\bullet)^*:=\partial(\bullet)/\partial \phi $.
Further, as shown by \citet{bijlaard54}, if $N_\mrf$ concentrated radial loads are applied at the middle of the cylinder and they are equally spaced along the circumferential direction, they can be expressed in terms of the double Fourier series as
\eqb{lll}
\begin{cases}
p_x = 0~,\\[4mm]
p_\phi = 0~,\\[4mm]
p_r = \ds\sum_{m=0}^{\infty}\sum_{n=1}^{\infty} p_{rmn}\,\cos(N_\mrf\,m\,\phi)\,\sin\left(\frac{\lambda\,x}{R}\right)~,
\end{cases}
\label{e:loadP}
\eqe
where $\lambda:= \ds\frac{n\,\pi\,R}{L}$, $L$ is the length of the cylinder, 
\eqb{lll}
\begin{cases}
p_{rmn} = \ds(-1)^{(n-1)/2} \,N_\mrf\,\ds\frac{P}{\pi R \,L}~,\quad &\mathrm{if}\; m = 0 \;\mathrm{and}\; n=1,\,3,\,5,\cdots~,\\[4mm]
p_{rmn} =  \ds(-1)^{(n-1)/2} \,N_\mrf\,\ds\frac{2\,P}{\pi R \,L}~,\quad &\mathrm{if}\; m \neq 0 \;\mathrm{and}\; n=1,\,3,\,5,\cdots~,\\[4mm]
p_{rmn} = 0~,\quad &\mathrm{otherwise}~,
\end{cases}
\label{e:loadP2}
\eqe
and $P$ is the magnitude of the loads. 

It should be noted that generally the representation of concentrated loads by a series such as Eq.~\eqref{e:loadP2} has numerical deficiencies. The $p_{rmn}$ terms alternate in sign and do not decay as $m\rightarrow\infty$. Although the series still converges in theory, it does not converge numerically due to inaccuracies caused by the Gibbs effect.  

In general, Eq.~\eqref{e:loadP} can represent any symmetric pressure distribution. If the cylinder is pinched by two opposing forces applied at the middle, the Fourier components can be found by setting $N_\mrf=2$, $ \phi=0 $ or $ \phi=\pi $ and $ x=L/2 $. The corresponding solution for Eq.~\eqref{e:pcylPDE} can be found by assuming the following ansatz for the displacement field
\eqb{lll}
u = \ds\sum_{m=0}^{\infty}\sum_{n=0}^{\infty} u_{mn}\,\cos(N_\mrf \,m\,\phi)\,\cos\left(\frac{\lambda\,x}{R}\right)~, \\[4mm]
v = \ds\sum_{m=1}^{\infty}\sum_{n=1}^{\infty} u_{mn}\,\sin(N_\mrf \,m\,\phi)\,\sin\left(\frac{\lambda\,x}{R}\right)~, \\[4mm]
w = \ds\sum_{m=0}^{\infty}\sum_{n=1}^{\infty} u_{mn}\,\cos(N_\mrf \,m\,\phi)\,\sin\left(\frac{\lambda\,x}{R}\right)~.
\label{e:dispF}
\eqe
As seen, Eq.~\eqref{e:dispF} satisfies the boundary conditions of the problem with rigid diaphragms or simple supports, i.e. $v = w = 0$ if $x = 0$ or $x = L$. Plugging Eq.~\eqref{e:dispF} into Eq.~\eqref{e:pcylPDE}, the following system of equations is obtained 
\eqb{lll}
\begin{cases}
\ds\left[\lambda^2 + \ds\frac{1-\nu}{2}\,N_\mrf^2\,m^2\,(1+k)\right]\,u_{mn} + \left[-\frac{1+\nu}{2}\,\lambda\,N_\mrf\,m\right]\,v_{mn}~ +\\[4mm]
~~~~~~~~~~~~~~~~~~~~~~~~~~~~~~~~~~~~~~ + \left[-\nu\,\lambda - k\,\Big(\lambda^3 - \ds\frac{1-\nu}{2}\,\lambda\,N_\mrf^2\,m^2\Big)\right]\,w_{mn} = 0~, \\[4mm]
\ds\left[-\ds\frac{1+\nu}{2}\,\lambda\,N_\mrf\,m\right]\,u_{mn} + \left[N_\mrf^2\,m^2 + \frac{1-\nu}{2}\,\lambda^2\,(1+3\,k)\right]\,v_{mn}~ +\\[4mm]
~~~~~~~~~~~~~~~~~~~~~~~~~~~~~~~~~~~~~~~~~~~~~~~~~ + \left[N_\mrf\,m + \ds\frac{3-\nu}{2}\,k\,\lambda^2\,N_\mrf\,m\right]\,w_{mn} = 0~,\\[6mm]
\ds\left[-\nu\,\lambda - k\,\Big(\lambda^3 - \frac{1-\nu}{2}\,\lambda\,N_\mrf^2\,m^2\Big)\right]\,u_{mn} + \left[N_\mrf\,m + \frac{3-\nu}{2}\,k\lambda^2\,N_\mrf\,m\right]\,v_{mn}~+ \\[4mm]
 ~~~~~~~~~~~~~~+ \Big[1 + k\,(\lambda^4 + 2\,\lambda^2\,N_\mrf^2\,m^2 + N_\mrf^4\,m^4 - 2\,N_\mrf^2\,m^2 +1)\Big]\,w_{mn} = \ds\frac{p_{rmn}\,R^2}{d}~,
 \end{cases}
\eqe
which can be solved numerically for the unknown coefficients of the Fourier series ($u_{mn}$, $v_{mn}$ and $w_{mn}$) in Eq.~\eqref{e:dispF}.
\section{Skew surface meshes}\label{s:mesh}
\begin{figure}[ht]
\begin{center} \unitlength1cm
\unitlength1cm
\begin{picture}(0,4.5)
\put(-8.0,0.5){\includegraphics[height=45mm]{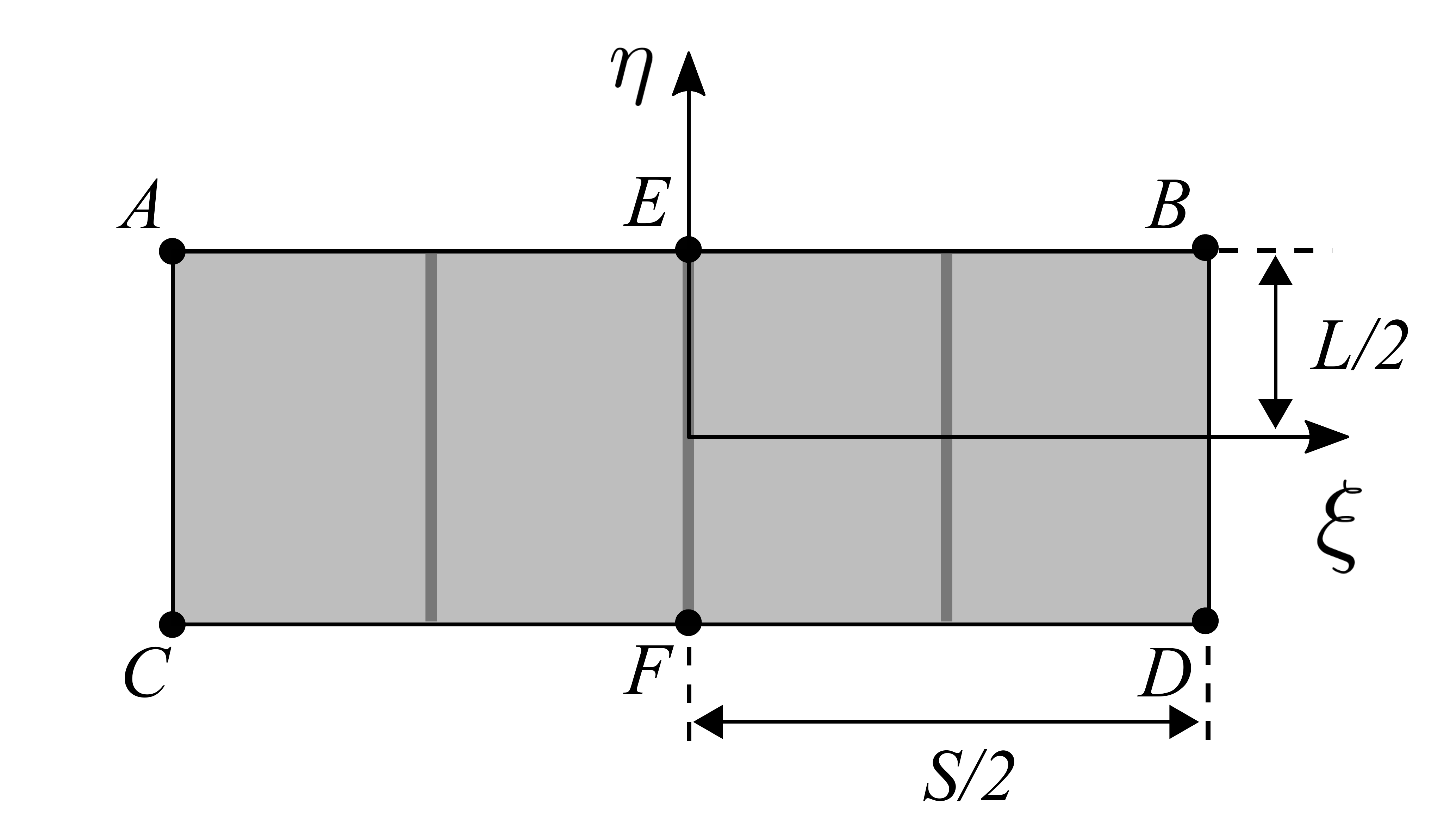}}
\put(-4.0,0.1){a.}
\put(0.0,0.5){\includegraphics[height=45mm]{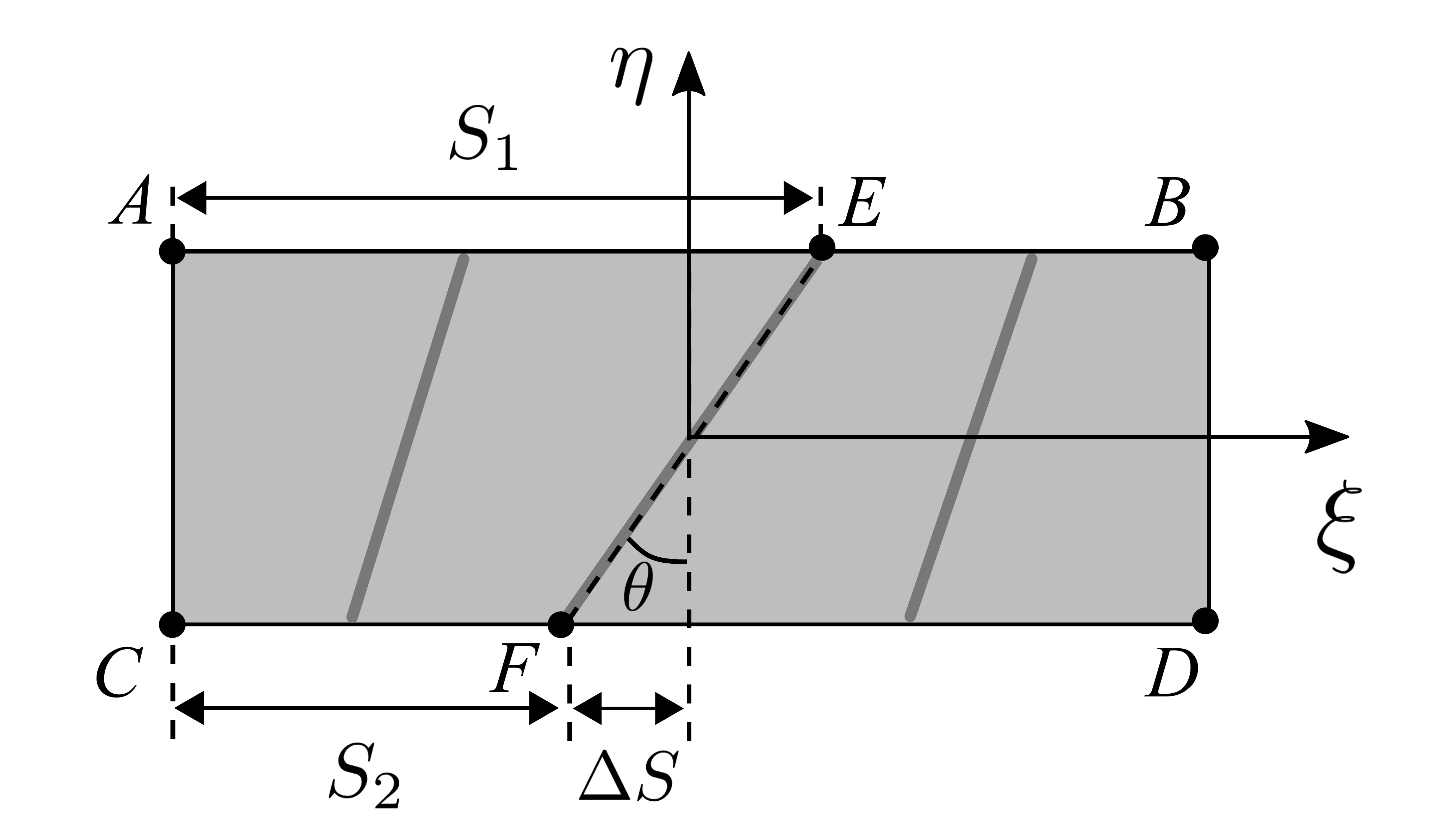}}
\put(4.0,0.1){b.}
\end{picture}
\caption{Distortion of a regular knot grid. a: Regular knot grid and b: skew knot grid. The control points are then recomputed using the knot insertion algorithm (see e.g.~\citet{hughes05})}
\label{f:mesh}
\end{center}
\end{figure}

This appendix explains the procedure to create skew meshes out of regular meshes for a single patch, which is used in Secs.~\ref{s:canbeam} and \ref{s:PB}. Fig.~\ref{f:mesh}a shows the knot grid of a regular rectangular mesh on a single patch. Assuming the knot space has the dimensions $ L\times S $ in general.  The skew mesh is obtained by two steps:

Step 1. Distorting the regular knot grid by the mapping
\eqb{lll}
\begin{cases}
Y_d(s,\eta) = Y(\xi,\eta)~,\\[3mm]
X_d(\xi,\eta) = X(\xi,\eta) -\ds\frac{2\Delta S}{L}\eta \left( 1 - \frac{4}{S^2}\xi^2\right) ~,
\end{cases}
\eqe
where $ (X,Y) $ and $ (X_d,Y_d) $ are the coordinates of the regular and distorted knot grids, respectively; $ \Delta S = L/2\,\tan\theta $, where $\theta$ is the angle of the distorted mid-line w.r.t.~the $\eta$-axis (see Fig.~\ref{f:mesh}b). The distortion or skewness ratio, $ r $, is defined by
\eqb{ll}
r:= \ds\frac{S_1-S_2}{S_1+S_2}=\frac{2\Delta S}{S}=\frac{L}{S}\tan\theta~,
\eqe
where $ r \in [0,1] $. 

Step 2. The positions of control points are recomputed with the corresponding distorted knot grid using the knot insertion algorithm of \citet{hughes05}. Here, the B{\'e}zier extraction operator is computed approximately at the center point of the knot grid.

\bibliographystyle{apalike}
\bibliography{ShellBib,bibliography}

\end{document}

%% file: neco_updated.tex
\newcommand{\bitm}{\begin{itemize}}
\newcommand{\eitm}{\end{itemize}}
\newcommand{\bnumr}{\begin{enumerate}}
\newcommand{\enumr}{\end{enumerate}}

\newcommand {\aab}{a^{\alpha\beta}}
\newcommand {\auab}{a_{\alpha\beta}}
\newcommand {\agd}{a^{\gamma\delta}}

\newcommand {\augd}{a_{\gamma\delta}}

\newcommand {\Aab}{A^{\alpha\beta}}
\newcommand {\Auab}{A_{\alpha\beta}}

\newcommand {\Mab}{M^{\alpha\beta}}

\newcommand {\bab}{b^{\alpha\beta}}

\newcommand {\buab}{b_{\alpha\beta}}

\newcommand {\bgd}{b^{\gamma\delta}}

\newcommand {\bugd}{b_{\gamma\delta}}

\newcommand {\tauab}{\tau^{\alpha\beta}}

\newcommand{\mrA}{\mathrm{A}}

\newcommand{\mrT}{\mathrm{T}}

\newcommand {\eqb}[1]{\begin{equation}\begin{array}{#1}}
\newcommand {\eqe}{\end{array}\end{equation}}

\newcommand {\esb}[1]{\begin{equation*}\begin{array}{#1}}
\newcommand {\ese}{\end{array}\end{equation*}}
\newcommand {\ds}{\displaystyle}

\newcommand {\pa}[2]{\frac{\partial{#1}}{\partial{#2}}}

\newcommand {\back}{\! \! \!}
\newcommand {\is}{\back &=& \back}
\newcommand {\dis}{\back &:=& \back}

\newcommand {\plus}{\back &+& \back}
\newcommand {\mi}{\back &-& \back}

\newcommand {\norm}[1]{\|#1\|}
\newcommand {\tr}{\mathrm{tr}\,}

\newcommand {\dif}{\mathrm{d}}

\newcommand {\II}{{I\kern-.3em I}}
\newcommand {\III}{{I\kern-.3em I\kern-.3em I}}

\newcommand {\mra}{\mathrm{a}}
\newcommand {\mrb}{\mathrm{b}}
\newcommand {\mrc}{\mathrm{c}}

\newcommand {\mre}{\mathrm{e}}
\newcommand {\mrf}{\mathrm{f}}

\newcommand {\mrn}{\mathrm{n}}

\newcommand {\mrq}{\mathrm{q}}

\newcommand {\mrs}{\mathrm{s}}

\newcommand {\md}{\mathbf{d}}

\newcommand {\mf}{\mathbf{f}}

\newcommand {\mk}{\mathbf{k}}

\newcommand {\mm}{\mathbf{m}}

\newcommand {\mq}{\mathbf{q}}

\newcommand {\mt}{\mathbf{t}}

\newcommand {\mx}{\mathbf{x}}

\newcommand {\ba}{\boldsymbol{a}}
\newcommand {\bb}{\boldsymbol{b}}

\newcommand {\bd}{\boldsymbol{d}}

\newcommand {\bff}{\boldsymbol{f}}
\newcommand {\bg}{\boldsymbol{g}}

\newcommand {\bi}{\boldsymbol{i}}

\newcommand {\bm}{\boldsymbol{m}}
\newcommand {\bn}{\boldsymbol{n}}

\newcommand {\bt}{\boldsymbol{t}}
\newcommand {\bu}{\boldsymbol{u}}
\newcommand {\bv}{\boldsymbol{v}}

\newcommand {\bx}{\boldsymbol{x}}

\newcommand {\bmu}{\boldsymbol{\mu}}

\newcommand {\bnu}{\mbox{\boldmath$\nu$}}

\newcommand {\bvphi}{\mbox{\boldmath$\varphi$}}

\newcommand {\bxi}{\mbox{\boldmath$\xi$}}

\newcommand {\btheta}{\mbox{\boldmath$\theta$}}

\newcommand {\mB}{\mathbf{B}}
\newcommand {\mC}{\mathbf{C}}
\newcommand {\mD}{\mathbf{D}}
\newcommand {\mE}{\mathbf{E}}
\newcommand {\mF}{\mathbf{F}}
\newcommand {\mG}{\mathbf{G}}

\newcommand {\mL}{\mathbf{L}}

\newcommand {\mN}{\mathbf{N}}

\newcommand {\mX}{\mathbf{X}}

\newcommand {\bA}{\boldsymbol{A}}
\newcommand {\bB}{\boldsymbol{B}}
\newcommand {\bC}{\boldsymbol{C}}

\newcommand {\bE}{\boldsymbol{E}}
\newcommand {\bF}{\boldsymbol{F}}
\newcommand {\bG}{\boldsymbol{G}}

\newcommand {\bI}{\boldsymbol{I}}

\newcommand {\bK}{\boldsymbol{K}}

\newcommand {\bM}{\boldsymbol{M}}
\newcommand {\bN}{\boldsymbol{N}}

\newcommand {\bQ}{\boldsymbol{Q}}
\newcommand {\bR}{\boldsymbol{R}}
\newcommand {\bS}{\boldsymbol{S}}
\newcommand {\bT}{\boldsymbol{T}}

\newcommand {\bX}{\boldsymbol{X}}

\newcommand {\bsig}{\mbox{\boldmath$\sigma$}}

\newcommand {\btau}{\mbox{\boldmath$\tau$}}

\newcommand {\bone}{\mathbf{1}}

\newcommand {\bbC}{\mathbb{C}}

\newcommand {\bbF}{\mathbb{F}}

\newcommand {\bbR}{\mathbb{R}}

\newcommand {\IR}{{\rm\kern.24em
   \vrule width.02em height1.53ex depth-.05ex
   \kern-.3em R}}
\newcommand {\ic}{{\rm\kern.20em
   \vrule width.02em height1.0ex depth-.05ex
   \kern-.22em c}}
\newcommand {\ia}{{\rm\kern.20em
   \vrule width.02em height1.05ex depth-.0ex
   \kern-.25em a}}
\newcommand {\IC}{{\rm\kern.24em
   \vrule width.02em height1.4ex depth-.05ex
   \kern-.26em C}}
\newcommand {\ID}{{\rm\kern.34em
   \vrule width.02em height1.5ex depth-.05ex
   \kern-.36em D}}
\newcommand {\IS}{{\rm\kern.24em
   \vrule width.02em height1.6ex depth.05ex
   \kern-.26em S}}
\newcommand {\IT}{{\rm\kern.50em
   \vrule width.02em height1.55ex depth-.05ex
   \kern-.52em T}}

\newcommand {\IE}{{\rm\kern.24em
   \vrule width.02em height1.55ex depth-.05ex
   \kern-.33em E}}
\newcommand {\IEa}{{\rm\kern.24em
   \vrule width.02em height1.55ex depth-.05ex
   \kern-.33em E}^{1}_{ijkl}}
\newcommand {\IEb}{{\rm\kern.24em
   \vrule width.02em height1.55ex depth-.05ex
   \kern-.33em E}^{2}_{ijkl}}

\newcommand {\sL}{\mathcal{L}}

\newcommand {\sP}{\mathcal{P}}

\newcommand {\sS}{\mathcal{S}}

\newcommand {\sV}{\mathcal{V}}

\newcommand {\vaua}{\ba_\alpha}

\newcommand {\vaub}{\ba_\beta}

\newcommand {\vaug}{\ba_\gamma}

\newcommand {\Ass}[2]{\kern 0.9ex \vrule width0.45em height0.2ex depth0ex \kern -2.1ex \bigwedge_{#1}^{#2}}
\newcommand {\ASS}[2]{\kern 1.45ex \vrule width0.5em height0.2ex depth0ex \kern -2.65ex \bigwedge_{#1}^{#2}}




%% file: solid_shell_30_arxiv_2025.bbl
\begin{thebibliography}{}

\bibitem[Becker et~al., 2011]{becker11}
Becker, G., Geuzaine, C., and Noels, L. (2011).
\newblock A one field full discontinuous galerkin method for
  {Kirchhoff-–Love} shells applied to fracture mechanics.
\newblock {\em Comp. Meth. Appl. Mech. Engrg.}, {\bf 200}(45–46):3223--3241.

\bibitem[Belytschko et~al., 1985]{belytschko85}
Belytschko, T., Stolarski, H., Liu, W.~K., Carpenter, N., and Ong, J.~S.
  (1985).
\newblock Stress projection for membrane and shear locking in shell finite
  elements.
\newblock {\em Comp. Meth. Appl. Mech. Engrg.}, {\bf 51}(1–3):221--258.

\bibitem[Benson et~al., 2010]{benson10}
Benson, D., Bazilevs, Y., Hsu, M., and Hughes, T. (2010).
\newblock Isogeometric shell analysis: The {Reissner-–Mindlin} shell.
\newblock {\em Comp. Meth. Appl. Mech. Engrg.}, {\bf 199}(5–8):276--289.
\newblock Computational Geometry and Analysis.

\bibitem[Benson et~al., 2011]{benson11}
Benson, D.~J., Bazilevs, Y., Hsu, M.-C., and Hughes, T. J.~R. (2011).
\newblock A large deformation, rotation-free, isogeometric shell.
\newblock {\em Comp. Methods Appl. Mech. Engrg.}, {\bf 200}(13-16):1367--1378.

\bibitem[Benson et~al., 2013]{benson13}
Benson, D.~J., Hartmann, S., Bazilevs, Y., Hsu, M.-C., and Hughes, T. J.~R.
  (2013).
\newblock Blended isogeometric shells.
\newblock {\em Comp. Methods Appl. Mech. Engrg.}, {\bf 255}:133--146.

\bibitem[Bertsekas, 1982]{dimitri82}
Bertsekas, D. (1982).
\newblock {\em Constrained optimization and Lagrange multiplier methods}.
\newblock Academic Press, New York.

\bibitem[Bijlaard, 1954]{bijlaard54}
Bijlaard, P. (1954).
\newblock Stresses from local loadings in cylindrical pressure vessels.
\newblock {\em Weld. J.}, {\bf 33}(12):615--623.

\bibitem[Bischoff and Ramm, 1997]{bischoff97}
Bischoff, M. and Ramm, E. (1997).
\newblock Shear deformable shell elements for large strains and rotations.
\newblock {\em Int. J. Numer. Meth. Engrg.}, {\bf 40}(23):4427--4449.

\bibitem[Bischoff et~al., 2004]{bischoff04}
Bischoff, M., Wall, W.~A., Bletzinger, K.-U., and Ramm, E. (2004).
\newblock Models and finite elements for thin-walled structures.
\newblock In Stein, E., de~Borst, R., and Hughes, T. J.~R., editors, {\em
  Encyclopedia of Computational Mechanics. Vol. 2: Solids and Structures.
  Chapter 3}. Wiley.

\bibitem[Borden et~al., 2011]{borden11}
Borden, M.~J., Scott, M.~A., Evans, J.~A., and Hughes, T. J.~R. (2011).
\newblock Isogeometric finite element data structures based on {B}ezier
  extraction of {NURBS}.
\newblock {\em Int. J. Numer. Meth. Engng.}, {\bf 87}:15--47.

\bibitem[Bouclier et~al., 2013a]{bouclier13}
Bouclier, R., Elguedj, T., and Combescure, A. (2013a).
\newblock Efficient isogeometric {NURBS}--based solid--shell elements: Mixed
  formulation and method.
\newblock {\em Comp. Meth. Appl. Mech. Engrg.}, {\bf 267}(0):86--110.

\bibitem[Bouclier et~al., 2013b]{bouclier13b}
Bouclier, R., Elguedj, T., and Combescure, A. (2013b).
\newblock On the development of nurbs-based isogeometric solid shell elements:
  2d problems and preliminary extension to 3d.
\newblock {\em Comput. Mech.}, {\bf 52}(5):1085--1112.

\bibitem[Bouclier et~al., 2015]{bouclier15}
Bouclier, R., Elguedj, T., and Combescure, A. (2015).
\newblock An isogeometric {locking-free} {NURBS-based} {solid-shell} element
  for geometrically nonlinear analysis.
\newblock {\em Int. J. Numer. Meth. Engrg.}, {\bf 101}(10):774--808.

\bibitem[Brunet and Sabourin, 2006]{brunet06}
Brunet, M. and Sabourin, F. (2006).
\newblock Analysis of a rotation-free 4-node shell element.
\newblock {\em Int. J. Numer. Meth. Engrg.}, {\bf 66}(9):1483--1510.

\bibitem[Canham, 1970]{canham70}
Canham, P.~B. (1970).
\newblock The minimum energy of bending as a possible explanation of the
  biconcave shape of the human red blood cell.
\newblock {\em J. Theoret. Biol.}, {\bf {26}}:61--81.

\bibitem[Ciarlet, 2005]{ciarlet05}
Ciarlet, P.~G. (2005).
\newblock An introduction to differential geometry with applications to
  elasticity.
\newblock {\em J. Elast.}, {\bf 78-79}:3--201.

\bibitem[Cirak and Long, 2010]{cirak10}
Cirak, F. and Long, Q. (2010).
\newblock Advances in subdivision finite elements for thin shells.
\newblock In De~Mattos~Pimenta, P. and Wriggers, P., editors, {\em New Trends
  in Thin Structures: Formulation, Optimization and Coupled Problems},
  volume~{\bf 519} of {\em CISM International Centre for Mechanical Sciences},
  pages 205--227. Springer Vienna.

\bibitem[Cirak and Ortiz, 2001]{cirak01}
Cirak, F. and Ortiz, M. (2001).
\newblock Fully {$C^1$--conforming} subdivision elements for finite deformation
  thin-shell analysis.
\newblock {\em Int. J. Numer. Meth. Engrg.}, {\bf 51}:813--834.

\bibitem[Cirak et~al., 2000]{cirak00}
Cirak, F., Ortiz, M., and Schr{\"{o}}der, P. (2000).
\newblock Subdivision surfaces: a new paradigm for thin-shell finite-element
  analysis.
\newblock {\em Int. J. Numer. Meth. Engrg.}, {\bf 47}(12):2039--2072.

\bibitem[Coleman and Noll, 1964]{coleman64}
Coleman, B.~D. and Noll, W. (1964).
\newblock The thermodynamics of elastic materials with heat conduction and
  viscosity.
\newblock {\em Arch. Ration. Mech. Anal.}, {\bf 13}:167--178.

\bibitem[De~Borst, 1991]{de91}
De~Borst, R. (1991).
\newblock The zero-normal-stress condition in plane-stress and shell
  elastoplasticity.
\newblock {\em Commun. Appl. Num. Meth.}, {\bf 7}(1):29--33.

\bibitem[Deng et~al., 2015]{deng15}
Deng, X., Korobenko, A., Yan, J., and Bazilevs, Y. (2015).
\newblock Isogeometric analysis of continuum damage in rotation-free composite
  shells.
\newblock {\em Comp. Meth. Appl. Mech. Engrg.}, {\bf 284}:349--372.
\newblock Isogeometric Analysis Special Issue.

\bibitem[Dornisch, 2015]{Dornisch-phd}
Dornisch, W. (2015).
\newblock {\em Interpolation of rotations and coupling of patches in
  isogeometric Reissner-Mindlin shell analyis}.
\newblock PhD thesis, RWTH Aachen University, Aachen, Germany.

\bibitem[Dornisch and Klinkel, 2014]{dornisch14}
Dornisch, W. and Klinkel, S. (2014).
\newblock Treatment of {Reissner-–Mindlin} shells with kinks without the need
  for drilling rotation stabilization in an isogeometric framework.
\newblock {\em Comp. Meth. Appl. Mech. Engrg.}, {\bf 276}:35--66.

\bibitem[Dornisch et~al., 2013]{dornisch13}
Dornisch, W., Klinkel, S., and Simeon, B. (2013).
\newblock Isogeometric {Reissner-–Mindlin} shell analysis with exactly
  calculated director vectors.
\newblock {\em Comp. Meth. Appl. Mech. Engrg.}, {\bf 253}:491--504.

\bibitem[Du et~al., 2015]{du215}
Du, X., Zhao, G., and Wang, W. (2015).
\newblock Nitsche method for isogeometric analysis of {Reissner-–Mindlin}
  plate with {non-conforming} {multi-patches}.
\newblock {\em Comp. Aid. Geom. Des.}, {\bf 35}-–{\bf 36}:121--136.
\newblock Geometric Modeling and Processing 2015.

\bibitem[Dvorkin et~al., 1995]{dvorkin95}
Dvorkin, E.~N., Pantuso, D., and Repetto, E.~A. (1995).
\newblock A formulation of the mitc4 shell element for finite strain
  elasto-plastic analysis.
\newblock {\em Comp. Meth. Appl. Mech. Engrg.}, {\bf 125}(1):17--40.

\bibitem[Echter et~al., 2013]{echter13}
Echter, R., Oesterle, B., and Bischoff, M. (2013).
\newblock A hierarchic family of isogeometric shell finite elements.
\newblock {\em Comp. Meth. Appl. Mech. Engrg.}, {\bf 254}:170--180.

\bibitem[Fl{\"u}gge, 1962]{flugge62}
Fl{\"u}gge, W. (1962).
\newblock {\em Stresses in shells}.
\newblock Springer-Verlag.

\bibitem[Goyal et~al., 2013]{goyal13}
Goyal, A., D{\"o}rfel, M., Simeon, B., and Vuong, A. (2013).
\newblock Isogeometric shell discretizations for flexible multibody dynamics.
\newblock {\em Multibody Syst. Dyn.}, {\bf 30}(2):139--151.

\bibitem[Green and Turkiyyah, 2005]{green05}
Green, S. and Turkiyyah, G.~M. (2005).
\newblock A rotation-free quadrilateral thin shell subdivision finite element.
\newblock {\em Commun. Numer. Meth. Engrg.}, {\bf 21}(12):757--767.

\bibitem[Guo and Ruess, 2015a]{guo15}
Guo, Y. and Ruess, M. (2015a).
\newblock Nitsche’s method for a coupling of isogeometric thin shells and
  blended shell structures.
\newblock {\em Comp. Meth. Appl. Mech. Engrg.}, {\bf 284}:881--905.
\newblock Isogeometric Analysis Special Issue.

\bibitem[Guo and Ruess, 2015b]{guo15b}
Guo, Y. and Ruess, M. (2015b).
\newblock Weak {Dirichlet} boundary conditions for trimmed thin isogeometric
  shells.
\newblock {\em Comput. Math. Appl.}, {\bf 70}(7):1425--1440.
\newblock High-Order Finite Element and Isogeometric Methods.

\bibitem[Hackl and Goodarzi, 2010]{Klaus10}
Hackl, K. and Goodarzi, M. (2010).
\newblock {\em Lecture note: An introduction to Linear Continuum Mechanics}.
\newblock Ruhr-University Bochum.

\bibitem[Helfrich, 1973]{helfrich73}
Helfrich, W. (1973).
\newblock Elastic properties of lipid bilayers: {T}heory and possible
  experiments.
\newblock {\em Z. Naturforsch.}, {\bf 28c}:693--703.

\bibitem[Hosseini et~al., 2014]{hosseini14}
Hosseini, S., Remmers, J.~J., Verhoosel, C.~V., and de~Borst, R. (2014).
\newblock An isogeometric continuum shell element for {non-linear} analysis.
\newblock {\em Comp. Meth. Appl. Mech. Engrg.}, {\bf 271}:1--22.

\bibitem[Hosseini et~al., 2013]{hosseini13}
Hosseini, S., Remmers, J. J.~C., Verhoosel, C.~V., and de~Borst, R. (2013).
\newblock An isogeometric solid--like shell element for nonlinear analysis.
\newblock {\em Int. J. Numer. Meth. Engrg.}, {\bf 95}(3):238--256.

\bibitem[Hughes and Carnoy, 1983]{hughes83}
Hughes, T.~J. and Carnoy, E. (1983).
\newblock Nonlinear finite element shell formulation accounting for large
  membrane strains.
\newblock {\em Comp. Meth. Appl. Mech. Engrg.}, {\bf 39}(1):69--82.

\bibitem[Hughes et~al., 2005]{hughes05}
Hughes, T. J.~R., Cottrell, J.~A., and Bazilevs, Y. (2005).
\newblock Isogeometric analysis: {CAD}, finite elements, {NURBS}, exact
  geometry and mesh refinement.
\newblock {\em Comp. Meth. Appl. Mech. Engrg.}, {\bf 194}:4135--4195.

\bibitem[Itskov, 2009]{itskov2009}
Itskov, M. (2009).
\newblock {\em Tensor Algebra and Tensor Analysis for Engineers}.
\newblock Springer-Verlag Berlin Heidelberg, 2$^{\text{nd}}$ edition.

\bibitem[Ivannikov et~al., 2014]{ivannikov14}
Ivannikov, V., Tiago, C., and Pimenta, P. (2014).
\newblock Meshless implementation of the geometrically exact
  {Kirchhoff-–Love} shell theory.
\newblock {\em Int. J. Numer. Meth. Engrg.}, {\bf 100}(1):1--39.

\bibitem[Kang and Youn, 2015]{kang15}
Kang, P. and Youn, S. (2015).
\newblock Isogeometric analysis of topologically complex shell structures.
\newblock {\em Finite Elem. Anal. Des.}, {\bf 99}:68--81.

\bibitem[Kiendl et~al., 2015a]{kiendl15b}
Kiendl, J., Auricchio, F., da~Veiga, L.~B., Lovadina, C., and Reali, A.
  (2015a).
\newblock Isogeometric collocation methods for the {Reissner-–Mindlin} plate
  problem.
\newblock {\em Comp. Meth. Appl. Mech. Engrg.}, {\bf 284}:489--507.
\newblock Isogeometric Analysis Special Issue.

\bibitem[Kiendl et~al., 2010]{kiendl10}
Kiendl, J., Bazilevs, Y., Hsu, M.-C., W{\"u}chner, R., and Bletzinger, K.-U.
  (2010).
\newblock The bending strip method for isogeometric analysis of
  {Kirchhoff-Love} shell structures comprised of multiple patches.
\newblock {\em Comput. Methods Appl. Mech. Engrg.}, {\bf
  199}(37-40):2403--2416.

\bibitem[Kiendl et~al., 2009]{kiendl09}
Kiendl, J., Bletzinger, K.-U., Linhard, J., and W{\"u}chner, R. (2009).
\newblock Isogeometric shell analysis with {Kirchhoff-Love} elements.
\newblock {\em Comput. Methods Appl. Mech. Engrg.}, {\bf 198}:3902--3914.

\bibitem[Kiendl et~al., 2015b]{kiendl15}
Kiendl, J., Hsu, M.-C., Wu, M.~C., and Reali, A. (2015b).
\newblock Isogeometric {Kirchhoff-–Love} shell formulations for general
  hyperelastic materials.
\newblock {\em Comp. Meth. Appl. Mech. Engrg.}, {\bf 291}:280--303.

\bibitem[Klinkel and Govindjee, 2002]{klinkel02}
Klinkel, S. and Govindjee, S. (2002).
\newblock Using finite strain 3d-material models in beam and shell elements.
\newblock {\em Engrg. Comput.}, {\bf 19}(8):902--921.

\bibitem[Lei et~al., 2015a]{lei15}
Lei, Z., Gillot, F., and Jezequel, L. (2015a).
\newblock Developments of the mixed grid isogeometric {Reissner-–Mindlin}
  shell: Serendipity basis and modified reduced quadrature.
\newblock {\em Eur. J. Mech. {A:Solids}}, {\bf 54}:105--119.

\bibitem[Lei et~al., 2015b]{lei15b}
Lei, Z., Gillot, F., and Jezequel, L. (2015b).
\newblock A multiple patches connection method in isogeometric analysis.
\newblock {\em Appl. Math. Model.}, {\bf 39}(15):4405--4420.

\bibitem[Macneal and Harder, 1985]{macneal85}
Macneal, R.~H. and Harder, R.~L. (1985).
\newblock A proposed standard set of problems to test finite element accuracy.
\newblock {\em Finite Elem. Anal. Des.}, {\bf 1}(1):3--20.

\bibitem[Morley and Morris, 1978]{morley78}
Morley, L. and Morris, A. (1978).
\newblock {\em Conflict Between Finite Elements and Shell Theory}.
\newblock Royal Aircraft Establishment (UK).

\bibitem[Munglani et~al., 2015]{munglani15}
Munglani, G., Vetter, R., Wittel, F., and Herrmann, H. (2015).
\newblock Orthotropic rotation-free thin shell elements.
\newblock {\em Comput. Mech.}, {\bf 56}(5):785--793.

\bibitem[Nagy et~al., 2013]{nagy13}
Nagy, A.~P., IJsselmuiden, S.~T., and Abdalla, M.~M. (2013).
\newblock Isogeometric design of anisotropic shells: Optimal form and material
  distribution.
\newblock {\em Comp. Meth. Appl. Mech. Engrg.}, {\bf 264}:145--162.

\bibitem[Nguyen et~al., 2014]{nguyen14nitsche}
Nguyen, V., Kerfriden, P., Brino, M., Bordas, S., and Bonisoli, E. (2014).
\newblock {Nitsche's} method for two and three dimensional {NURBS} patch
  coupling.
\newblock {\em Comput. Mech.}, {\bf 53}(6):1163--1182.

\bibitem[Nguyen-Thanh et~al., 2011]{thanh11}
Nguyen-Thanh, N., Kiendl, J., Nguyen-Xuan, H., W{\"u}chner, R., Bletzinger,
  K.-U., Bazilevs, Y., and Rabczuk, T. (2011).
\newblock Rotation free isogeometric thin shell analysis using {PHT}--splines.
\newblock {\em Comput. Methods Appl. Mech. Engrg.}, {\bf
  200}(47-48):3410--3424.

\bibitem[Nguyen-Thanh et~al., 2015]{nguyen15}
Nguyen-Thanh, N., Valizadeh, N., Nguyen, M., Nguyen-Xuan, H., Zhuang, X.,
  Areias, P., Zi, G., Bazilevs, Y., Lorenzis, L.~D., and Rabczuk, T. (2015).
\newblock An extended isogeometric thin shell analysis based on
  {Kirchhoff–-Love} theory.
\newblock {\em Comp. Meth. Appl. Mech. Engrg.}, {\bf 284}:265--291.
\newblock Isogeometric Analysis Special Issue.

\bibitem[Noels and Radovitzky, 2008]{noels08}
Noels, L. and Radovitzky, R. (2008).
\newblock A new discontinuous galerkin method for {Kirchhoff-–Love} shells.
\newblock {\em Comp. Meth. Appl. Mech. Engrg.}, {\bf 197}(33–40):2901--2929.

\bibitem[O{\~{n}}ate and Z{\'{a}}rate, 2000]{onate00}
O{\~{n}}ate, E. and Z{\'{a}}rate, F. (2000).
\newblock Rotation-free triangular plate and shell elements.
\newblock {\em Int. J. Numer. Meth. Engrg.}, {\bf 47}(1-3):557--603.

\bibitem[Reali and Gomez, 2015]{reali15}
Reali, A. and Gomez, H. (2015).
\newblock An isogeometric collocation approach for {Bernoulli-–Euler} beams
  and {Kirchhoff} plates.
\newblock {\em Comp. Meth. Appl. Mech. Engrg.}, {\bf 284}:623--636.
\newblock Isogeometric Analysis Special Issue.

\bibitem[{Riffnaller-Schiefer} et~al., 2016]{riffnaller16}
{Riffnaller-Schiefer}, A., Augsd{\" o}rfer, U., and Fellner, D. (2016).
\newblock Isogeometric shell analysis with {NURBS} compatible subdivision
  surfaces.
\newblock {\em Appl. Math. Comput.}, {\bf 272}, Part 1:139--147.
\newblock Subdivision, Geometric and Algebraic Methods, Isogeometric Analysis
  and Refinability.

\bibitem[Roohbakhshan et~al., 2016]{biomembrane}
Roohbakhshan, F., Duong, T.~X., and Sauer, R.~A. (2016).
\newblock A projection method to extract biological membrane models from {3D}
  material models.
\newblock {\em J. Mech. Behav. Biomed. Mater.}, {\bf 58}:90--104.

\bibitem[Sauer, 2014]{droplet}
Sauer, R.~A. (2014).
\newblock Stabilized finite element formulations for liquid membranes and their
  application to droplet contact.
\newblock {\em Int. J. Numer. Meth. Fluids}, {\bf 75}(7):519--545.

\bibitem[Sauer and Duong, 2015]{shelltheo}
Sauer, R.~A. and Duong, T.~X. (2015).
\newblock On the theoretical foundations of solid and liquid shells.
\newblock {\em Math. Mech. Solids}, doi:10.1177/1081286515594656.

\bibitem[Sauer et~al., 2014]{membrane}
Sauer, R.~A., Duong, T.~X., and Corbett, C.~J. (2014).
\newblock A computational formulation for solid and liquid membranes based on
  curvilinear coordinates and isogeometric finite elements.
\newblock {\em Comput. Methods Appl. Mech. Engrg.}, {\bf 271}:48--68.

\bibitem[Sauer et~al., 2017]{liquidshell}
Sauer, R.~A., Duong, T.~X., Mandadapu, K.~K., and Steigmann, D.~J. (2017).
\newblock A stabilized finite element formulation for liquid shells and its
  application to lipid bilayers.
\newblock {\em J. Comput. Phys.}, {\bf 330}:436--466.

\bibitem[Schillinger et~al., 2012]{schill12}
Schillinger, D., Dede, L., Scott, M.~A., Evans, J.~A., Borden, M.~J., Rank, E.,
  and Hughes, T. J.~R. (2012).
\newblock An isogeometric design-through-analysis methodology based on adaptive
  hierarchical refinement of nurbs, immersed boundary methods, and {T-spline}
  {CAD} surfaces.
\newblock {\em Comput. Methods Appl. Mech. Engrg.}, {\bf 249}:116--150.

\bibitem[Scott et~al., 2011]{scott11}
Scott, M.~A., Borden, M.~J., Verhoosel, C.~V., Sederberg, T.~W., and Hughes, T.
  J.~R. (2011).
\newblock Isogeometric finite element data structures based on {B}\'ezier
  extraction of {T}-splines.
\newblock {\em Int. J. Numer. Meth. Engng.}, {\bf 88}(2):126--156.

\bibitem[Simo and Fox, 1989]{simo89}
Simo, J.~C. and Fox, D.~D. (1989).
\newblock On a stress resultant geometrically exact shell model. {P}art {I}:
  {F}ormulation and optimal parameterization.
\newblock {\em Comput. Meth. Appl. Mech. Engrg.}, {\bf 72}:267--304.

\bibitem[Simo et~al., 1990]{simo90}
Simo, J.~C., Fox, D.~D., and Rifai, M.~S. (1990).
\newblock On a stress resultant geometrically exact shell model. {P}art {III}:
  {C}omputational aspects of the nonlinear theory.
\newblock {\em Comput. Meth. Appl. Mech. Engrg.}, {\bf 79}:21--70.

\bibitem[Steigmann, 2013]{steigman13}
Steigmann, D. (2013).
\newblock Koiter's shell theory from the perspective of three-dimensional
  nonlinear elasticity.
\newblock {\em J. Elast.}, {\bf 111}(1):91--107.

\bibitem[Steigmann, 1999a]{steigmann99b}
Steigmann, D.~J. (1999a).
\newblock Fluid films with curvature elasticity.
\newblock {\em Arch. Rat. Mech. Anal.}, {\bf 150}:127--152.

\bibitem[Steigmann, 1999b]{steigmann99}
Steigmann, D.~J. (1999b).
\newblock On the relationship between the {C}osserat and {K}irchhoff-{L}ove
  theories of elastic shells.
\newblock {\em Math. Mech. Solids}, {\bf 4}:275--288.

\bibitem[Stolarski et~al., 2013]{stolarski13}
Stolarski, H., Gilmanov, A., and Sotiropoulos, F. (2013).
\newblock Nonlinear rotation-free three-node shell finite element formulation.
\newblock {\em Int. J. Numer. Meth. Engrg.}, {\bf 95}(9):740--770.

\bibitem[Sze et~al., 2004]{sze04}
Sze, K., Liu, X., and Lo, S. (2004).
\newblock Popular benchmark problems for geometric nonlinear analysis of
  shells.
\newblock {\em Finite Elem. Anal. Des.}, {\bf 40}(11):1551--1569.

\bibitem[Tepole et~al., 2015]{tepole15}
Tepole, A.~B., Kabaria, H., Bletzinger, K.-U., and Kuhl, E. (2015).
\newblock Isogeometric {Kirchhoff-–Love} shell formulations for biological
  membranes.
\newblock {\em Comp. Meth. Appl. Mech. Engrg.}, {\bf 293}:328--347.

\bibitem[Thai et~al., 2012]{thai12}
Thai, C.~H., Nguyen-Xuan, H., Nguyen-Thanh, N., Le, T.-H., Nguyen-Thoi, T., and
  Rabczuk, T. (2012).
\newblock Static, free vibration, and buckling analysis of laminated composite
  {Reissner-–Mindlin} plates using nurbs-based isogeometric approach.
\newblock {\em Int. J. Numer. Meth. Engrg.}, {\bf 91}(6):571--603.

\bibitem[Ugural, 2009]{ugural09}
Ugural, A.~C. (2009).
\newblock {\em Stresses in beams, plates, and shells}.
\newblock CRC Press, Boca Raton.

\bibitem[Uhm and Youn, 2009]{uhm09}
Uhm, T.-K. and Youn, S.-K. (2009).
\newblock T-spline finite element method for the analysis of shell structures.
\newblock {\em Int. J. Numer. Meth. Engrg.}, {\bf 80}(4):507--536.

\bibitem[Wriggers, 2008]{wriggers-fee}
Wriggers, P. (2008).
\newblock {\em Nonlinear Finite Element Methods}.
\newblock Springer, Berlin.

\bibitem[Yang et~al., 2000]{yang00}
Yang, H. T.~Y., Saigal, S., Masud, A., and Kapania, R.~K. (2000).
\newblock A survey of recent shell finite elements.
\newblock {\em Int. J. Numer. Meth. Engng}, {\bf 47}:101--127.

\bibitem[Z\'{a}rate and O\~{n}ate, 2012]{zarate12}
Z\'{a}rate, F. and O\~{n}ate, E. (2012).
\newblock Extended rotation-free shell triangles with transverse shear
  deformation effects.
\newblock {\em Comput. Mech.}, {\bf 49}(4):487--503.

\end{thebibliography}
